\documentclass[preprint, prd, aps, showpacs, preprintnumbers, amsmath, amssymb]{revtex4-1}
\usepackage{graphics, epsfig, subfigure}
\usepackage{diagbox}
\usepackage[usenames]{color}
\usepackage{color}
\usepackage{graphicx}
\usepackage{amsfonts}
\usepackage{indentfirst}
\usepackage{booktabs}
\usepackage{hyperref}
\hypersetup{hypertex=ture, backref=true, colorlinks=ture, linkcolor=blue, anchorcolor=blue, citecolor=blue}
\usepackage{float}
\usepackage{multirow}

\begin{document}
\renewcommand{\baselinestretch}{1.3}
\newcommand\beq{\begin{equation}}
\newcommand\eeq{\end{equation}}
\newcommand\beqn{\begin{eqnarray}}
\newcommand\eeqn{\end{eqnarray}}
\newcommand\nn{\nonumber}
\newcommand\fc{\frac}
\newcommand\lt{\left}
\newcommand\rt{\right}
\newcommand\pt{\partial}

\title{\Large \bf Hybrid Proca-boson stars}
\author{Tian-Xiang Ma, 
Chen Liang,  Jie Yang\footnote{yangjiev@lzu.edu.cn, corresponding author
}
and Yong-Qiang Wang\footnote{yqwang@lzu.edu.cn, corresponding author
}
}

\affiliation{ $^{1}$Lanzhou Center for Theoretical Physics, 
Key Laboratory of Theoretical Physics of Gansu Province, 
	School of Physical Science and Technology, Lanzhou University, Lanzhou 730000, China\\
	$^{2}$Institute of Theoretical Physics $\&$ Research Center of Gravitation, 
	Lanzhou University, Lanzhou 730000, China}

\begin{abstract}
	In this paper, we construct a hybrid boson star model that contains a complex scalar field and a Proca field. The scalar field is in the ground state, while the Proca field is in the first excited state. We numerically solve the model and obtain solution families of different coexisting states by considering both synchronized and nonsynchronized cases. By examining the relation between ADM mass and synchronized frequency $\tilde{\omega}$ or nonsynchronized frequency $\tilde{\omega}_P$, we identify several types of solution families for the hybrid boson stars.
	In addition to solutions that intersect the scalar field and the Proca field at each end, 
	there are also several types of multi-branch coexisting state solutions. 
	The characteristics of various solutions are analyzed and discussed in detail. 	
	We calculate the binding energy $E$ of the hybrid Proca-boson stars and provide the relationship between $E$ and both synchronized frequency $\tilde{\omega}$ and nonsynchronized frequency $\tilde{\omega}_P$. Furthermore, we obtain the stability of the corresponding hybrid  star solution families from these analyses  above.
\end{abstract}

\maketitle

\section{Introduction}\label{Sec1}

The latest cosmological data suggest that about $26\%$ of cosmic content is dark matter(DM)~\cite{raveri2017partially}. However, the basic nature of DM is ambiguous. Among the many assumptions there are several different views, some people believe that DM consists of weakly interacting massive particles~\cite{hindmarsh_dark_2005}, or the mass of these dark matter corresponding is derived from the primordial black holes~\cite{carr_primordial_2020}. In addition to these two models, there is a novel idea that boson stars could also be a candidate for dark matter~\cite{eby_boson_2016,chen_new_2021,sharma_boson_2008}, namely macroscopic Bose-Einstein condensates formed by super-light bosons/fields under gravity. The gravitational structure and mass distribution of different scalar dark matter in the universe can be explained by changing the mass of the scalar field or introducing the self-interaction term~\cite{ryan_spinning_1997} to change some properties of the boson star. The boson star model has become one of the important candidates for dark matter since it plays an important role in studying the dynamics of early star clusters~\cite{torres_supermassive_2000}, the rotation curves of low surface brightness galaxies~\cite{swaters_high-resolution_2000}, the kinematics of the galactic center~\cite{mayer_formation_2004}, and the formation of supermassive black holes~\cite{annulli_response_2020}. 

The study of the boson-star model can be traced back to the 1960s. D. J. Kaup coupled the complex scalar field with the four-dimensional Einstein gravity~\cite{kaup_klein-gordon_1968}, and then obtained the spherical symmetric solution of the Einstein-Klein-Gordon equation. In the same period, R. Ruffini and S. Bonazzola solved the model of coupling real scalar field and gravity~\cite{ruffini_systems_1969}, they also obtained the same solution. Later, the soliton solution of the object formed by the scalar field under the gravitational interaction was called boson star. Since then, various studies have been conducted around the boson star model. The self-interacting boson star can be obtained by adding the self-interacting term (quartic term or sextic term) to the Lagrangian density~\cite{li_self-interacting_2021,kling_profiles_2018,schunck_boson_2000,sanchis-gual_self-interactions_2022}. A charged boson star is obtained by coupling a complex scalar field with an electromagnetic field~\cite{jetzer_stability_1989,jetzer_charged_1993,jetzer_charged_1989,garcia_charged_2016,kumar_boson_2014}. The Newtonian boson star is obtained by solving the Einstein-Klein-Gordon equation in the Newtonian limit~\cite{harrison_numerical_2002,harrison_numerical_2002,silveira_boson_1995}. In addition, others have studied rotating boson stars with angular momentum~\cite{li2020rotating,yoshida_rotating_1997,kleihaus_rotating_2005,siemonsen_stability_2021}. Gravity can also be coupled to a field with a non-zero spin. In 2015, Brito et al. studied the static solution of a system with a Proca field (spin 1) coupled to gravity, called a Proca star~\cite{brito_Proca_2016}. Soon after, I. Salazarlandea and F. Garciaka constructed models of charged Proca stars~\cite{garcia_charged_2016}. In addition to coupling to the boson fields, Finster et al.~\cite{finster_particle-like_1999} also constructed a spherically symmetric Dirac star coupled by two spin 1/2 spinor fields and Einstein’s gravity. Ref.~\cite{guerra_axion_2019,zeng_rotating_2021,delgado_rotating_2020} considers a system coupled by an axion field and a complex scalar field, called an axion boson star (ABS). Their work greatly enriched the boson-star model and allowed the development of this research field to flourish. In recent years, with the development of astrophysics, boson stars are also considered to be one of the candidates for dark matter, in addition, it has been widely used in black hole shadow simulation~\cite{cunha_chaotic_2016,grould_comparing_2017} and analysis of gravitational wave signals~\cite{croon_boson_2019,dietrich_full_2019,bezares_gravitational_2018,bustillo_searching_2022}. 

Other recent studies have shown systems in which gravity is coupled to multiple matter fields, called multi-state boson stars. In Ref.~\cite{bernal_multi-state_2010}, Bernal et al. constructed a system consisting of two complex scalar fields, the ground state and the first excited state, respectively. Later, Ref.~\cite{li2020rotating,li_self-interacting_2021} studied the rotating multi-state boson stars. The matter field of spherically symmetric boson stars can also be extended to an odd number of complex scalar fields~\cite{alcubierre_ell-boson_2018,alcubierre_dynamical_2019}. 

Some recent work has shown that Proca star plays an important role in the simulation of black hole shadow~\cite{cunha_chaotic_2016,vincent_imaging_2016} and gravitational wave signal analysis~\cite{bustillo_searching_2022}, etc. Multi-field models including Proca field were also studied in Ref.~\cite{delgado_kerr_2021}. The multi-field boson stars model studied by previous people is mostly composed of two complex scalar fields~\cite{li_self-interacting_2021}. In the Ref.~\cite{liang_dirac-boson_2022}, a spherically symmetric boson star solution coupled by a complex scalar field and two Fermi fields (spin 1/2) is also studied. The main work of this paper is to study the coexisting state solutions coupled by two boson fields with different spins (spin 0 and 1). The aim of this work is to solve Einstein-Proca-Klein-Gordon equation numerically, construct a spherically symmetric boson stars composed of a Proca field and a complex scalar field, and study the properties of its solutions. 

This paper is organized as follows. In Sec.~\ref{Sec2}, we propose a four-dimensional Einstein gravity model coupled with a complex scalar field and a Proca field. In Sec.~\ref{Sec3}, the boundary conditions of Proca boson stars are studied. In Sec.~\ref{Sec4}, we show the numerical results obtained by solving the model, and show the properties of the polymorphic solutions in two different cases. In Sec.~\ref{Sec5}, We summarize and describe possible future work. 

\section{The model setup}\label{Sec2}
We consider the minimal coupling of Proca field and complex scalar field to 3+1 dimensional Einstein gravity. The action is given by
\begin{equation}\label{equ1}
	S=\int \sqrt{-g} d^{4} x\left(\frac{R}{16 \pi G}+\mathcal{L}_{S}+\mathcal{L}_{P}\right), 
\end{equation}
where $G$ is the gravitational constant, $R$ is the Ricci scalar, $\mathcal{L}_{S}$ and $\mathcal{L}_{P}$ respectively represent the Lagrangian of the scalar field and Proca field, and their specific forms are
\begin{equation}\label{equ2}
	\mathcal{L}_{S}=-g^{\alpha \beta} \bar{\Phi}_{, \alpha} \Phi_{, \beta}-\mu^{2} \bar{\Phi} \Phi, 
    \quad \mathcal{L}_{P}=-\frac{1}{4} {\mathcal{F}}_{\alpha \beta} \overline{{\mathcal{F}}}^
    {\alpha \beta}-\frac{1}{2} \mu^{2} {\mathcal{A}}_{\alpha} \overline{\mathcal{A}}^{\alpha}, 
\end{equation}
where $\Phi$ and $\mathcal{A}$ are complex scalar and Proca fields respectively, $\bar{\Phi}$ and $\bar{\mathcal{A}}$ are complex conjugates of their corresponding fields, $\mathcal{F}=d \mathcal{A}$. 

The corresponding dynamic tensors can be obtained from Lagrangian, where $T_{\alpha \beta}^{S}$ and $T_{\alpha \beta}^{P}$ represent the dynamic tensors of scalar field and Proca field respectively
\begin{equation}\label{equ3}
	T_{\alpha \beta}^{S}=\bar{\Phi}_{, \alpha} \Phi_{, \beta}+\bar{\Phi}_{, \beta} \Phi_{, \alpha}-
	g_{\alpha \beta}\left[\frac{1}{2} g^{\gamma \delta}\left(\bar{\Phi}_{, \gamma} \Phi_{, \delta}+
	\bar{\Phi}_{, \delta} \Phi_{, \gamma}\right)+\mu^{2} \bar{\Phi} \Phi\right], 
\end{equation}
\begin{equation}\label{equ4}
	T_{\alpha \beta}^{P}=\frac{1}{2}\left(\mathcal{F}_{\alpha \sigma} \overline{\mathcal{F}}_{\beta \gamma}+
	\overline{\mathcal{F}}_{\alpha \sigma} \mathcal{F}_{\beta \gamma}\right) g^{\sigma \gamma}-
	\frac{1}{4} g_{\alpha \beta} \mathcal{F}_{\sigma \tau} \overline{\mathcal{F}}^{\sigma \tau}+
	\frac{1}{2} \mu^{2}\left[\mathcal{A}_{\alpha} \overline{\mathcal{A}}_{\beta}+
	\overline{\mathcal{A}}_{\alpha} \mathcal{A}_{\beta}-g_{\alpha \beta} \mathcal{\mathcal { A }}
	_{\sigma} \overline{\mathcal{A}}^{\sigma}\right]. 
\end{equation}
	
The field equation is obtained by the variation of the Lagrange
\begin{equation}\label{equ5}
	R_{\alpha \beta}-\frac{1}{2} g_{\alpha \beta} R=8 \pi G\left(T_{\alpha \beta}^{S}+
	T_{\alpha \beta}^{P}\right), 
\end{equation}
\begin{equation}\label{equ6}
	\nabla^{2} \Phi-\mu_{S}^{2} \Phi=0, 
\end{equation}
\begin{equation}\label{equ7}
	\nabla_{\alpha} \mathcal{F}^{\alpha \beta}-\mu_{p}^{2} \mathcal{A}^{\beta}=0. 
\end{equation}
	
The action of the matter fields are invariant under the $U(1)$ transformation $\Phi\rightarrow e^{i\alpha}\Phi$, $\mathcal{A}^{\beta}\rightarrow e^{i\alpha}\mathcal{A}^{\beta}$ with a constant $\alpha$. According to Noether's theorem, there are conserved currents corresponding to these two matter fields:
\begin{equation}\label{equ8}
	J_S^{\mu} = -i\left(\Phi^*\partial^\mu\Phi - \Phi\partial^\mu\Phi^*\right), 
	\qquad J_P^{\mu} = \frac{i}{2}\left[\overline{\mathcal{F}}^{\alpha \beta} \mathcal{A}_
	{\beta}-\mathcal{F}^{\alpha \beta} \overline{\mathcal{A}}_{\beta}\right] . 
\end{equation}

We can integrate the timelike component of these conserved currents on a spacelike hypersurface $\varSigma$, then there obtain the Noether charges:
\begin{equation}\label{equ9}
	Q_S = \int_{\varSigma}J_S^t\, , \qquad Q_P = \int_{\varSigma}J_P^t\, . 
\end{equation}

We use ansatz corresponding to the static spherically symmetric Schwarzschild metric of the following form to solve Proca boson star
\begin{equation}\label{equ10}
	ds^2 = -N(r)\sigma^2(r)dt^2 + \frac{dr^2}{N(r)} + r^2\left(d\theta^2 + \sin^2\theta d\varphi^2\right), 
\end{equation}
where $N(r) = 1 - {2m(r)}/{r}$, function $m(r)$ and $\sigma(r)$ only depend on the radial variable $r$. In addition, for the static spherically symmetric system, we use the following ansatz of the scalar and Proca fields~\cite{brito_Proca_2016}:
\begin{equation}\label{equ11}
	\Phi = \phi(r)e^{-i\omega_St}, 
\end{equation}
\begin{equation}\label{equ12}
	\mathcal{A}=[F(r) d t+i G(r) d r] e^{-i \omega_P t}
\end{equation}
where $\phi(r)$, $F(r)$ and $G(r)$ are real functions. Besides, the constants $\omega_S$ and $\omega_P$ are the frequency of the scalar and Proca fields, respectively. When  $\omega_S$ and $\omega_P$ meet $\omega_S = \omega_P = \omega$, we call $\omega$ as the synchronized frequency. When $\omega_S\ne\omega_P$, these two frequencies are called nonsynchronized frequencies. 
	
Substituting the above ansatz into the field equations~(\ref{equ5}--\ref{equ7}), we can get the following equations 
for $\phi(r)$, $F(r)$, $G(r)$, $m(r)$ and $\sigma(r)$:
	
\begin{equation}\label{equ13}
	\phi^{\prime\prime}+\left(\frac{2}{r} + \frac{N^\prime}{N} + \frac{\sigma^\prime}{\sigma}\right)\phi^\prime + \left(\frac{\omega_S^2}{N\sigma^2} - \mu_S^2\right)\frac{\phi}{N} = 0\, , 
\end{equation}
\begin{equation}\label{equ14}
	\frac{d}{d r}\left\{\frac{r^{2}\left[F^{\prime}-\omega_P G\right]}{\sigma}\right\}=\frac{\mu_P^{2} r^{2} F}{\sigma N}, 
\end{equation}
\begin{equation}\label{equ15}
	\omega_P G-F^{\prime}=\frac{\mu_P^{2} \sigma^{2} N G}{\omega_P}, 
\end{equation}
\begin{equation}\label{equ16}
	m^\prime = r^2N\phi_n^{\prime2} + \left(\mu_S^2 + \frac{\omega_S^2}{N\sigma}\right)r^2\phi^2 + \frac{r^2\left( F^{\prime}-\omega_P G\right)^{2}}{2 \sigma^{2}}+\frac{\mu_P^{2} r^2}{2}\left(G^{2} N+\frac{F^{2}}{N \sigma^{2}}\right), 
\end{equation}
\begin{equation}\label{equ17}
	\frac{\sigma^\prime}{\sigma} = 2r\left(\phi^{\prime2} + \frac{\omega_S^2\phi^2}{N^2\sigma^2}\right) + \mu_P^{2}r\left(G^{2}+\frac{F^{2}}{N^{2} \sigma^{2}}\right). 
\end{equation}
Also, the specific forms of the Noether charges obtained from Eq.~(\ref{equ8}) and Eq.~(\ref{equ9}) are
\begin{equation}\label{equ18}
	Q_S = 8\pi\int_0^\infty r^2\frac{\omega_S\phi^2}{N\sigma}dr\, , \qquad Q_P = 8\pi\int_0^\infty r^2\frac{\left(w G-F^{\prime}\right) G}{\sigma}. 
\end{equation}
	
\section{Boundary conditions}\label{Sec3}
In order to solve the equations of ordinary differential equations obtained in the previous section, we need to give corresponding boundary conditions for each unknown function. Since they are asymptotically flat solutions, the metric functions $m(r)$ and $\sigma(r)$ need to satisfy the boundary conditions:
\begin{equation}\label{equ19}
m(0) = 0, \qquad \sigma(0) = \sigma_0, \qquad m(\infty) = M, \qquad \sigma(\infty) = 1, 
\end{equation}
where $M$ and $\sigma_0$ are currently unknown, the values of these two quantities can be obtained after finding the solution of the differential equation system. 
For the matter field functions, at infinity we require
\begin{equation}\label{equ20}
\phi(\infty) = 0, \qquad F(\infty) = 0, \qquad G(\infty) = 0. 
\end{equation}
Additionally, by considering the form of the field equation~(\ref{equ13}--\ref{equ15}) at the origin, we can obtain the following boundary conditions satisfied by the field functions:
\begin{equation}\label{equ21}
\left. \frac{d\phi(r)}{dr}\right|_{r = 0} = 0, 
\left. \frac{dF(r)}{dr}\right|_{r = 0} = 0, 
\qquad G(0) = 0, \qquad . 
\end{equation}

\section{Numerical results}\label{Sec4}
To facilitate numerical calculations, we use dimensionless quantities:
\begin{equation}\label{equ22}
\begin{split}
\tilde{r} = r/\rho, \quad \tilde{\phi} = \frac{\sqrt{4\pi}}{M_{Pl}}\phi, 
\quad \tilde{\omega}_S = \omega_S\rho, \quad \tilde{\mu}_S = \mu_S\rho, 
\quad\\ \tilde{F} = \frac{\sqrt{4\pi}}{M_{Pl}}F, \quad \tilde{G} = \frac{\sqrt{4\pi}}{M_{Pl}}G, 
\quad \tilde{\omega}_P = \omega_P\rho, \quad \tilde{\mu}_P = \mu_P\rho, 
\end{split}
\end{equation}
where $M_{Pl} = 1/\sqrt{G}$ is the Planck mass, 
$\rho$ is a positive constant whose dimension is length, 
we let the constant $\rho$ be $1/\mu_S$. Additionally, we introduce a new radial variable
\begin{equation}\label{equ23}
x = \frac{\tilde{r}}{1+\tilde{r}}. 
\end{equation}
where the radial coordinate $\tilde{r}\in[0, \infty)$, so $x\in[0, 1]$. We numerically solve the system of differential equations based on the finite element method, using 1000 grid points in the integration region $0 \le x \le 1$. The iterative method we use is the Newton-Raphson method, and to ensure that the calculation results are correct, we require the relative error to be less than $10^{-5}$. 

In the model constructed by us, the ground state scalar field function $\phi$ has no node in the radial direction, so it is represented by $S_0$, and the script is the total number of radial nodes of the field function. The first excited state field function $F$ and $G$ of Proca field have a total of one node in the radial direction, so it is represented by $P_1$, and the meaning of the script is the same as that of the scalar field. So in this model we represent the coexisting state of the scalar field and Proca field in terms of $S_0P_1$. Next, we will analyze the different classifications of solution families of Proca-boson stars in detail. 

\subsection{Synchronized frequency}
According to the characteristics of mixed state solutions, we divide the synchronized frequency solution families into three categories: the one-branch solution family, the two-branch solution family and the multi-branch solution family. There is a one-to-one correspondence between the mixed state one-branch solution and the synchronized frequency $\tilde{\omega}$, but for the two-branch solution, when $\tilde{\omega}$ is valued in some ranges, one $\tilde{\omega}$ corresponds to two different solutions; Similarly, multi-branch solutions will have one $\tilde{\omega}$ corresponding to more than two solutions. According to our numerical results, when $0. 808<\tilde{\mu}_P<1$, the mixed state solution is of one-branch type. When $0. 801<\tilde{\mu}_P\le0. 808$, the mixed state solution is of multi-branch type. When $0. 772\le\tilde{\mu}_P\le0. 801$, the mixed state solution is of two-branch type. We will explore these families of solutions in detail next. 
\subsubsection{One-Branch}
\begin{figure}[!htbp]
    \begin{center}
    \includegraphics[height=.24\textheight]{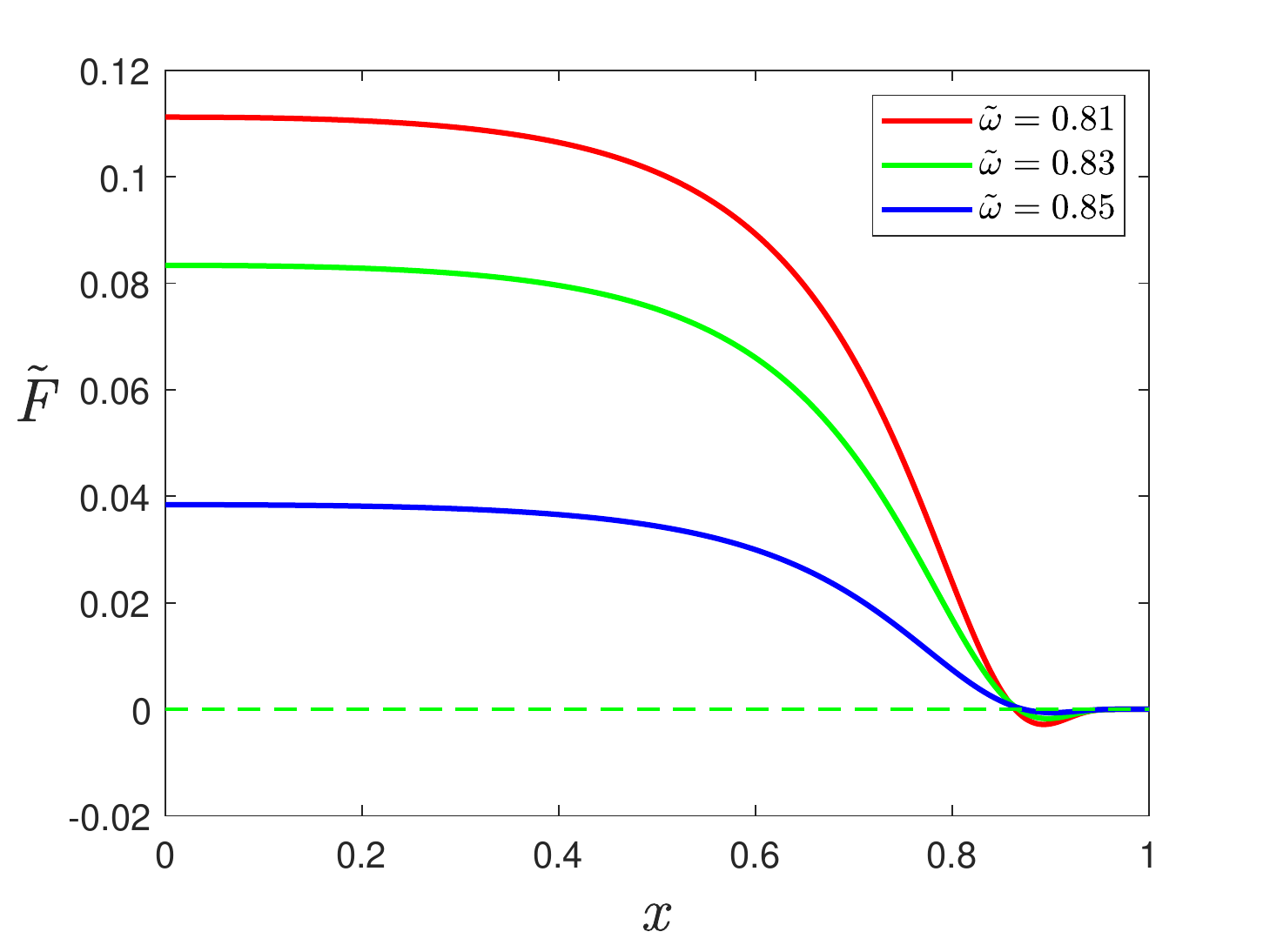}
    \includegraphics[height=.24\textheight]{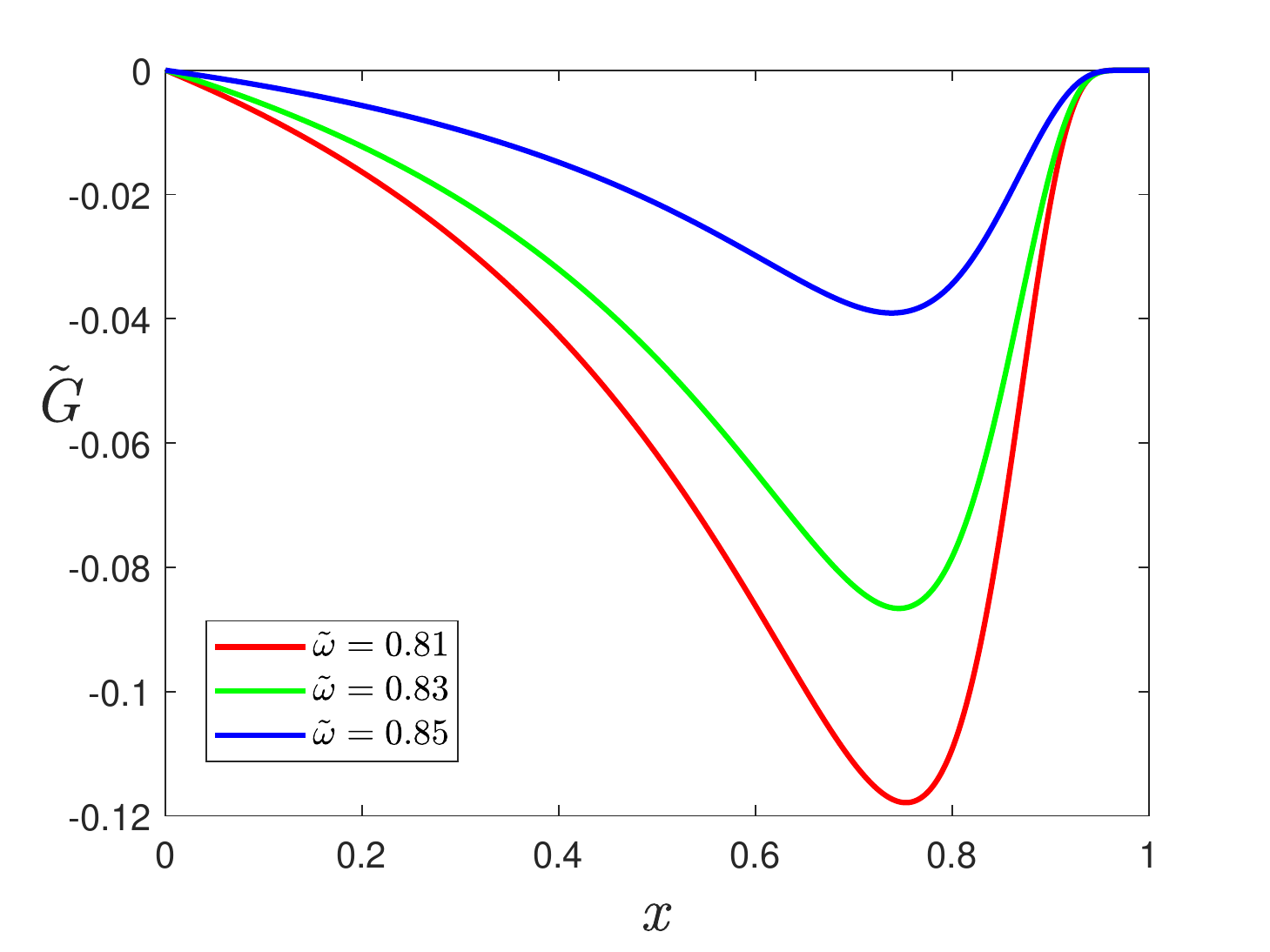}
    \includegraphics[height=.24\textheight]{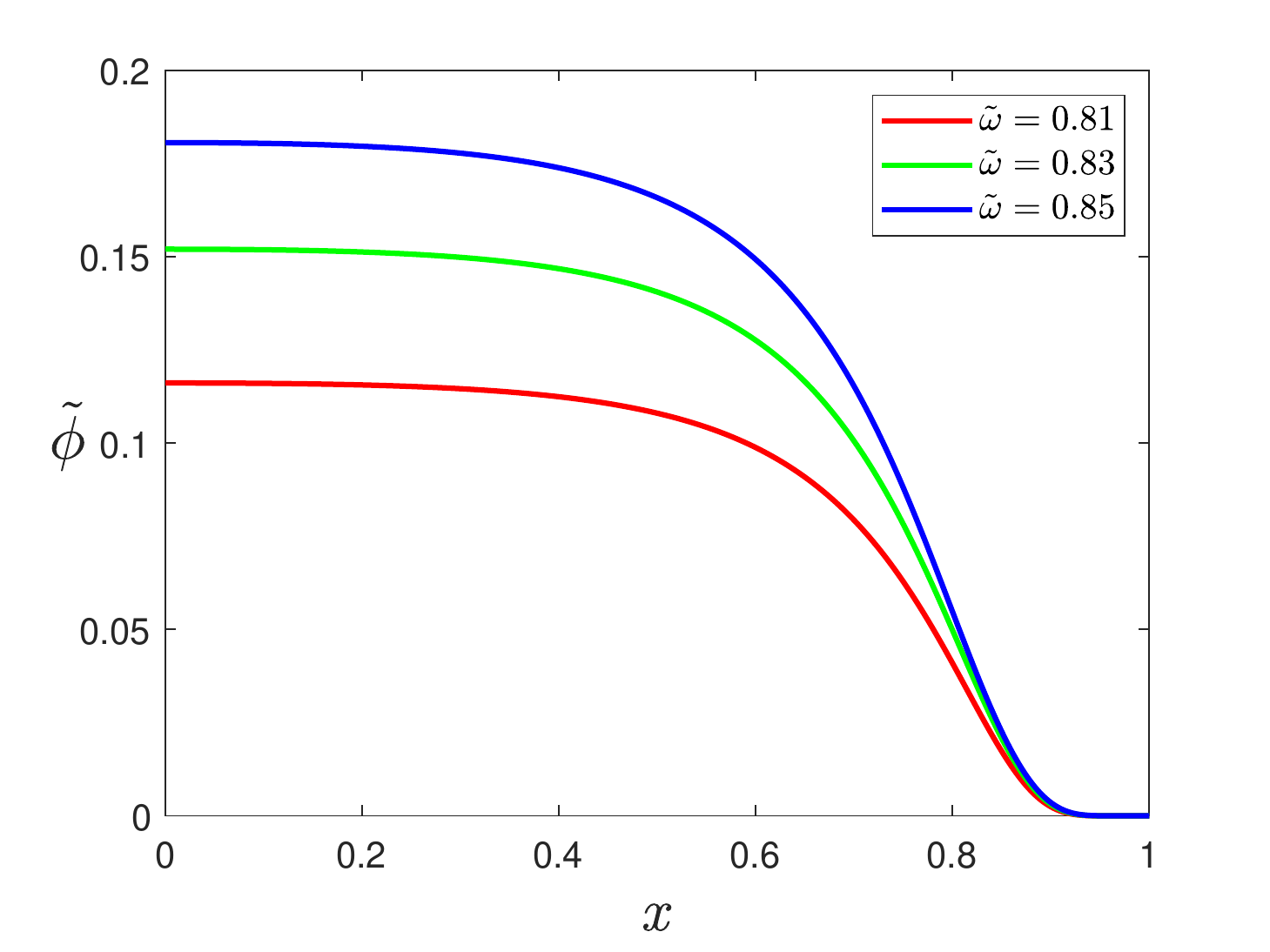}
    \end{center}
    \vspace{-2em}
    \caption{Proca field functions $\tilde{F}$ and $\tilde{G}$(top panels)
    and scalar field function $\tilde{\phi}$(bottom panel) 
    as functions of $x$ for $\tilde{\omega} = \tilde{\omega}_S = \tilde{\omega}_P 
    = 0. 81, 0. 83, 0. 85$. All solutions have $\tilde{\mu}_P= 0. 92$ and $\tilde{\mu}_S=1$. }
    \label{field-one-synchronized}
\end{figure}

The relation between the field function $\tilde{F}$, $\tilde{G}$, $\tilde{\phi}$ and the synchronized frequency $\tilde{\omega}$ is shown in Fig.~\ref{field-one-synchronized}. For scalar field functions, $\left\lvert \tilde{\phi} \right\rvert _{max}$ increases as $\tilde{\omega}$ increases; For Proca field functions, $\left\lvert \tilde{F} \right\rvert _{max}$ and $\left\lvert \tilde{G} \right\rvert _{max}$ decrease as $\tilde{\omega}$ increases. According to the analysis of Fig.~\ref{field-one-synchronized}, the function $\tilde{F}$ has one node, $\tilde{G}$ and $\tilde{\phi}$ have no node, which means that the mixed state is $S_0 P_1$ state, that is, the corresponding scalar field is in the ground state, and Proca field is in the first excited state. Next we will examine the properties of the $S_0 P_1$ state in detail. 
 
\begin{figure}[!htbp]
    \begin{center}
    \hbox to\linewidth{\hss
    \resizebox{9cm}{6.5cm}{\includegraphics{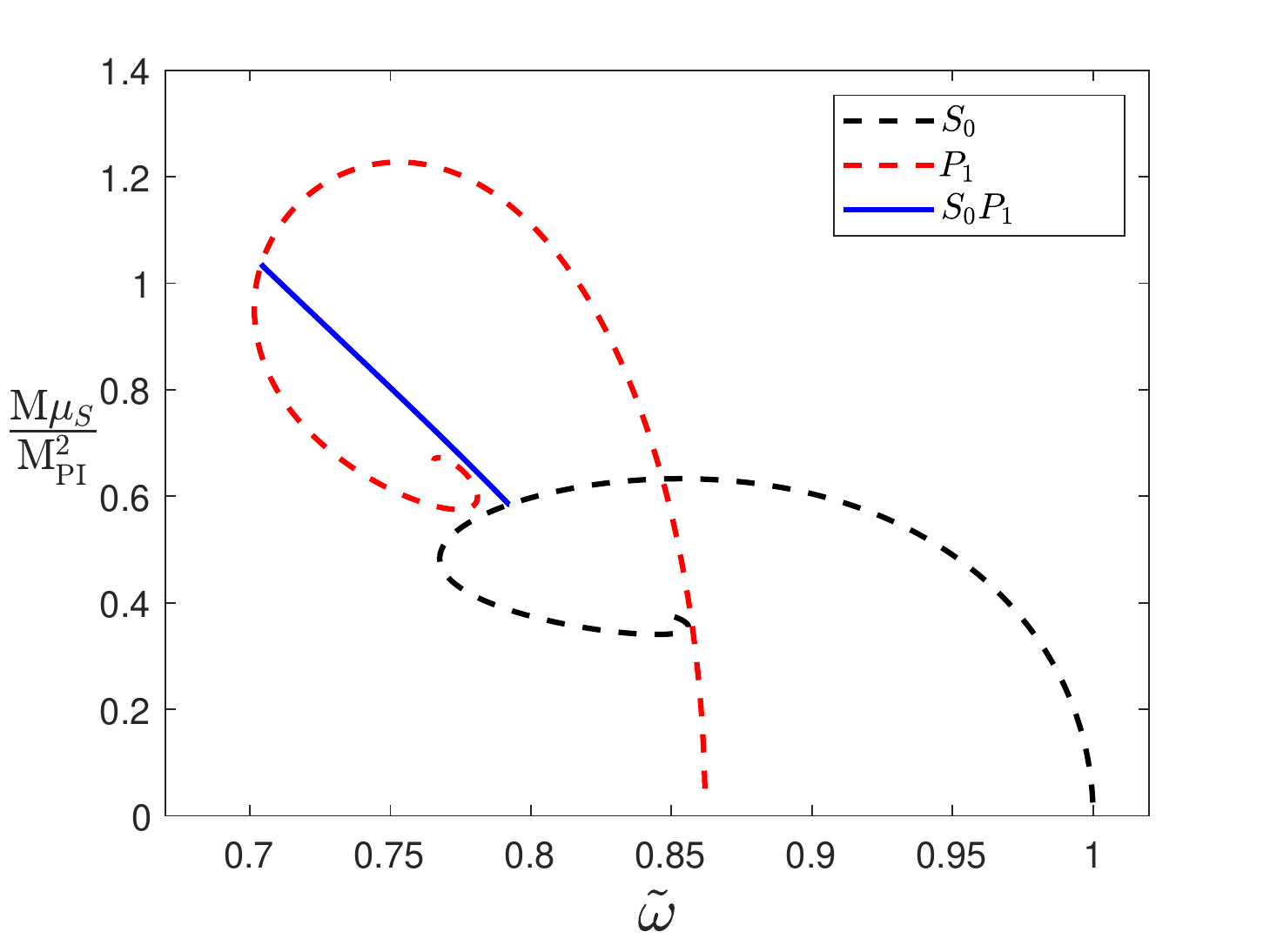}}
    \resizebox{9cm}{6.5cm}{\includegraphics{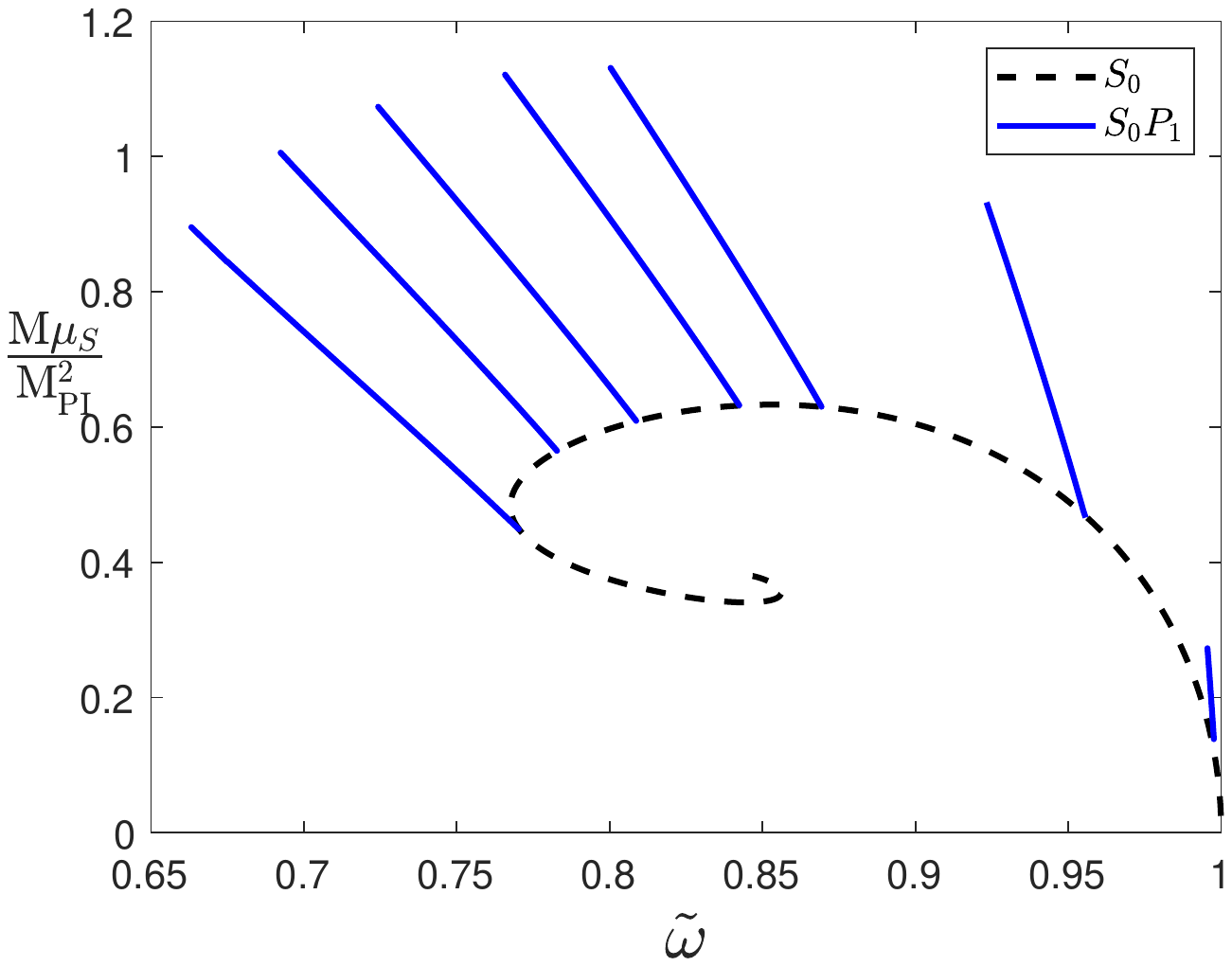}}
    \hss}
    \end{center}
    \vspace{-2em}
    \caption{Left: The ADM mass $M$ as a function of the synchronized frequency $\tilde{\omega}$. 
    The black dashed line represents the $S_0$ state solutions with $\tilde{\mu}_S = 1$, 
    the red dashed line represents the $P_1$ state solutions with $\tilde{\mu}_P=0. 862$, 
    and the blue line denote the coexisting state $S_0P_1$ with $\tilde{\mu}_S = 1$ and $\tilde{\mu}_P=0. 862$. 
    Right:The ADM mass $M$ as a function of the synchronized frequency $\tilde{\omega}$. 
    The black dashed line represents the $S_0$ state solutions with $\tilde{\mu}_S = 1$. 
    The blue lines from left to right represent the coexisting state $S_0P_1$ with 
    $\tilde{\mu}_P=0. 81, 0. 85, 0. 88, 0. 91, 0. 93, 0.98, 0.999$ and 
    all solutions have $\tilde{\mu}_S=1$. }
    \label{ADM-one-synchronized}
\end{figure}

Fig.~\ref{ADM-one-synchronized} shows the relationship between ADM mass and synchronized frequency $\tilde{\omega}$, where we take different values of $\tilde{\mu}_P$ to obtain a one-branch solution for different mixed states. The black dashed line represents the $S_0$ state solutions of the boson stars with $\tilde{\mu}_S = 1$, the red dashed line represents the $P_1$ state solutions of the Proca stars, and the blue line denotes the coexisting state $S_0P_1$. The relationship between the ADM mass and the synchronized frequency is similar to the case of the $^{1}S^{2}S$ or $^{1}S^{2}P$ of the RMSBSs in Ref.~\cite{li2020rotating}. It can be seen from the image that $S_0P_1 $ has only one branch, the ADM mass decreases with the increase of frequency, and the two ends of the blue line fall on the black and red spiral dashed lines respectively. As can be seen from Fig.~\ref{field-one-synchronized}, $\left\lvert \tilde{\phi} \right\rvert _{max}$ decreases when the frequency decreases, and when the frequency decreases to the minimum value, the mixed-state solution falls on the Proca single field curve. At this time, the ADM mass reaches the maximum value, the amplitude of the scalar field function decreases to 0, only the Proca field remains, and the mixed star becomes a Proca star. Similarly, when the frequency increases to the maximum value, the ADM mass reaches the minimum value, and $\left\lvert \tilde{F} \right\rvert _{max}$ and $\left\lvert \tilde{G} \right\rvert _{max}$ decreases to 0, leaving only the scalar field, and the mixed star becomes a boson star. 

As can be seen from Fig.~\ref{ADM-one-synchronized}, when we gradually reduce $\tilde{\mu_{P}}$, the intersection point of the mixed state and two single-field curves will gradually move from the larger part of the synchronized frequency of the first branch to the inflection point of the first branch, and then pass the inflection point and move along the second branch to the part of the increased synchronized frequency. The synchronized frequency range of the mixed state solution will also gradually increase until the multi-branch solution appears. 

In Table \ref{table1}, we show the existence range of the synchronized frequency $\tilde{\omega}$ and the range of $M$ when the one-branch solution family takes several different values of the Proca field mass $\tilde{\mu}_P$. When $\tilde{\mu}_P$ increases, the existence domain of $\tilde{\omega}$ decreases, $M_{max}$ and $M_{min}$ both increase first and then decrease. The existence range of $M$ does not change obviously when $\tilde{\mu}_P$ is small. When $\tilde{\mu}_P$ is large, the existence domain of $M$ rapidly decreases with the decrease of the existence domain of $\tilde{\omega}$.

\subsubsection{Multi-Branch}
\begin{figure}[!htbp]
    \begin{center}
    \includegraphics[height=.173\textheight]{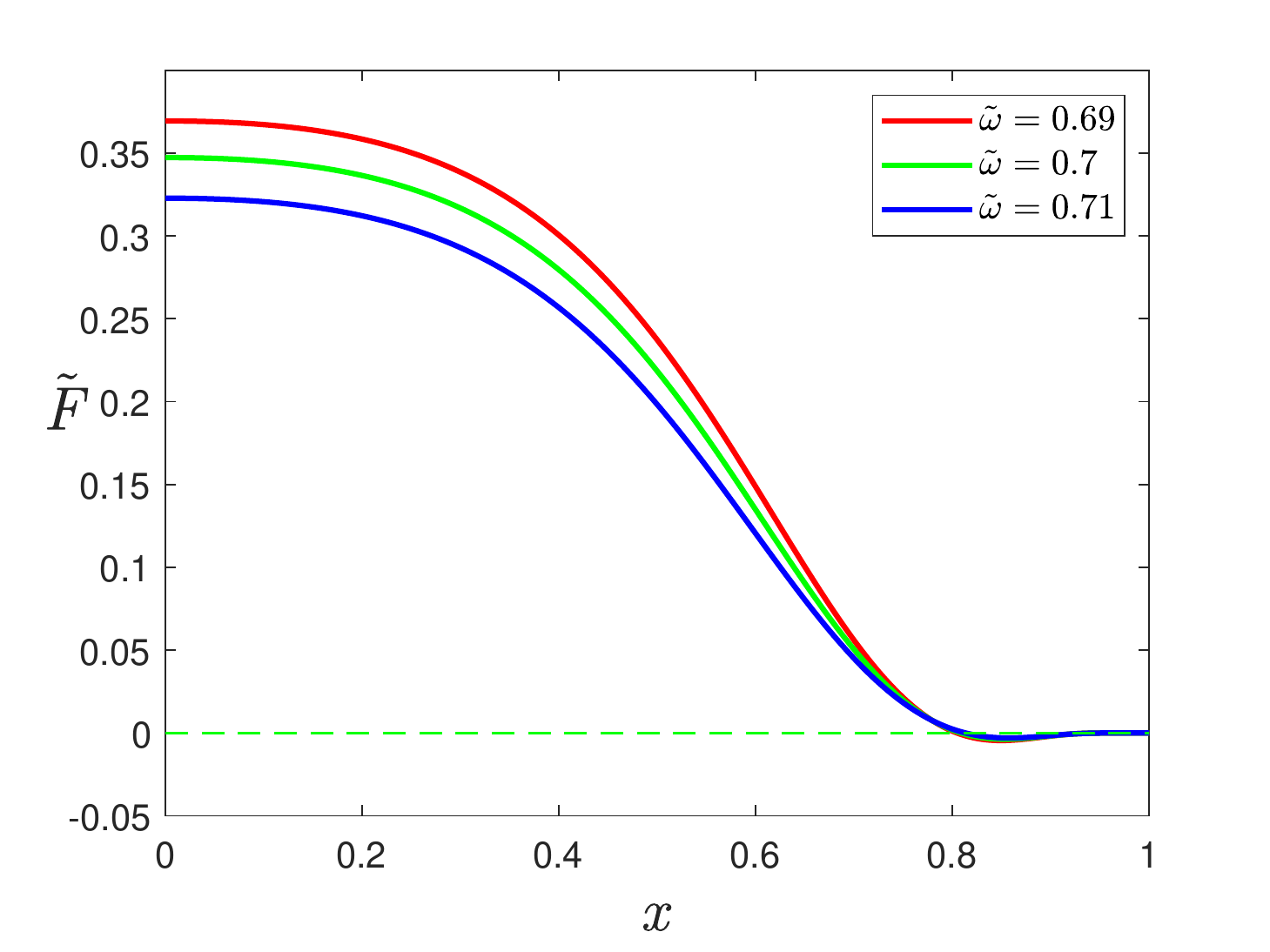}
    \includegraphics[height=.173\textheight]{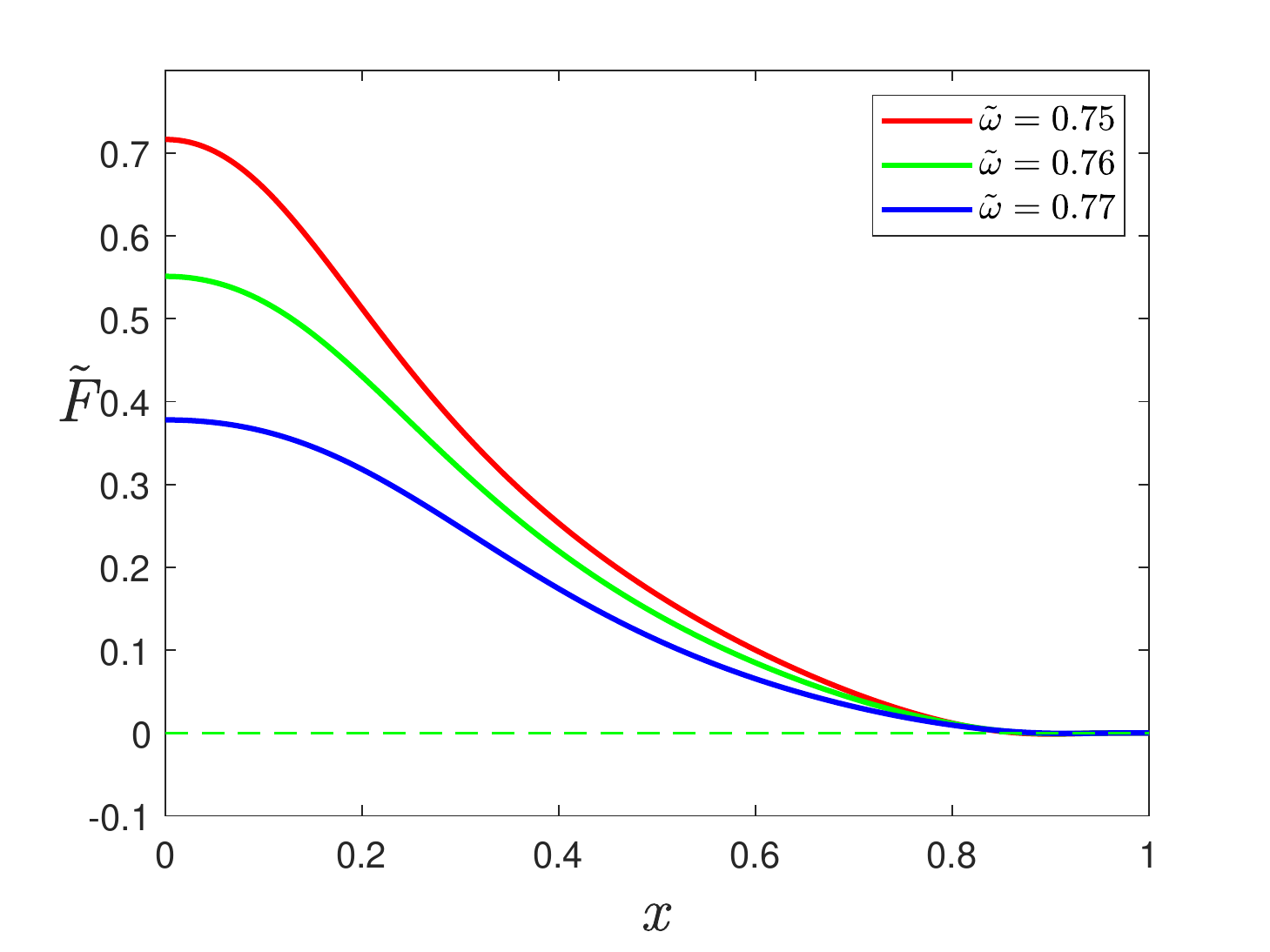}
    \includegraphics[height=.173\textheight]{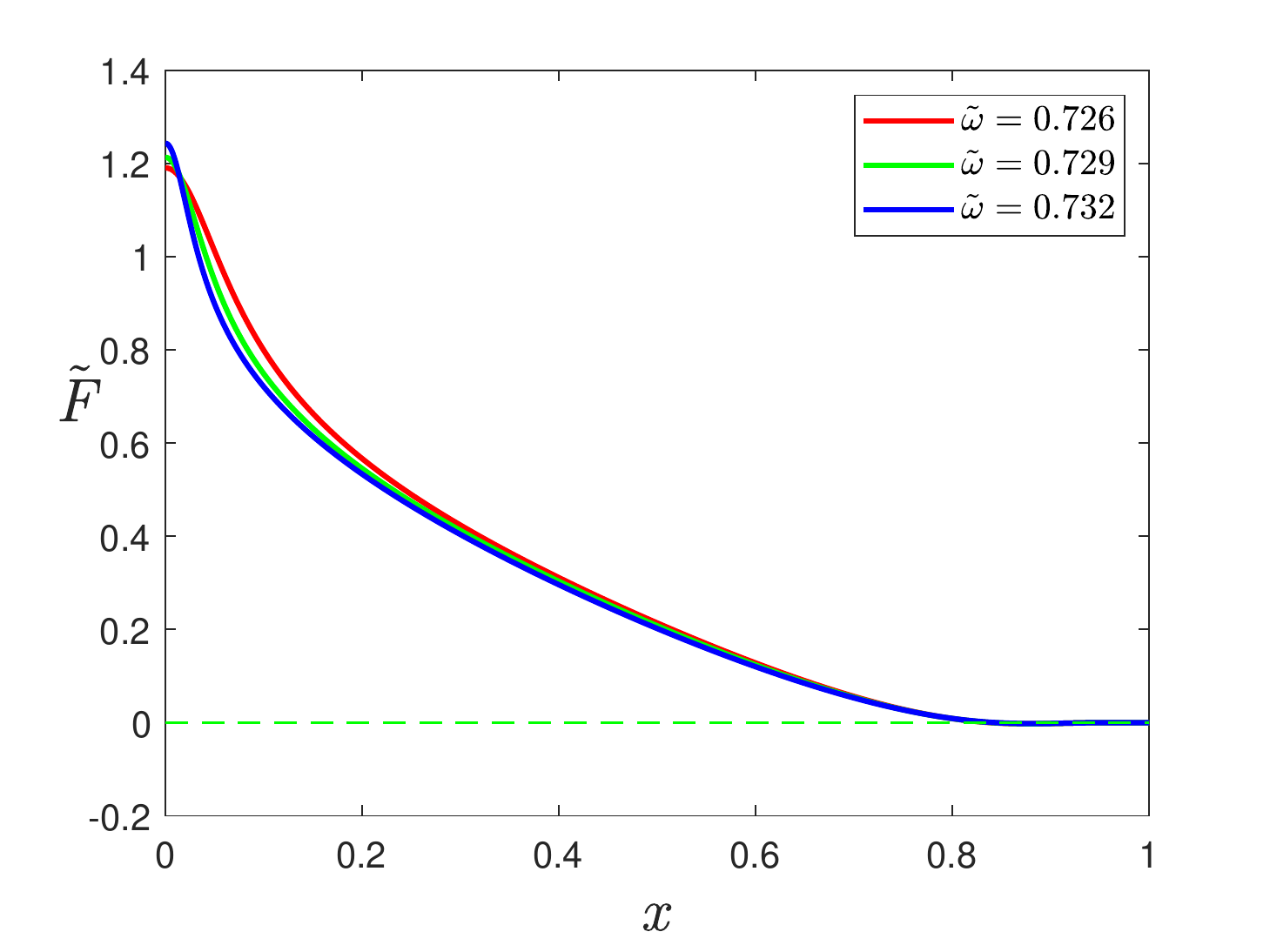}
    \includegraphics[height=.173\textheight]{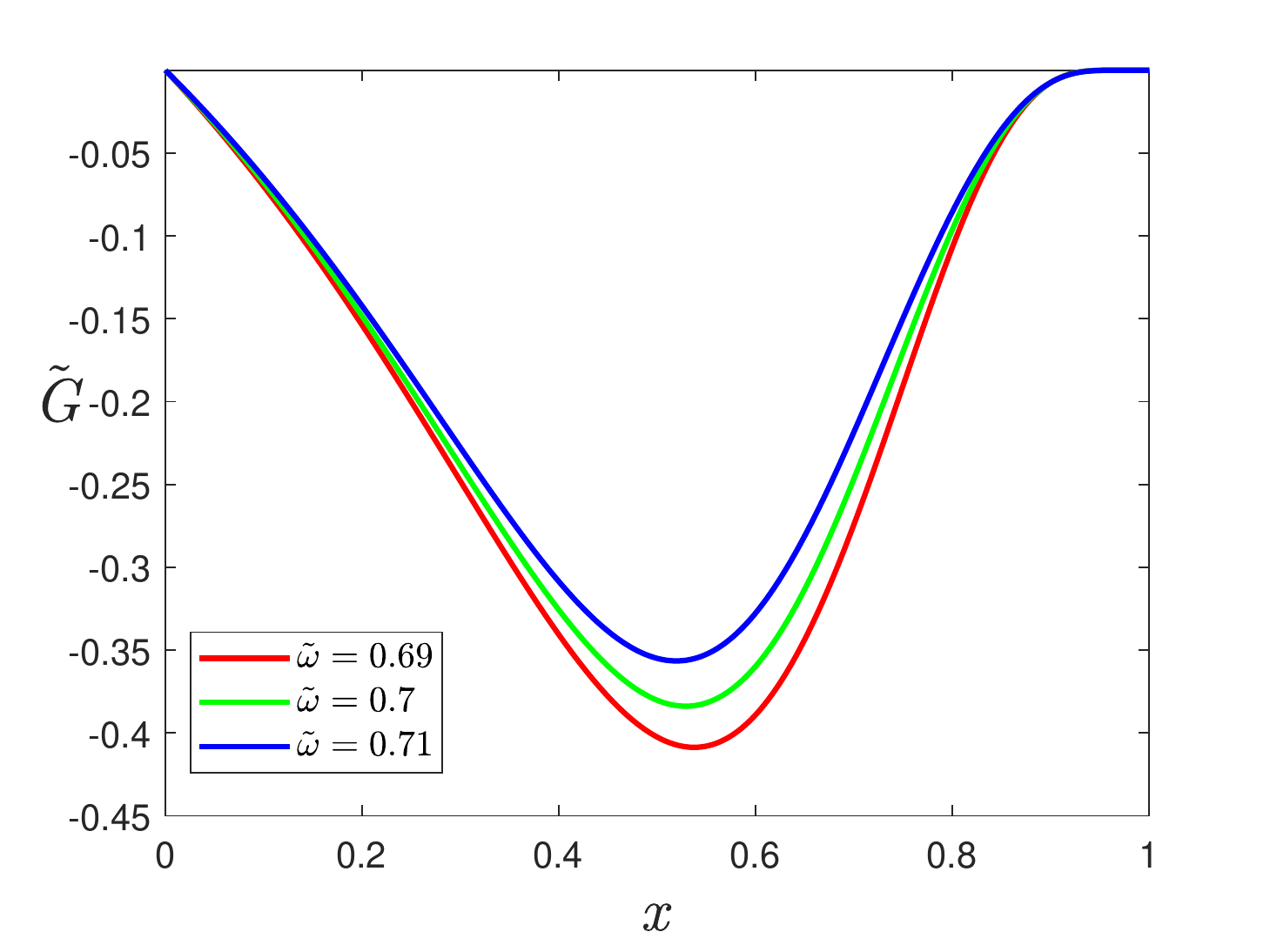}
    \includegraphics[height=.173\textheight]{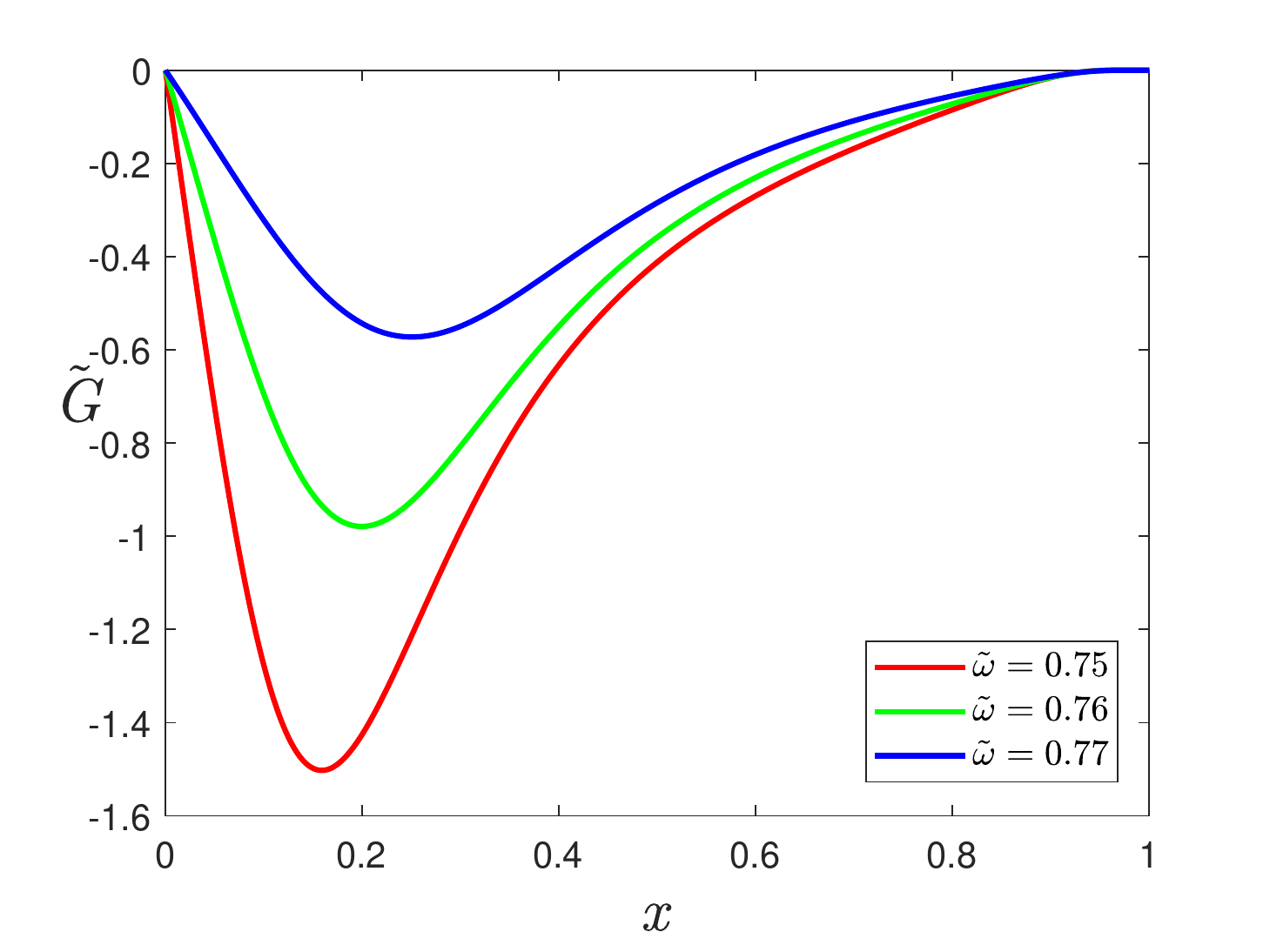}
    \includegraphics[height=.173\textheight]{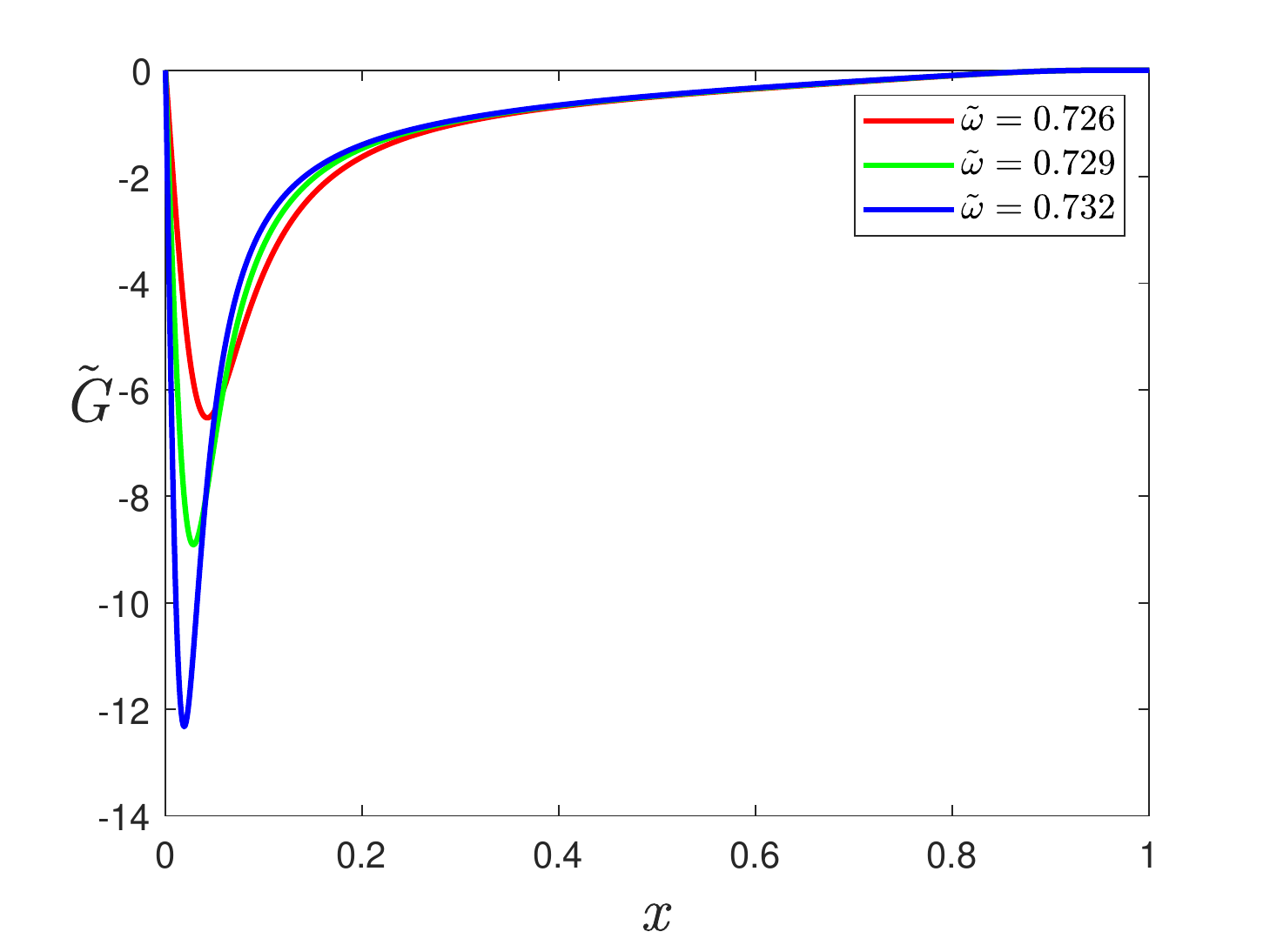}
    \includegraphics[height=.173\textheight]{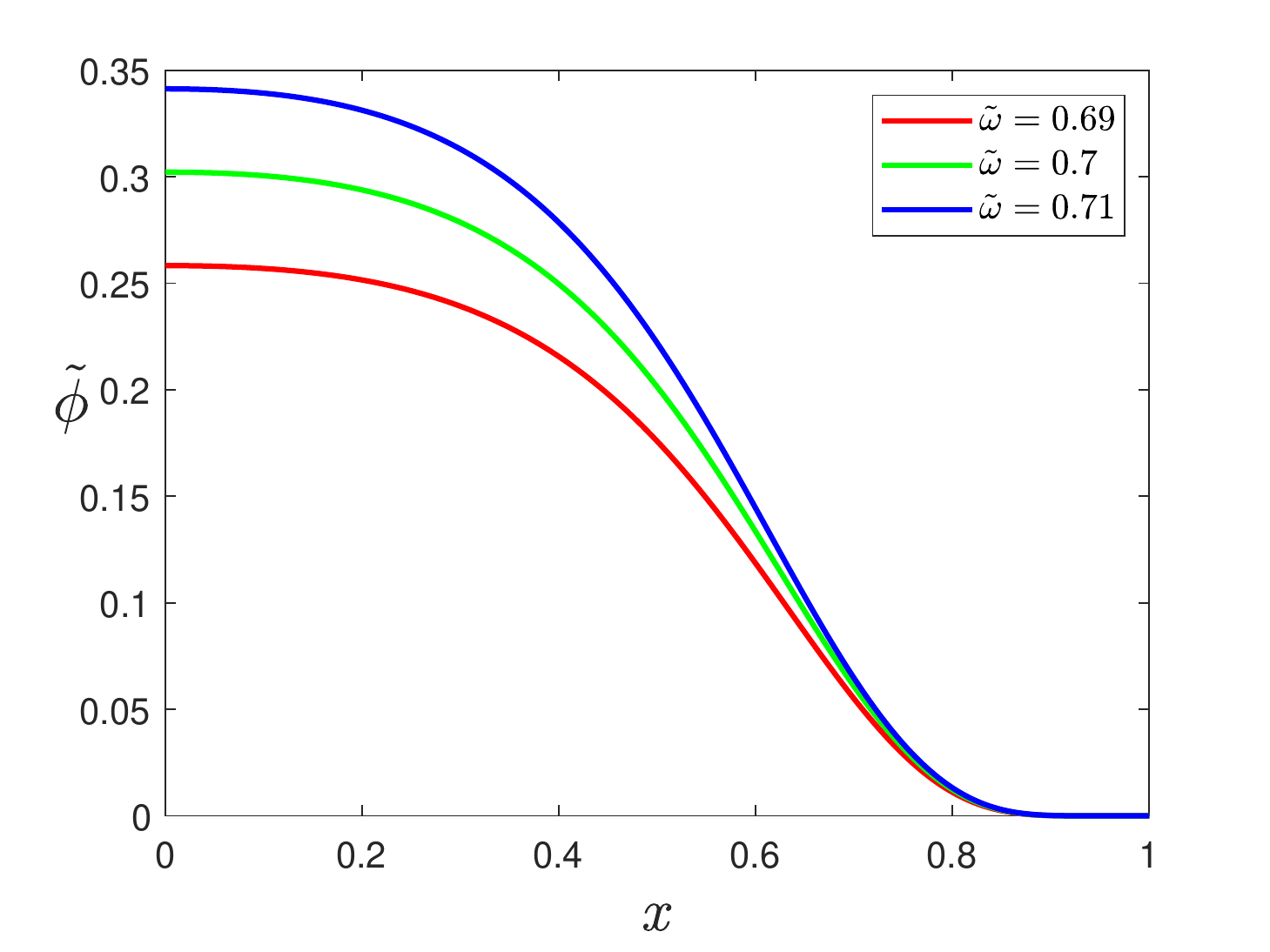}
    \includegraphics[height=.173\textheight]{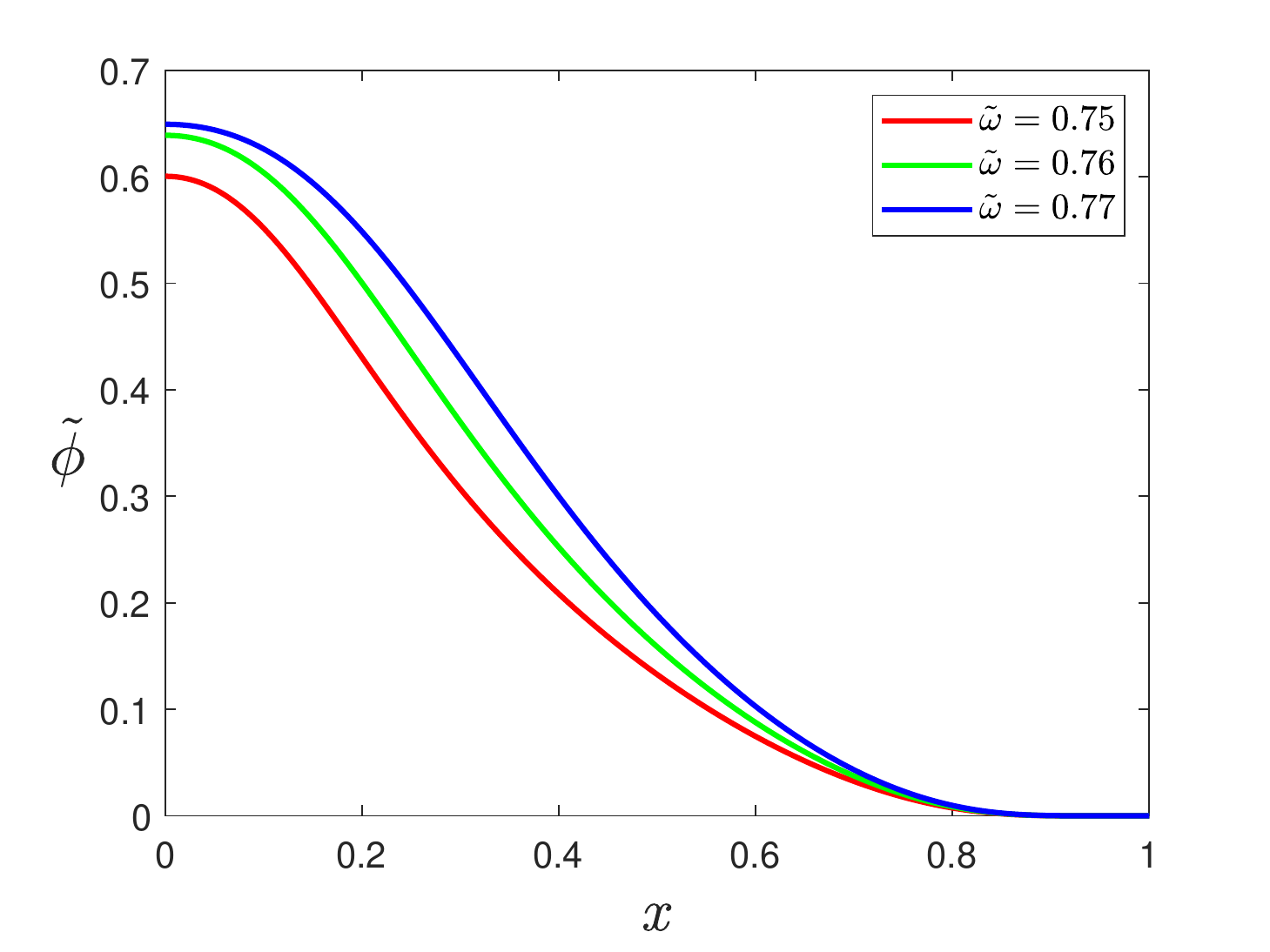}
    \includegraphics[height=.173\textheight]{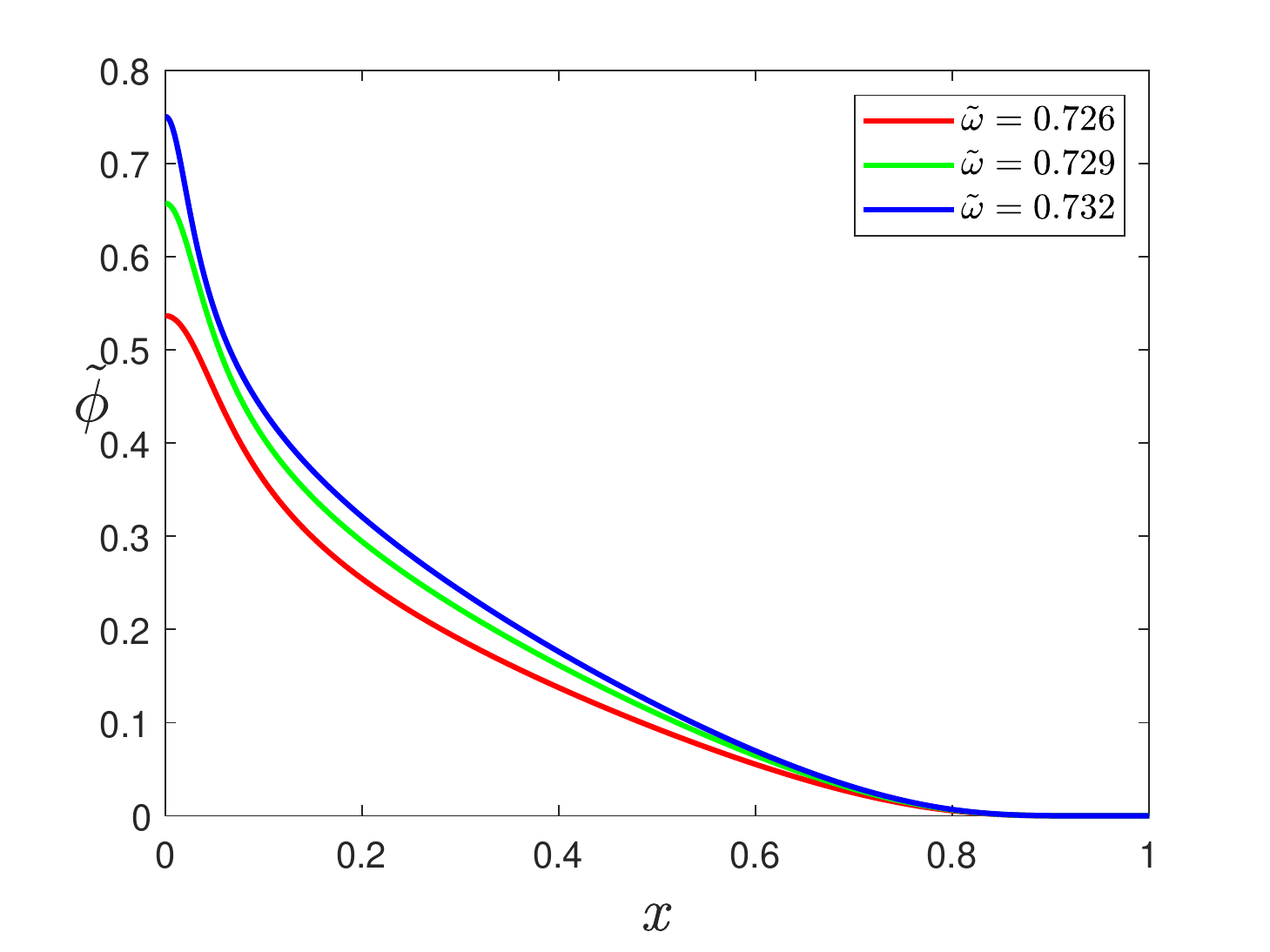}
    \end{center}
    \vspace{-2em}
    \caption{Proca field functions $\tilde{F}$(left panel) and $\tilde{G}$(middle panel)
    and scalar field function $\tilde{\phi}$(right panel) 
    as functions of $x$ with several values of synchronized frequency
    $\tilde{\omega}$, where the field functions on the first, 
    second and third branches are  located in the first row, 
    second row and third row. 
    All solutions have $\tilde{\mu}_P= 0. 808$ and $\tilde{\mu}_S=1$. }
    \label{field-multi-synchronized}
\end{figure}

Fig.~\ref{field-multi-synchronized} shows the relationship between field functions $\tilde{F}$, $\tilde{G}$, $\tilde{\phi}$ and synchronized frequency $\tilde{\omega}$ of multi-branch solutions. The graphs in the first row, second row and third row represent the first branch, second two-branch solution and third branch respectively. For the first and second branches, the field function $\left\lvert \tilde{F} \right\rvert _{max}$, $\left\lvert \tilde{G} \right\rvert _{max}$, $\left\lvert \tilde{\phi} \right\rvert _{max}$ varies with the synchronized frequency $\tilde{\omega}$ in the same way as one-branch solutions. For the third branch, the field function $\left\lvert \tilde{F} \right\rvert _{max}$, $\left\lvert \tilde{G} \right\rvert _{max}$, $\left\lvert \tilde{\phi} \right\rvert _{max}$increases with the increase of synchronized frequency. 

\begin{figure}[!htbp]
    \begin{center}
    \hbox to\linewidth{\hss
    \resizebox{9cm}{6.5cm}{\includegraphics{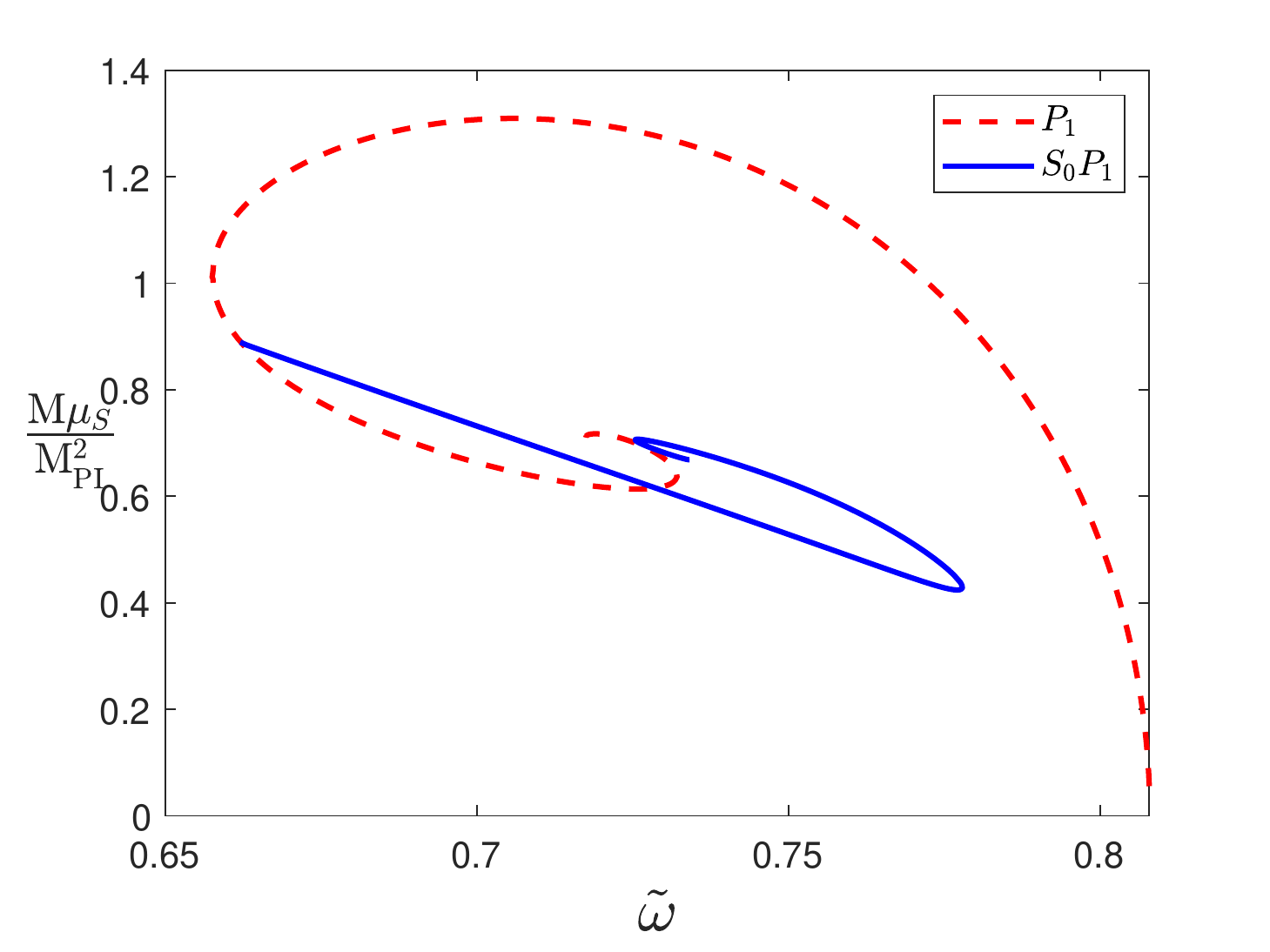}}
    \resizebox{9cm}{6.5cm}{\includegraphics{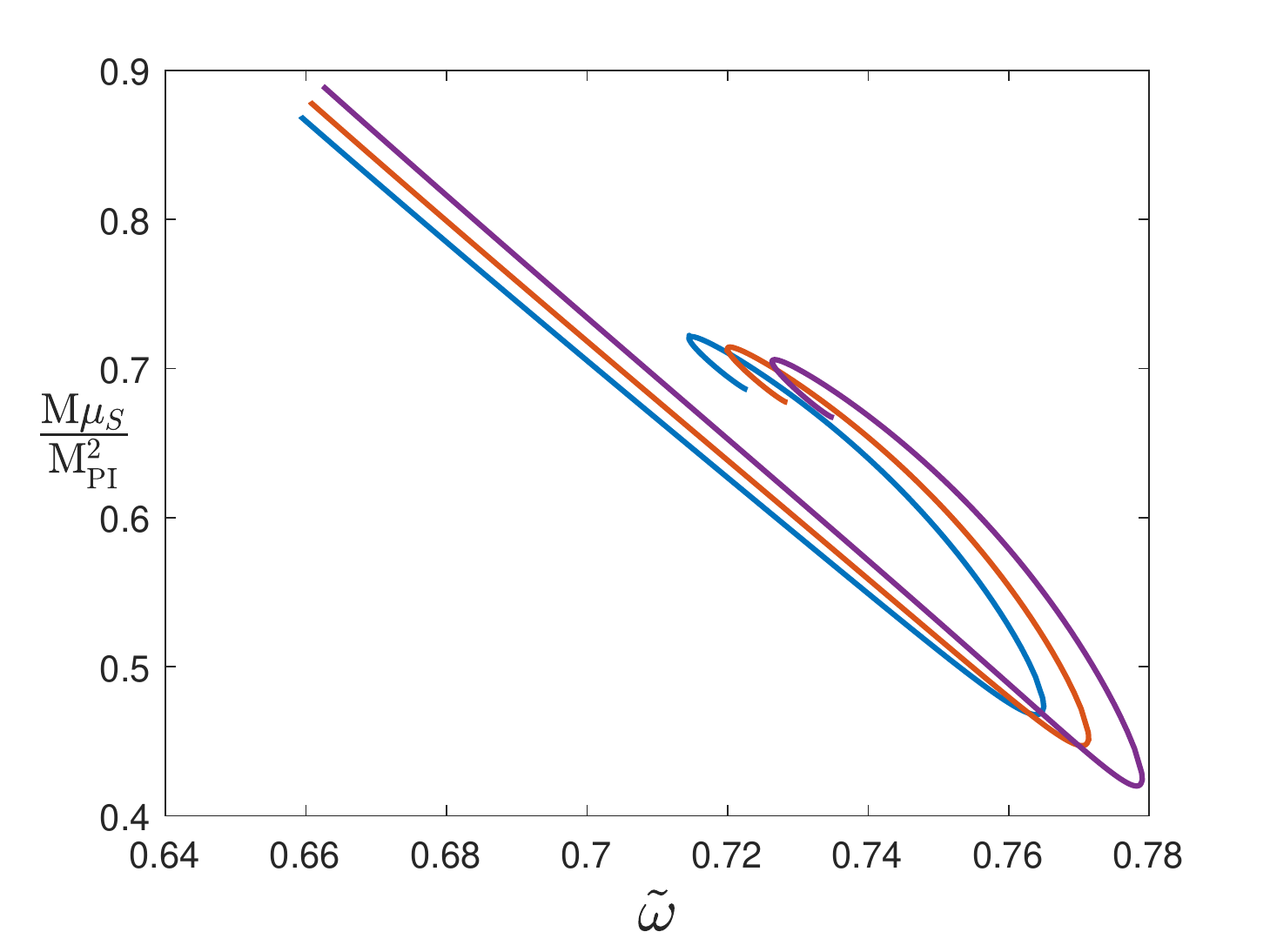}}
    \hss}
    \end{center}
    \vspace{-2em}
    \caption{Left:The ADM mass $M$ as a function of the synchronized frequency $\tilde{\omega}$. 
    The black dashed line represents the $S_0$ state solutions with $\tilde{\mu}_S = 1$, 
    the red dashed line represents the $P_1$ state solutions with $\tilde{\mu}_P=0. 808$, 
    and the blue line denote the coexisting state $S_0P_1$ with $\tilde{\mu}_S = 1$ and $\tilde{\mu}_P=0. 808$. 
    Right:The ADM mass $M$ as a function of the synchronized frequency $\tilde{\omega}$. 
    The light blue, red, purple lines denote the coexisting state $S_0P_1$, 
    with $\tilde{\mu}_P=0. 802, 0. 805, 0. 806$. 
    The black dashed line represents the $S_0$ state solutions with $\tilde{\mu}_S = 1$. 
    All solutions have $\tilde{\mu}_S=1$. }
    \label{ADM-multi-synchronized}
\end{figure}

The relationship between ADM mass and synchronized frequency can be seen from Fig.~\ref{ADM-multi-synchronized}. In Fig.~\ref{ADM-one-synchronized}, when $\tilde{\mu}_P=0. 81$, the mixed state solution is a line almost tangent to the second branch of the scalar field curve. If we continue to reduce the value of $\tilde{\mu}_P$, the right end of the mixed state solution will not fall on the scalar field curve, but will appear as a spiral. In this kind of solution family, the left end of the first branch of the mixed state is similar to the one-branch case, which still starts from the Proca field helix, but for mixed state there are still some differences, after passing the inflection point of the first branch, the second branch is extended, and then the next inflection point is passed again, and the third branch is extended, and finally the helix is formed. There is no Proca field or scalar field disappearing in the mixed state. In short, the multi-branch solution family is different from both one-branch solution family, and more complex than one-branch solution family. 

In Table \ref{table2}, we show the existence domain of the synchronized frequency $\tilde{\omega}$ with different values of $\tilde{\mu}_P$ and the value range of $M$ in the multi-branch solution family.With the increase of $\tilde{\mu}_P$, the existence domain of $B_1$ and $B_2$ gradually increases, while the existence domain of $B_3$ has little change. $M_{max}$ shows a gradual increase trend, while $M_{min}$ shows a gradual decrease trend. The value range of $M$ increases with the increase of the existence domain of $\tilde{\omega}$.

\subsubsection{Two-Branch}
\begin{figure}[!htbp]
    \begin{center}
    \includegraphics[height=.24\textheight]{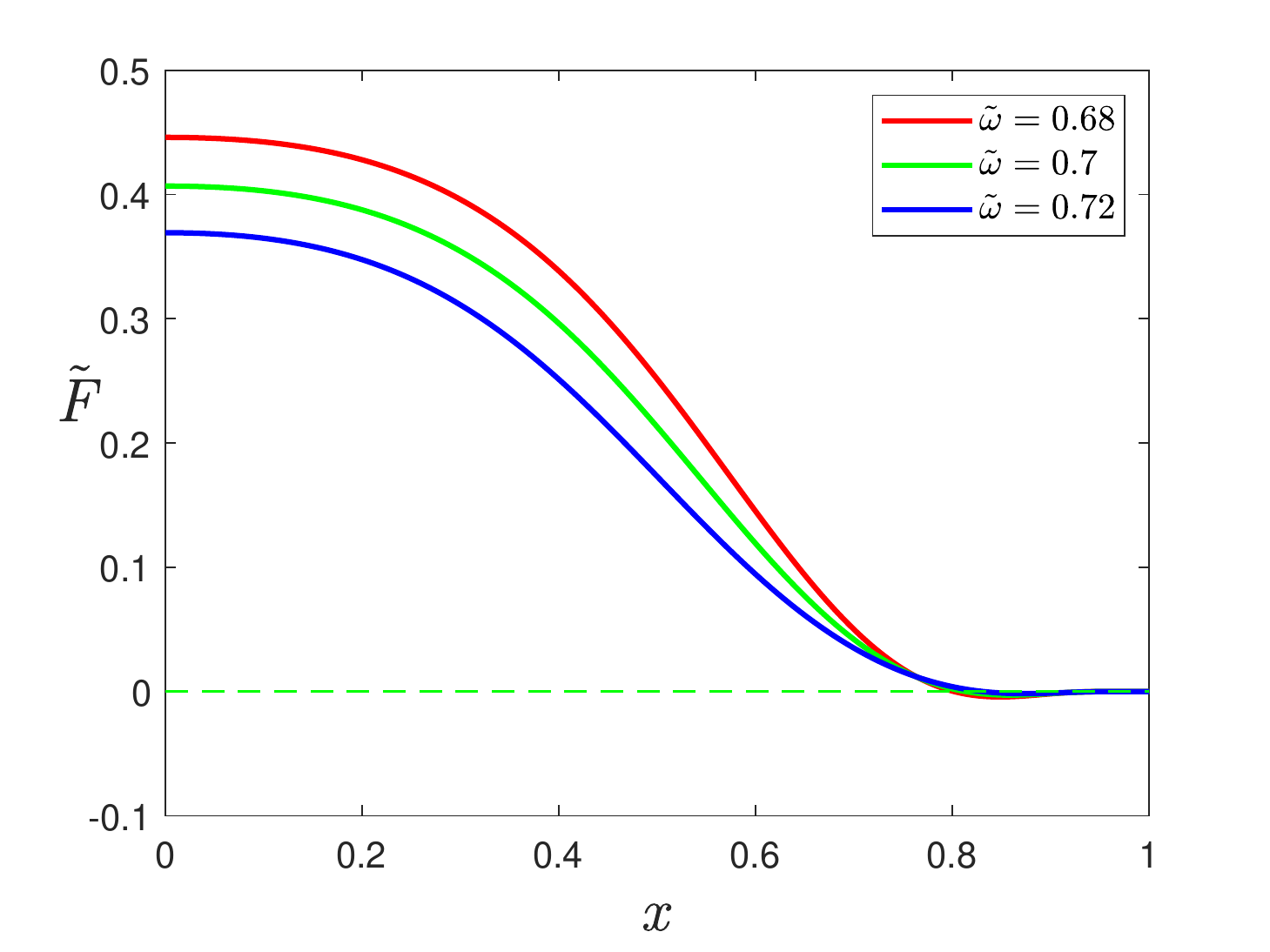}
    \includegraphics[height=.24\textheight]{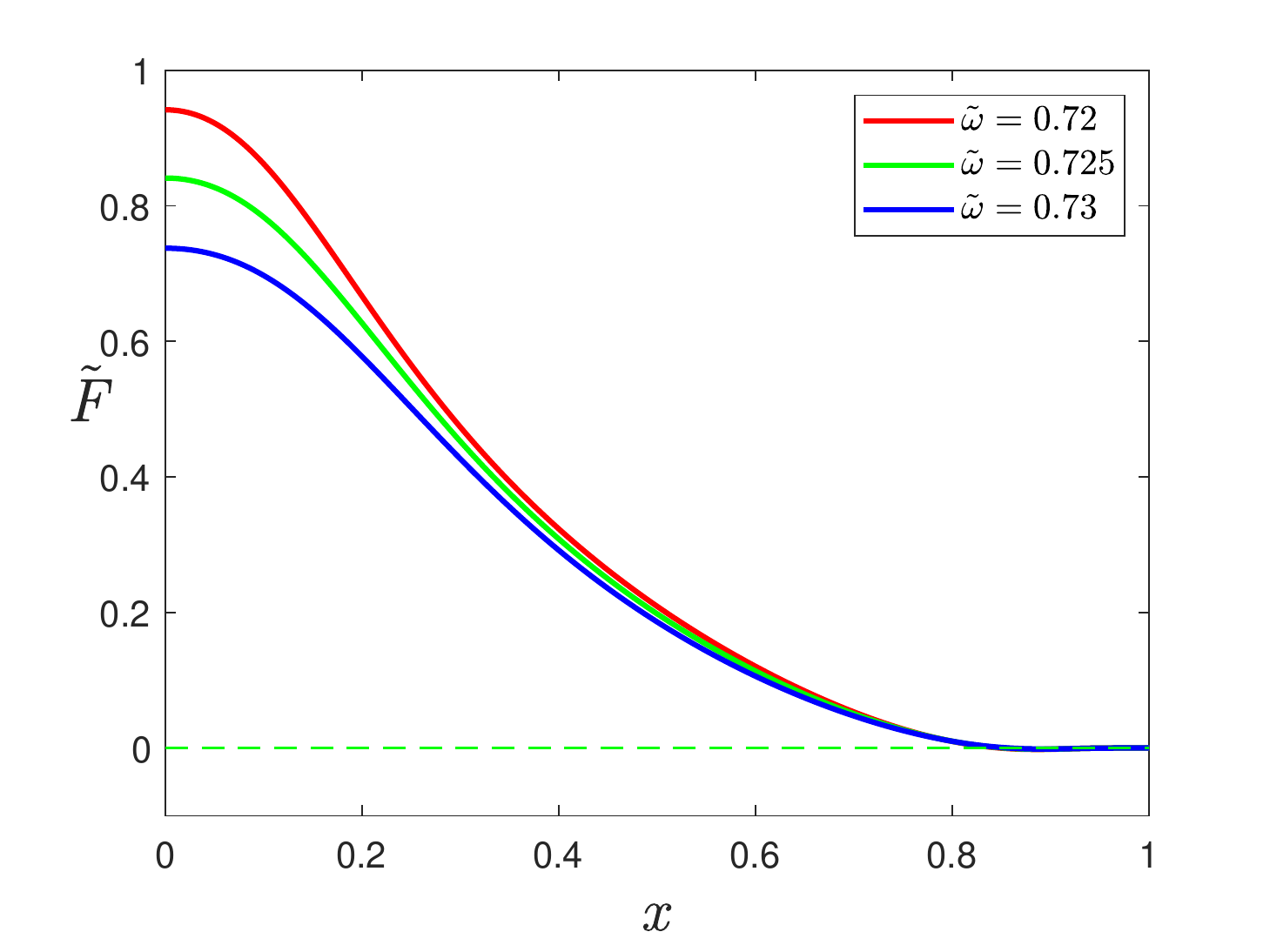}
    \includegraphics[height=.24\textheight]{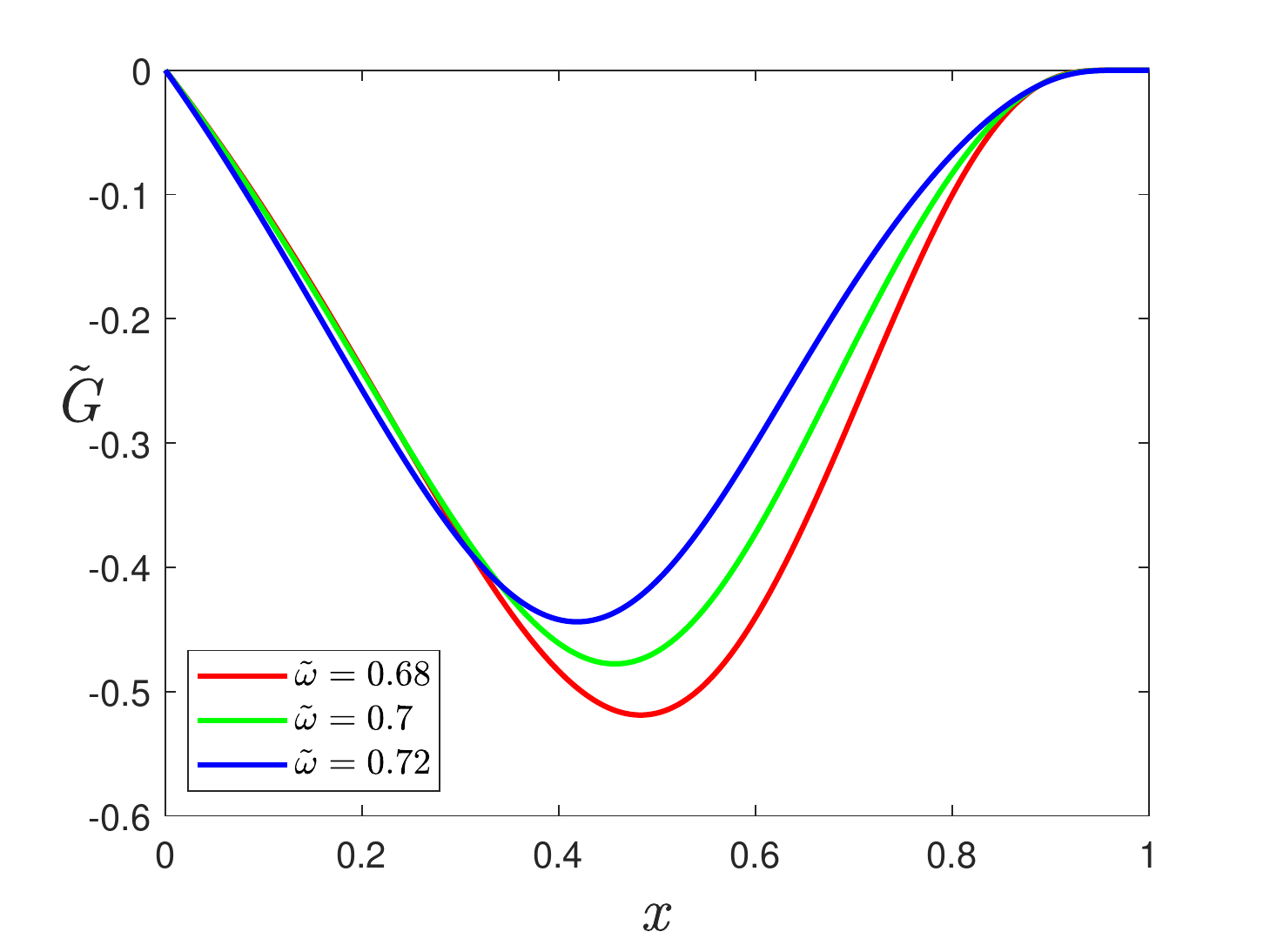}
    \includegraphics[height=.24\textheight]{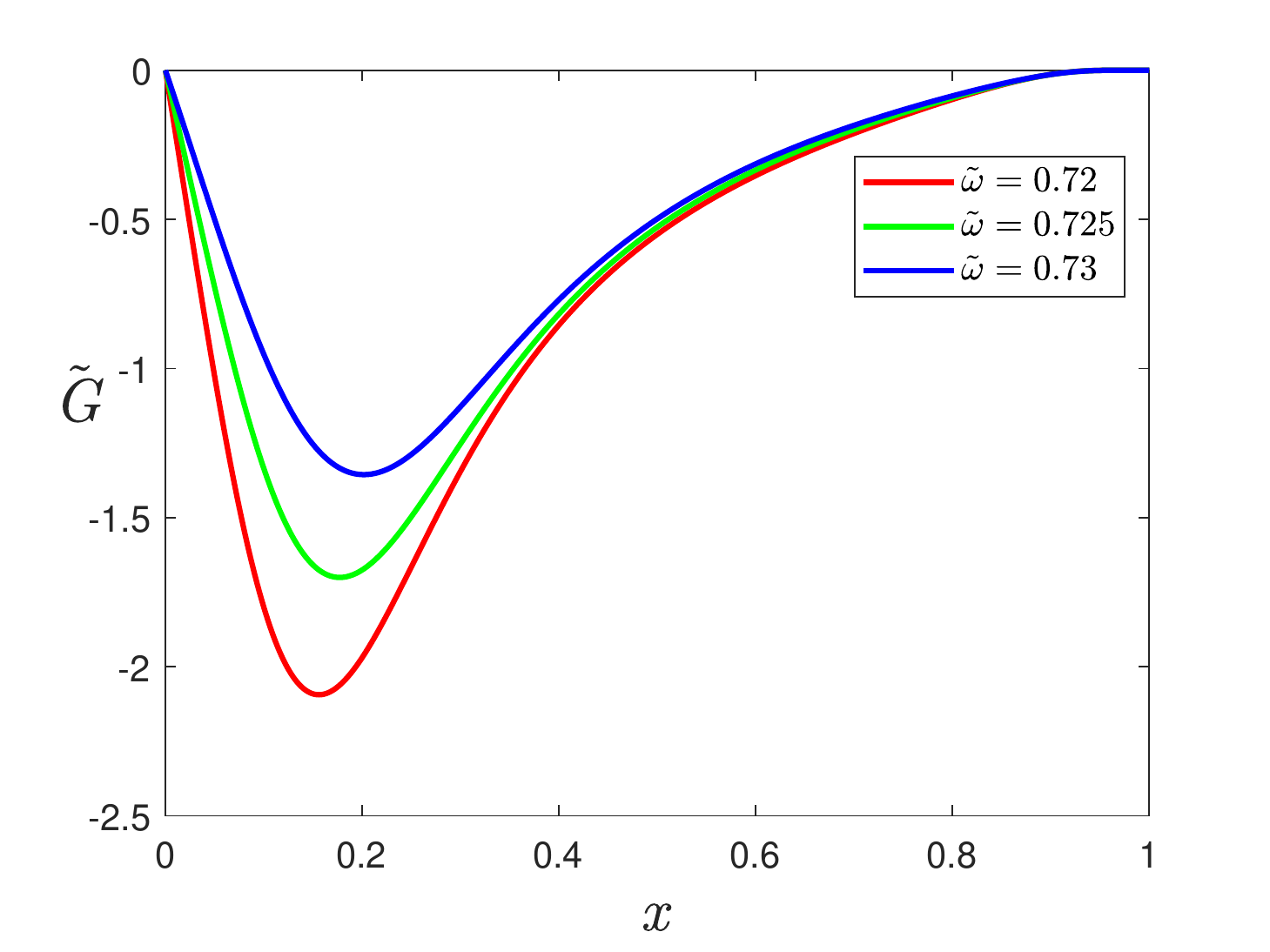}
    \includegraphics[height=.24\textheight]{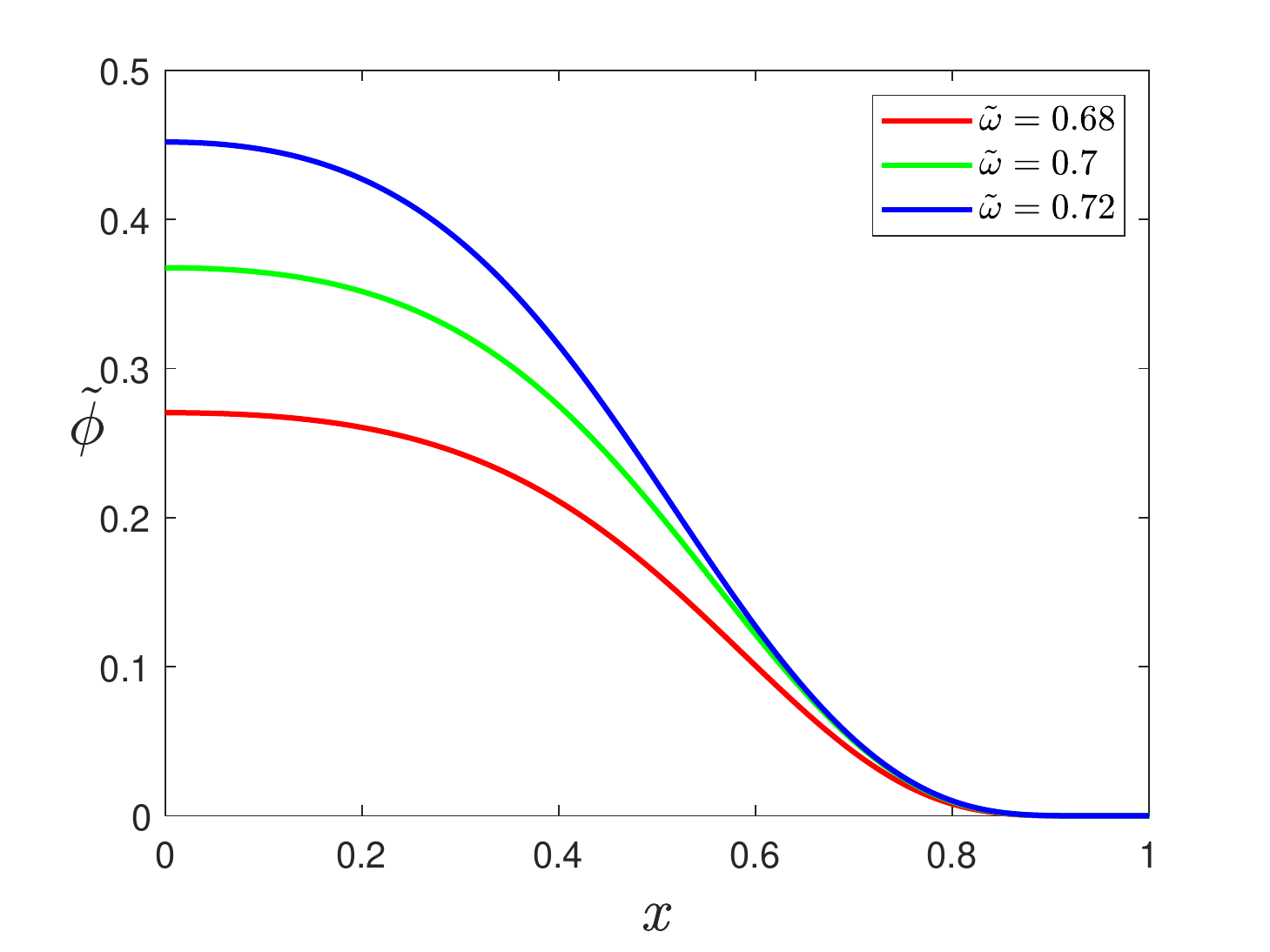}
    \includegraphics[height=.24\textheight]{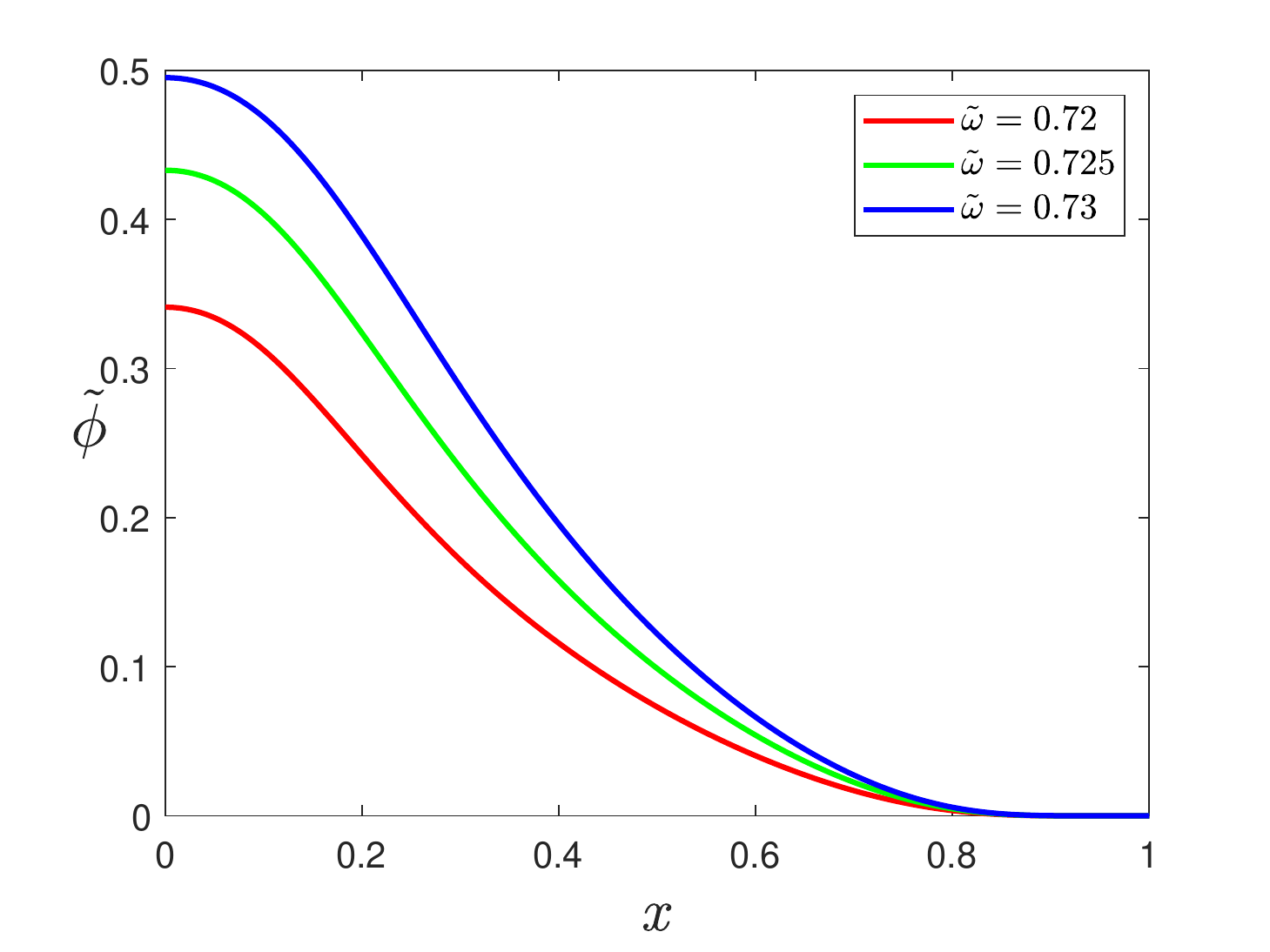}
    \end{center}
    \vspace{-2em}
    \caption{Proca field functions $\tilde{F}$(top panel) and $\tilde{G}$(middle panel)
    and scalar field function $\tilde{\phi}$(bottom panel) 
    as functions of $x$ with several values of synchronized frequency
    $\tilde{\omega}$, where the field functions on the first and 
    second branches are located in the first column and second column. 
    All solutions have $\tilde{\mu}_P= 0. 79$ and $\tilde{\mu}_S=1$. }
    \label{field-two-synchronized}
\end{figure}

Fig.~\ref{field-two-synchronized} shows the relationship between the field functions $\tilde{F}$, $\tilde{G}$, $\tilde{\phi}$ and the synchronized frequency $\tilde{\omega}$ of two-branch solutions. The graphs in the first column and the second column represent the first branch and the second two-branch solution respectively. The two-branch solution is similar to the one-branch solution, $\left\lvert \tilde{F} \right\rvert _{max}$ and $\left\lvert \tilde{G} \right\rvert _{max}$ decreases with the increasing of $\tilde{\omega}$, $\left\lvert \tilde{\phi} \right\rvert _{max}$ increases as the synchronized frequency $\tilde{\omega}$ increases. For the second branch, the field function $\left\lvert \tilde{F} \right\rvert _{max}$, $\left\lvert \tilde{G} \right\rvert _{max}$, $\left\lvert \tilde{\phi} \right\rvert _{max}$ shows the same trend as the first branch. 

\begin{figure}[!htbp]
    \begin{center}
    \hbox to\linewidth{\hss
    \resizebox{9cm}{6.5cm}{\includegraphics{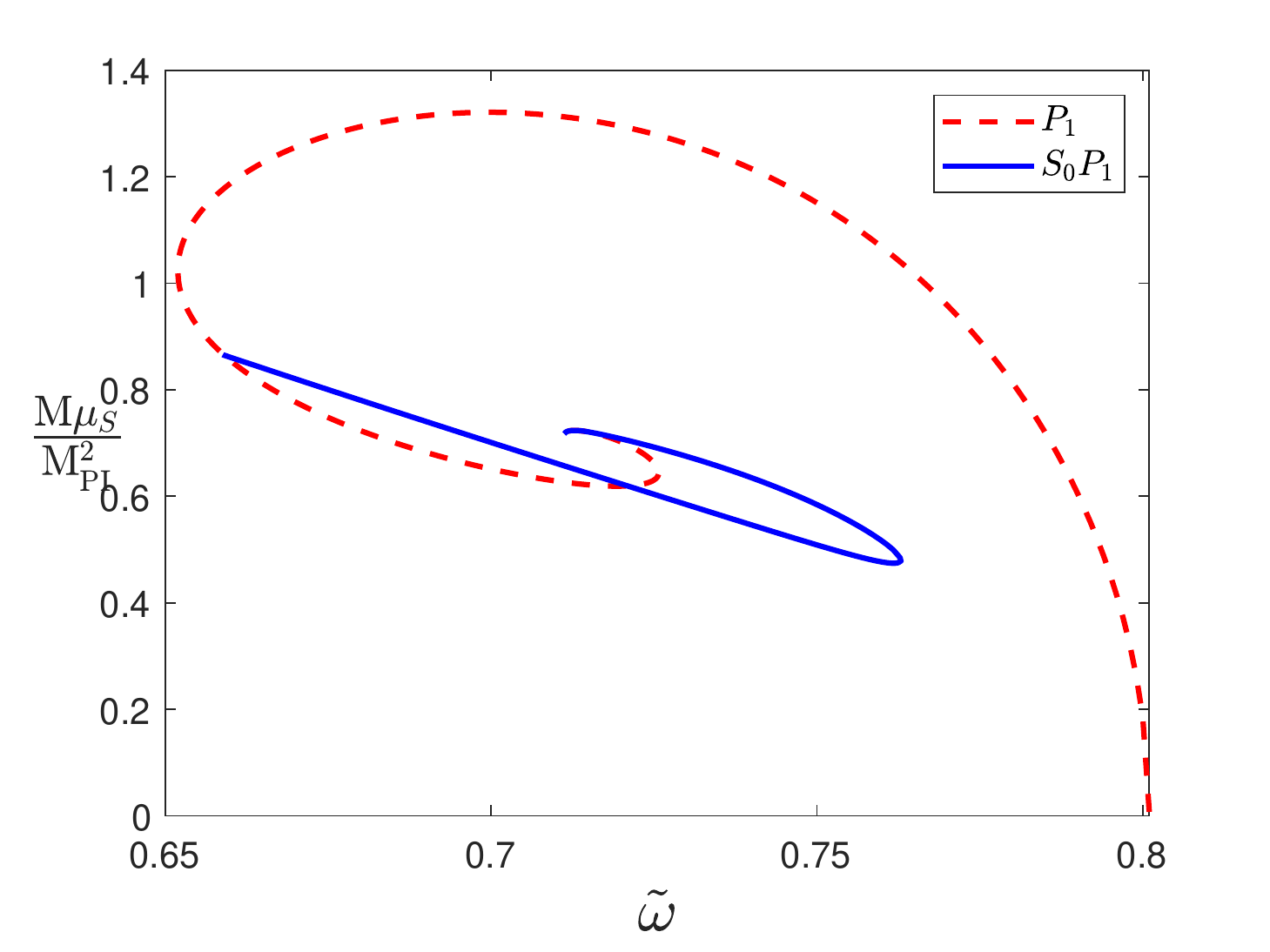}}
    \resizebox{9cm}{6.5cm}{\includegraphics{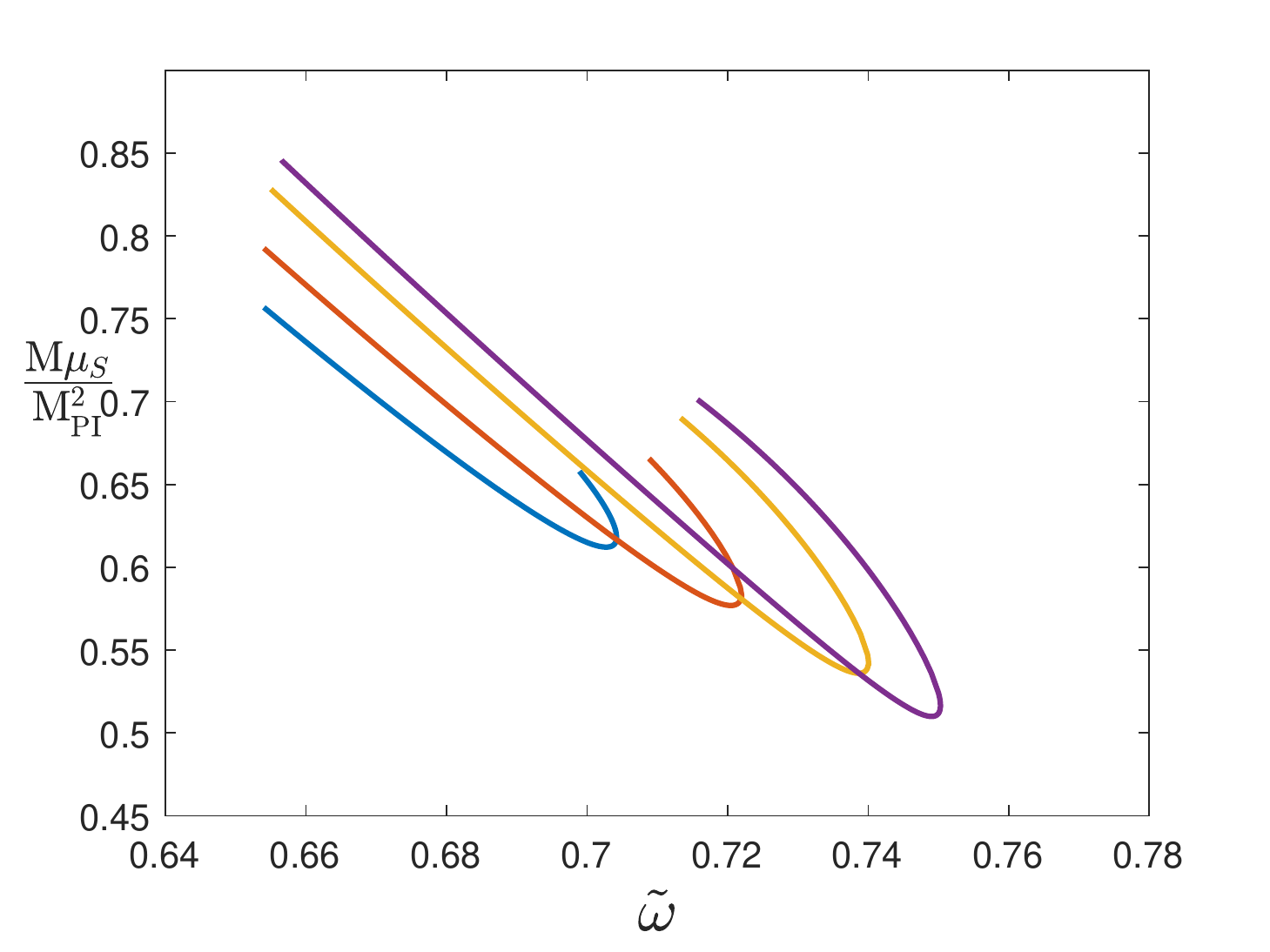}}
    \hss}
    \end{center}
    \vspace{-2em}
    \caption{Left:The ADM mass $M$ as a function of the synchronized frequency $\tilde{\omega}$. 
    The black dashed line represents the $S_0$ state solutions with $\tilde{\mu}_S = 1$, 
    the red dashed line represents the $P_1$ state solutions with $\tilde{\mu}_P=0. 801$, 
    and the blue line denote the coexisting state $S_0P_1$ with $\tilde{\mu}_S = 1$ and $\tilde{\mu}_P=0. 801$. 
    Right:The ADM mass $M$ as a function of the synchronized frequency $\tilde{\omega}$. 
    The light blue, red, orange, purple lines denote the coexisting state $S_0P_1$, 
    with $\tilde{\mu}_P=0. 772, 0. 781, 0. 79, 0. 795$. 
    The black dashed line represents the $S_0$ state solutions with $\tilde{\mu}_S = 1$. 
    All solutions have $\tilde{\mu}_S=1$. }
    \label{ADM-two-synchronized}
\end{figure}

Similar to Fig.~\ref{ADM-one-synchronized}, we show the relationship between ADM mass and synchronized frequency $\tilde{\omega}$ in Fig.~\ref{ADM-two-synchronized}. Unlike one-branch solutions, mixed state solutions are two-branch solutions when $\tilde{\mu}_P$ is small. This solution is different from the form of one-branch solution in which two ends fall on two single-field helices respectively. When the synchronized frequency increases to a certain value, there exists a synchronized frequency corresponding to two different solutions, and the scalar field function $\tilde{\phi}$ never vanishes. The mixed state curve still starts from the Proca field, but the right endpoint does not fall on the scalar field single field curve, but appears an inflection point. After crossing the inflection point, as the synchronized frequency decreases, the mixed state solution moves along the second branch. When the final synchronized frequency reaches the minimum, the mixed state curve intersects with the spiral of Proca single field again. At this time, the mixed state will only have Proca field, and the mixed star will become a Proca star. 

In Table \ref{table3}, we show the existence range of the synchronized frequency $\tilde{\omega}$ and the range of $M$ when the two-branch solution family takes several different values of the Proca field mass $\tilde{\mu}_P$. As $\tilde{\mu}_P$ decreases, the existence domains of both $B_1$ and $B_2$ become narrower. When $\tilde{\mu}_P=0.772$, the second branch is very narrow and almost disappears. $M_{max}$ shows a gradually decreasing trend, while $M_{min}$ shows a gradually increasing trend, so the value range of $M$ also decreases.

\subsection{Nonsynchronized frequency}
Similar to the case of synchronized frequency, we divide the mixed state solutions in the case of nonsynchronized frequencies into three categories:the one-branch-A solution family, multi-branch solution family, and one-branch-B solution family. When $0. 768 < \tilde{\mu}_P \le 0. 939$, the mixed-state solution is one-branch-A solution family. When $0. 729< \tilde{\mu}_P \le 0. 768$, the mixed-state multi-branch solution family. When $0. 7109 \le \tilde{\mu}_P \le 0. 729$, the mixed state solution is one-branch-B solution family. However, there are some differences in details between the classification of solution families in the case of nonsynchronized frequencies and that in the case of the synchronized frequency. The three solution families here do not have the very obvious critical case in the case of the same frequency.We will discuss the properties of these three types of solutions in detail below. 
\subsubsection{One-Branch-A}
\begin{figure}[!htbp]
    \begin{center}
    \includegraphics[height=.24\textheight]{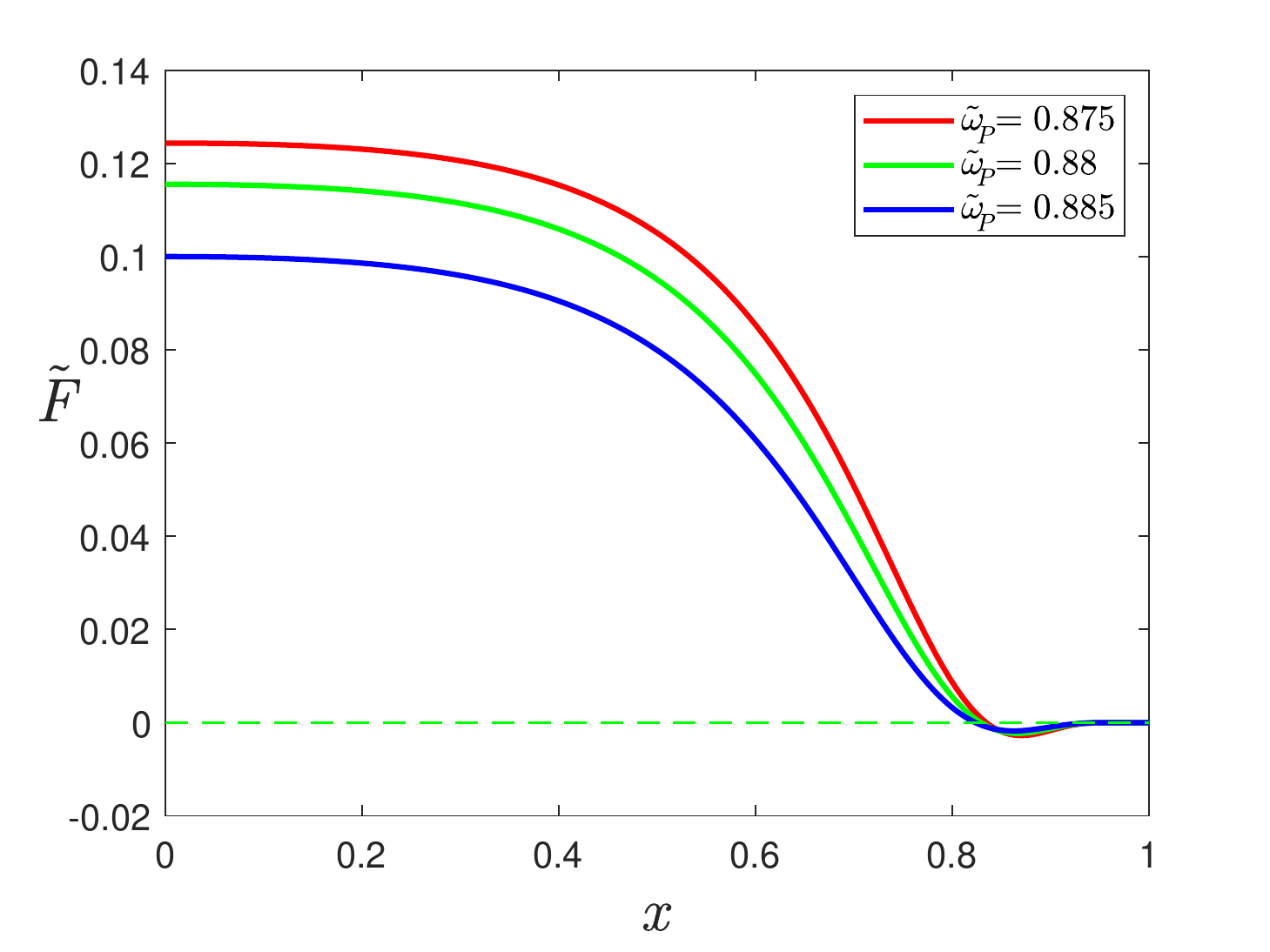}
    \includegraphics[height=.24\textheight]{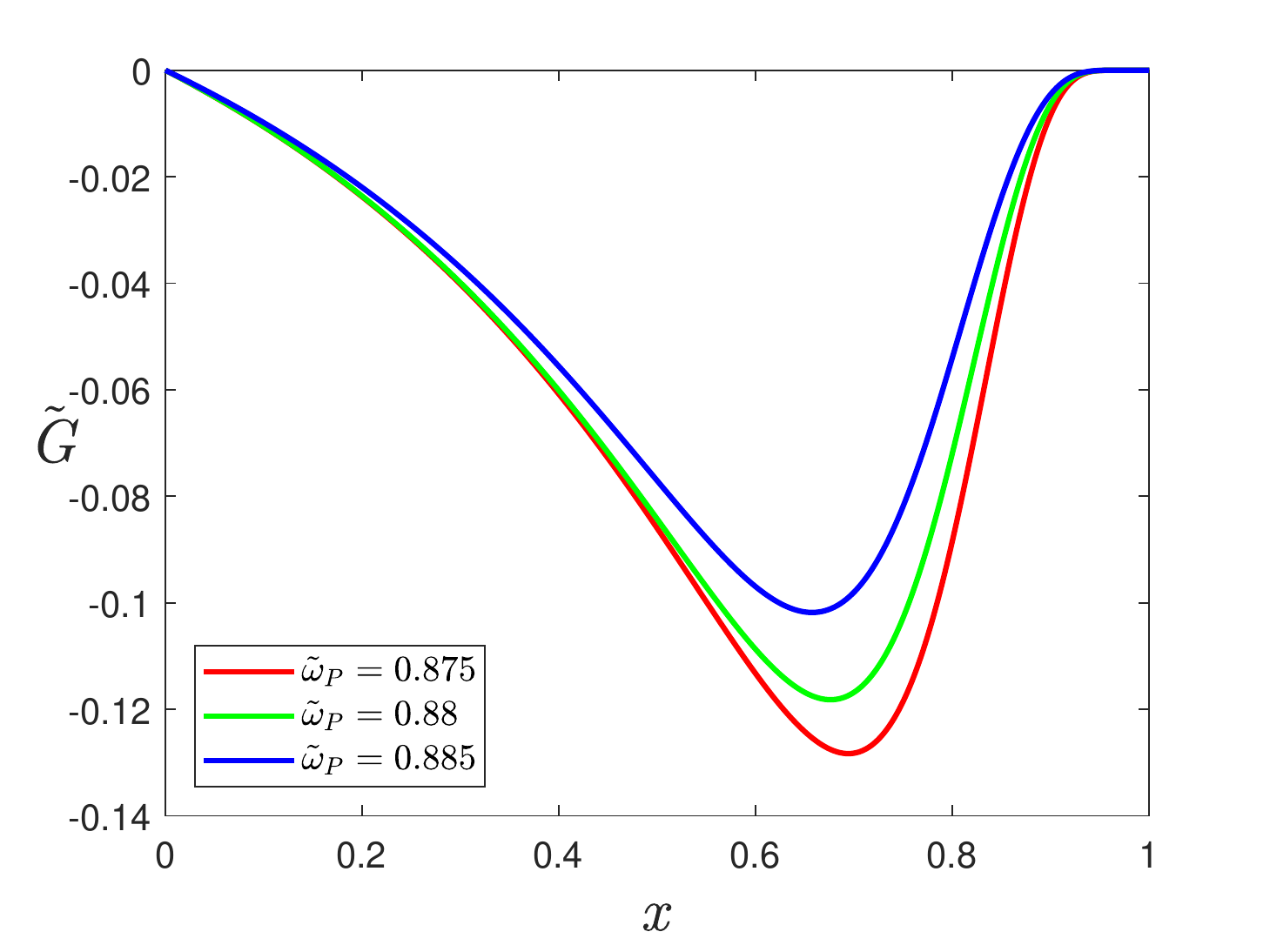}
    \includegraphics[height=.24\textheight]{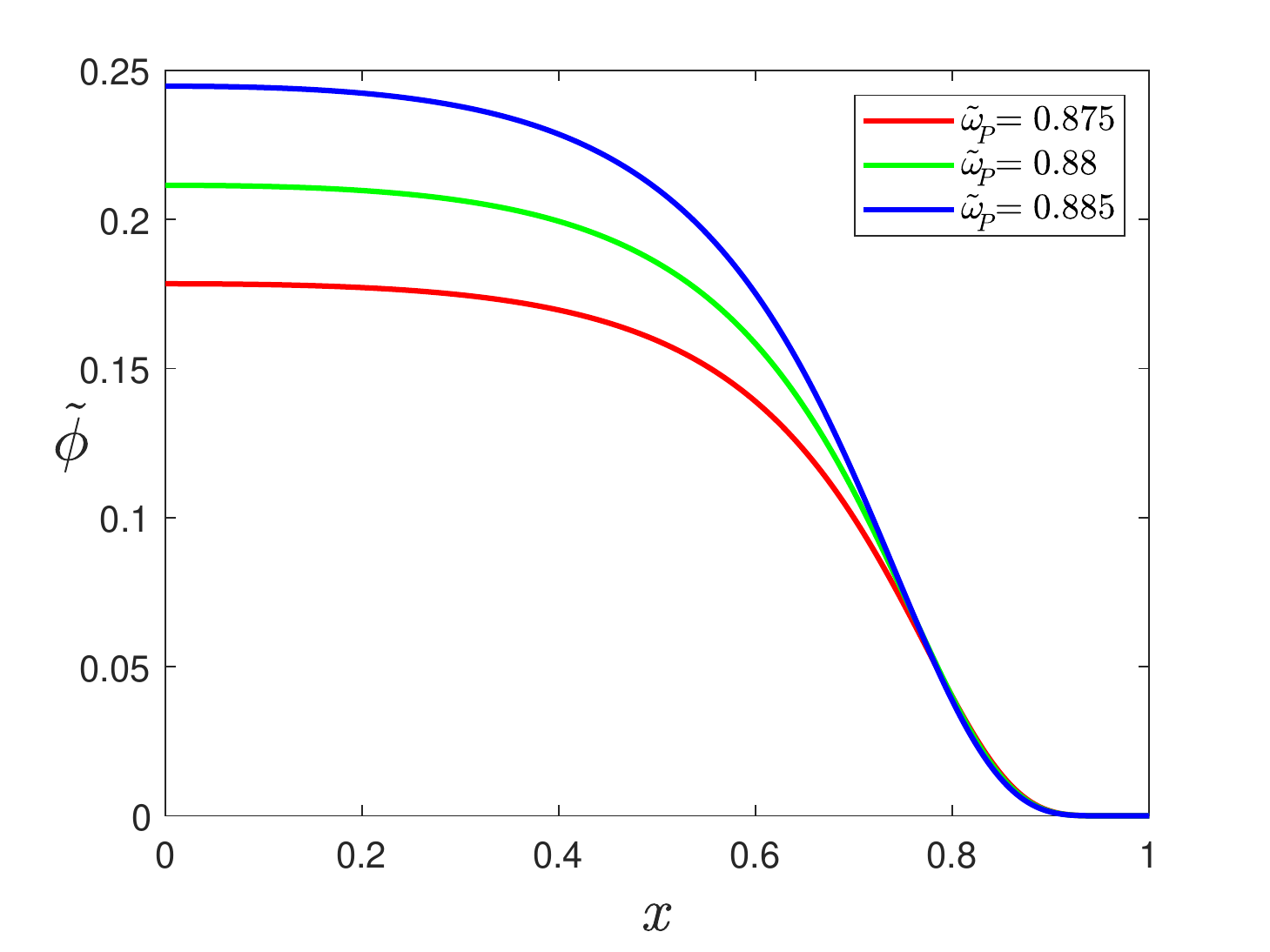}
    \end{center}
    \vspace{-2em}
    \caption{Proca field function $\tilde{F}$ and $\tilde{G}$(top panel) 
    and scalar field functions 
    $\tilde{\phi}$(bottom panel) as functions of $x$ with 
    $\tilde{\omega}_P = 0. 875, 0. 88, 0. 885$. All solutions have $\tilde{\omega}_S= 0. 8$ 
    and $\tilde{\mu}_S = \tilde{\mu}_P = 1$. }
    \label{field-one1-nonsynchronized}
\end{figure}

For the one-branch-A class solution, the image of the field function $\tilde{F}$, $\tilde{G}$, $\tilde{\phi}$ is shown in Fig.~\ref{field-one1-nonsynchronized}, similar to the one-branch solution in the case of the synchronized frequency. For the scalar field function, $\left\lvert \tilde{\phi} \right\rvert _{max}$ increases as the nonsynchronized frequency $\tilde{\omega}_P$ increases. For the Proca field function, $\left\lvert \tilde{F} \right\rvert _{max}$ and $\left\lvert \tilde{G} \right\rvert _{max}$ with nonsynchronized frequency $\tilde{\omega}_P$. 

\begin{figure}[!htbp]
    \begin{center}
    \hbox to\linewidth{\hss
    \resizebox{9cm}{6.5cm}{\includegraphics{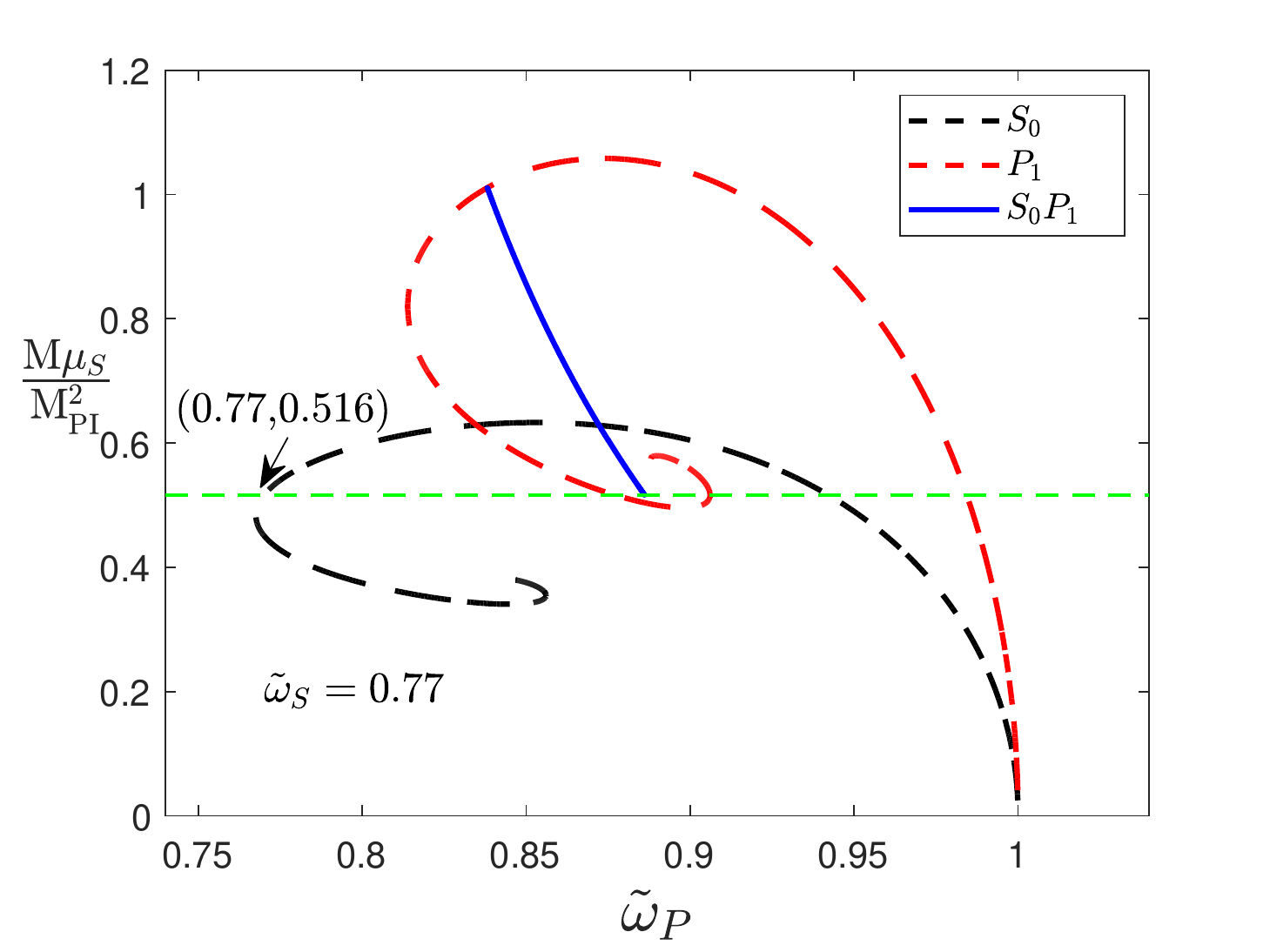}}
    \resizebox{9cm}{6.5cm}{\includegraphics{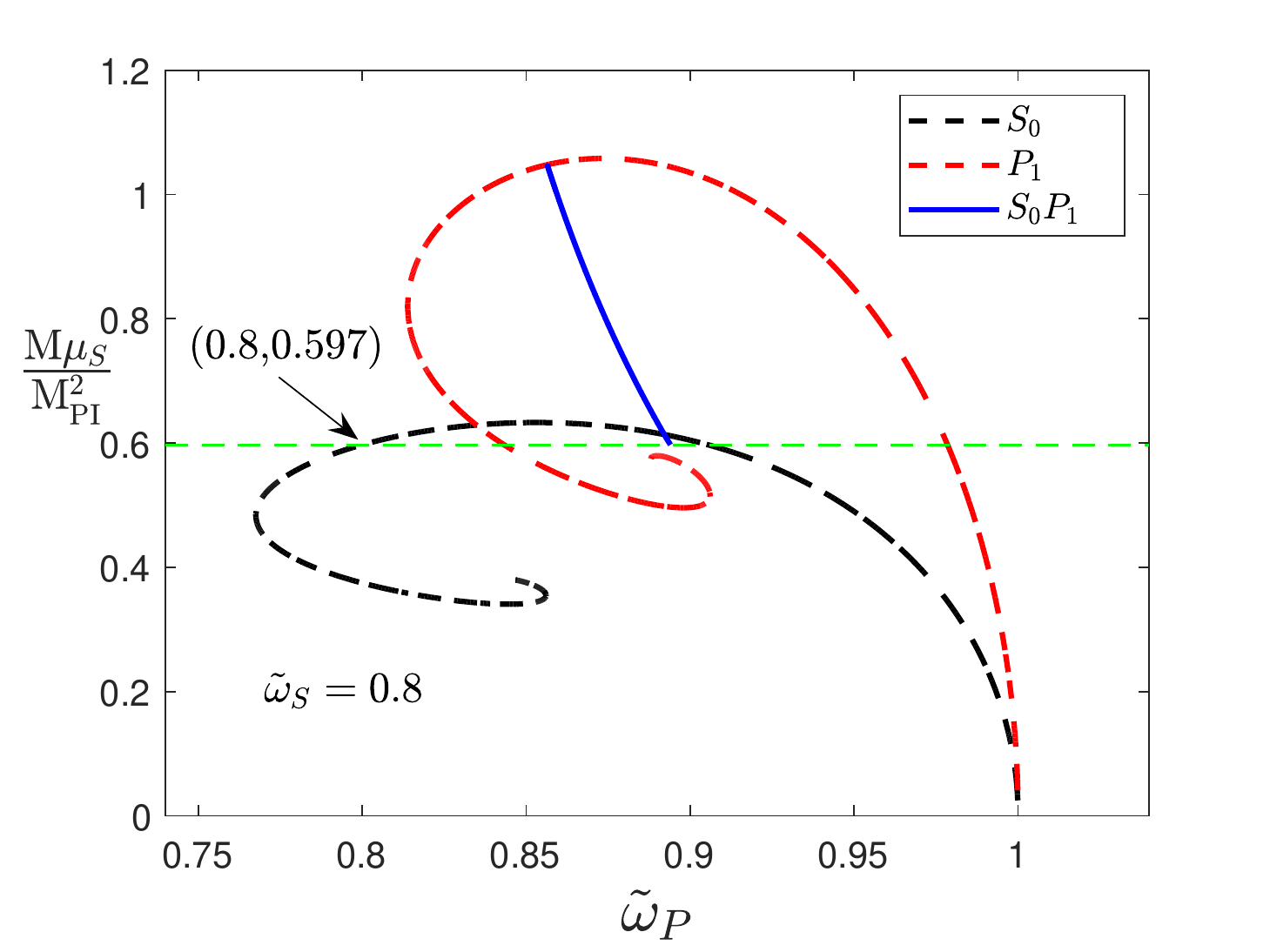}}
    \hss}
    \hbox to\linewidth{\hss
    \resizebox{9cm}{6.5cm}{\includegraphics{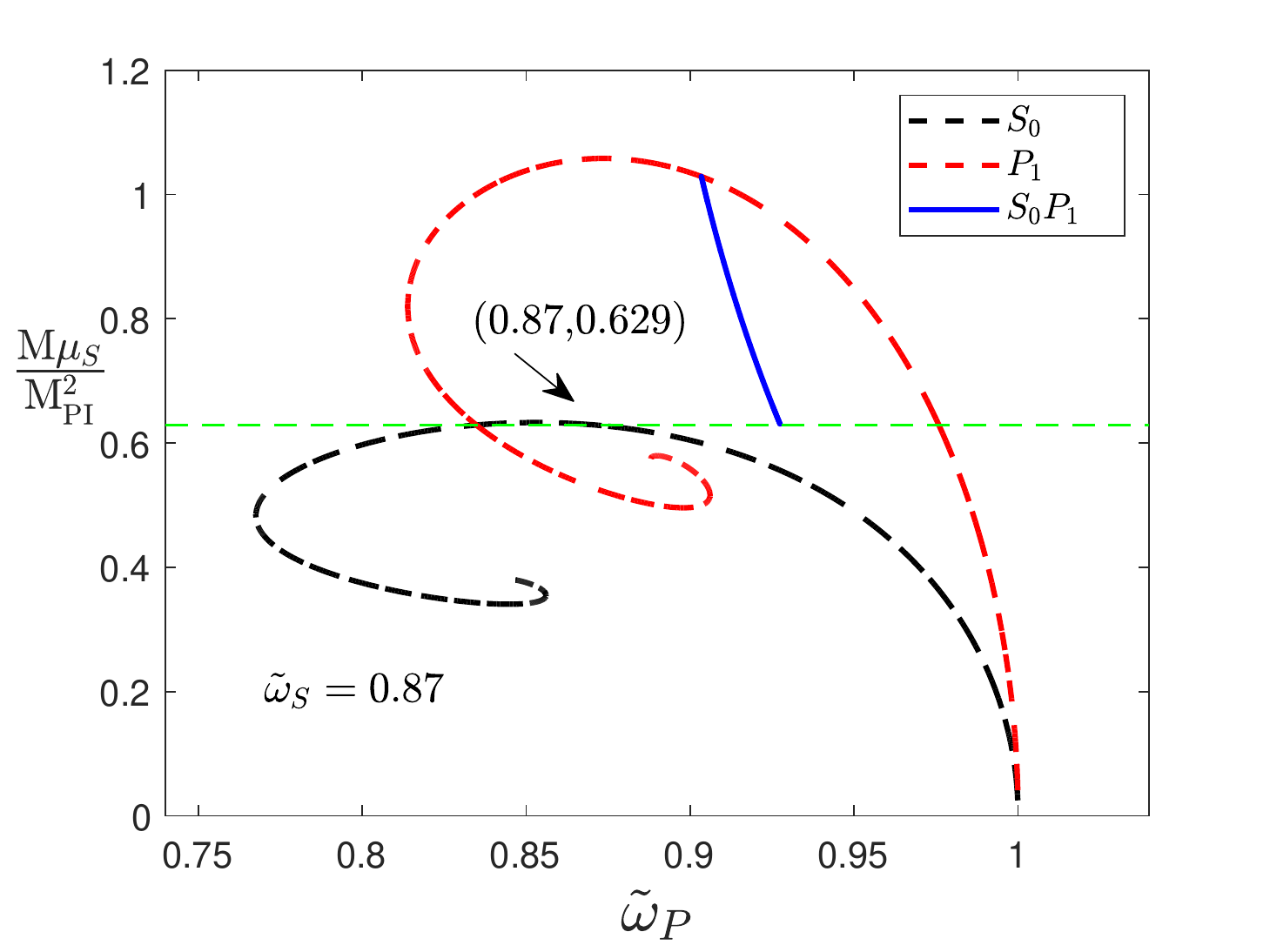}}
    \resizebox{9cm}{6.5cm}{\includegraphics{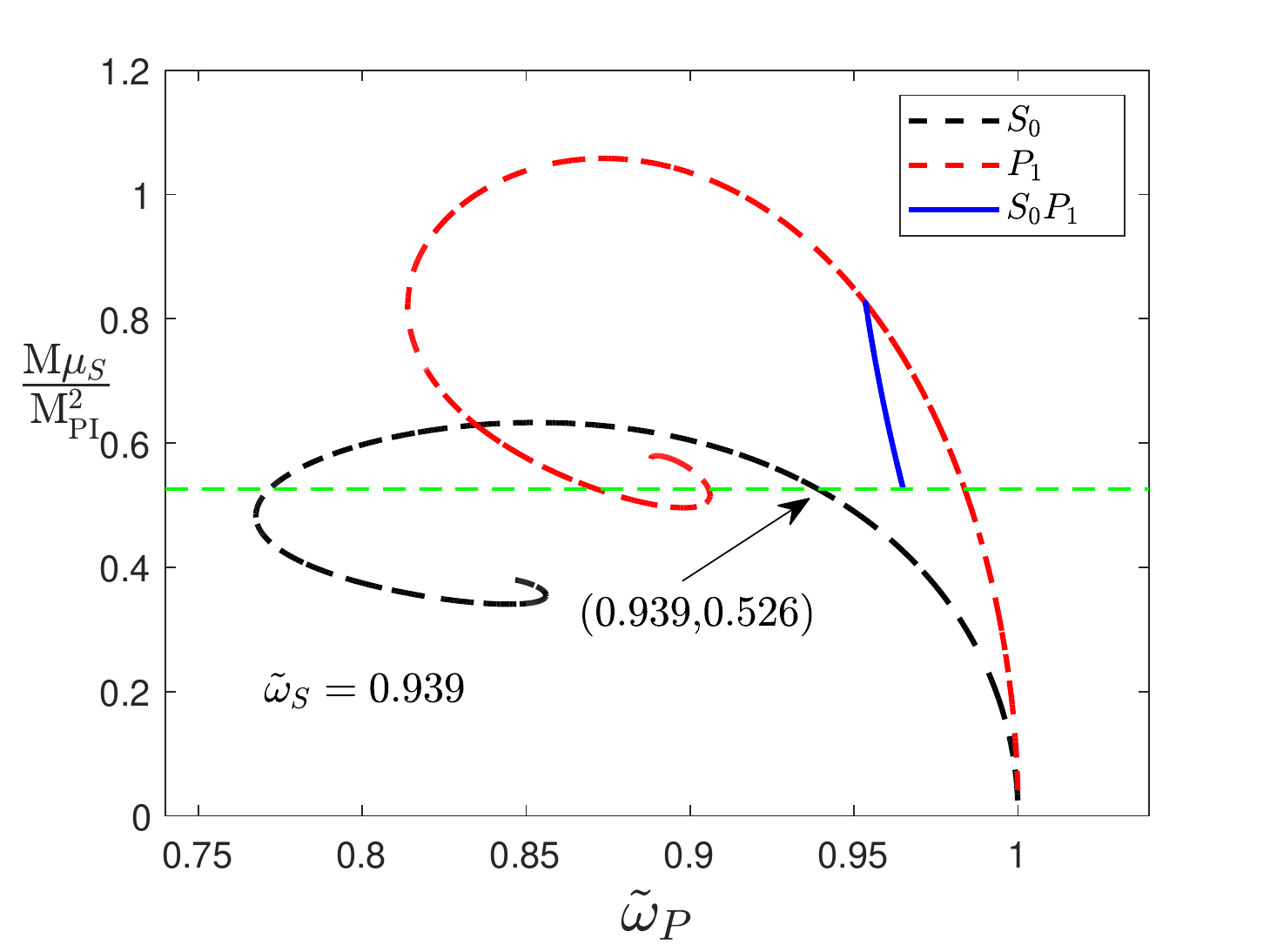}}
    \hss}
    \end{center}
    \vspace{-2em}
    \caption{The ADM mass $M$ of the Proca-boson stars as a function of the nonsynchronized frequency 
    $\tilde{\omega_P}$ for $\tilde{\omega}_S=0. 77, 0. 8, 0. 87, 0. 939$. 
    The black dashed line represents the $S_0$ state solutions and
    red dashed line represents the $P_1$ state solutions, 
    the blue line denote the coexisting state $S_0P_1$. All solutions 
    have $\tilde{\mu}_S=\tilde{\mu}_P=1$. }
    \label{ADM-one1-nonsynchronized}
\end{figure}

In Fig.~\ref{ADM-one1-nonsynchronized}, we show the relationship between the ADM mass $M$ and the nonsynchronized frequency $\tilde{\omega}_P$ when the scalar field frequency $\tilde{\omega}_S$ takes different values. This solution is similar to the one-branch solution in the case of the synchronized frequency, the left end of the mixed state solution curve still falls on the single field helix of Proca field, and the ADM mass value of the right end is equal to the ADM mass of the single field of the scalar field when the corresponding value of $\tilde{\omega}_S$ is taken in $S_0P_1$. In other words, when the nonsynchronized frequency $\tilde{\omega}_P$ reaches its maximum value, the Proca field vanishes and the mixed star becomes a boson star. Moreover, with the decrease of $\tilde{\omega}_S$, the existence range of the mixed state solution gradually increases until there is no solution. 

In Table IV, we show the existence domain of nonsynchronized frequency $\tilde{\omega}_P$ with different values of scalar field frequency $\tilde{\omega}_S$ and the values of $M_{max}$ and $M_{min}$.With the increase of $\tilde{\omega}_S$, the existence domain of $\tilde{\omega}_P$ gradually becomes narrower, and the existence domain of $\tilde{\omega}_P$ becomes quite narrow when $\tilde{\omega}_S=0.939$. In addition, with the increase of $\tilde{\omega}_S$, $M_{max}$ and $M_{min}$ first increase and then decrease.

\subsubsection{Multi-Branch}
\begin{figure}[!htbp]
    \begin{center}
    \includegraphics[height=.24\textheight]{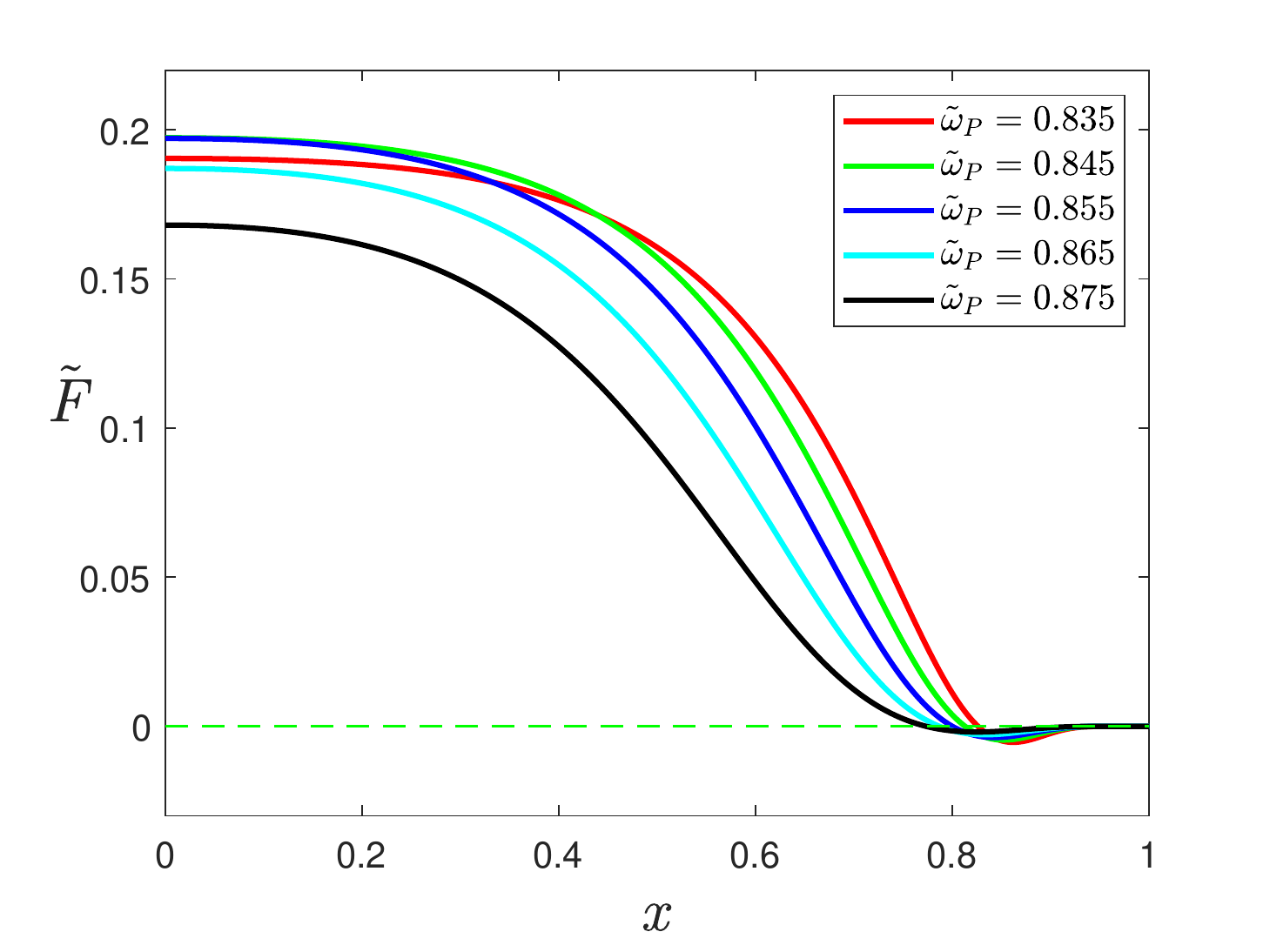}
    \includegraphics[height=.24\textheight]{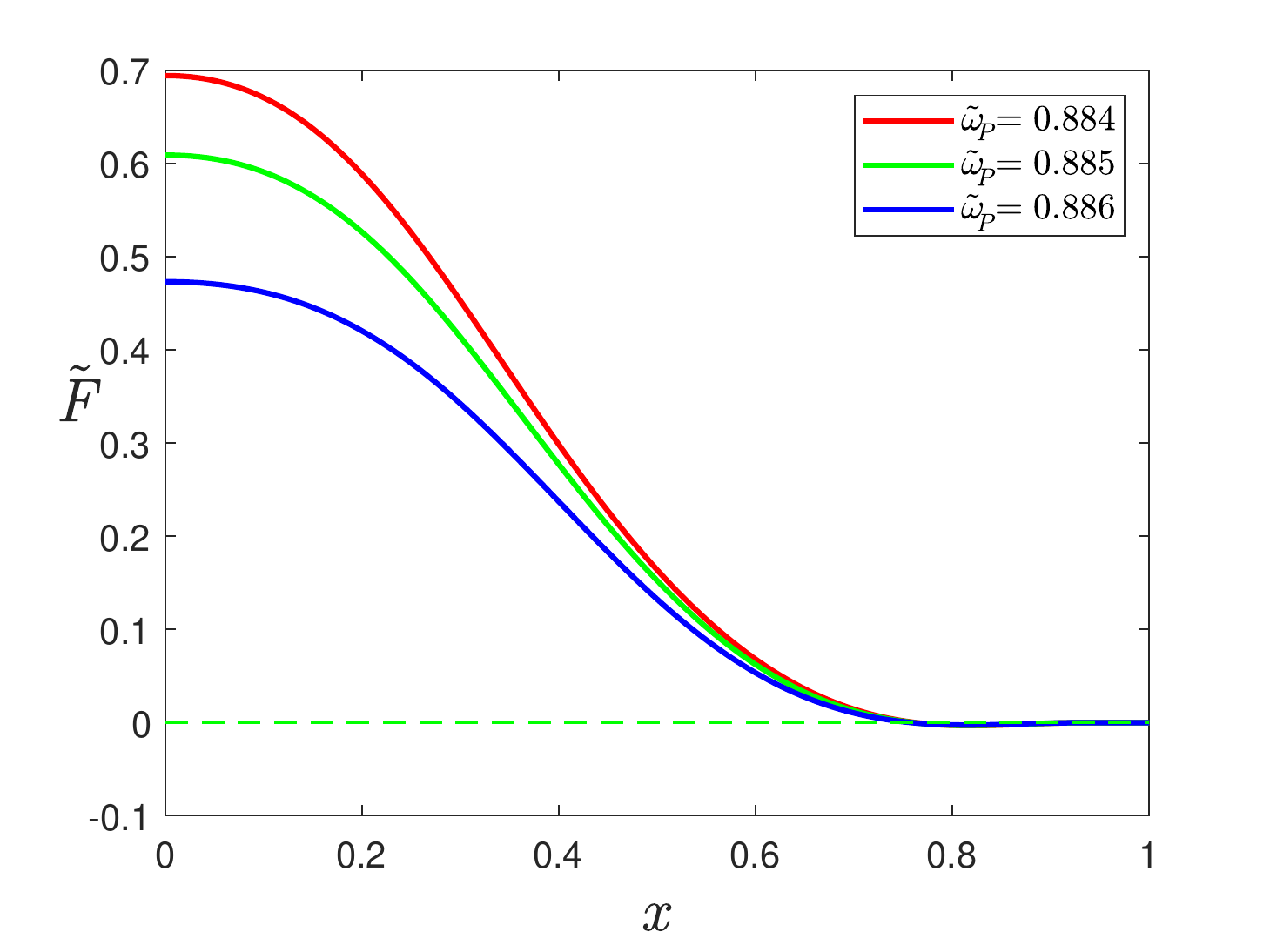}
    \includegraphics[height=.24\textheight]{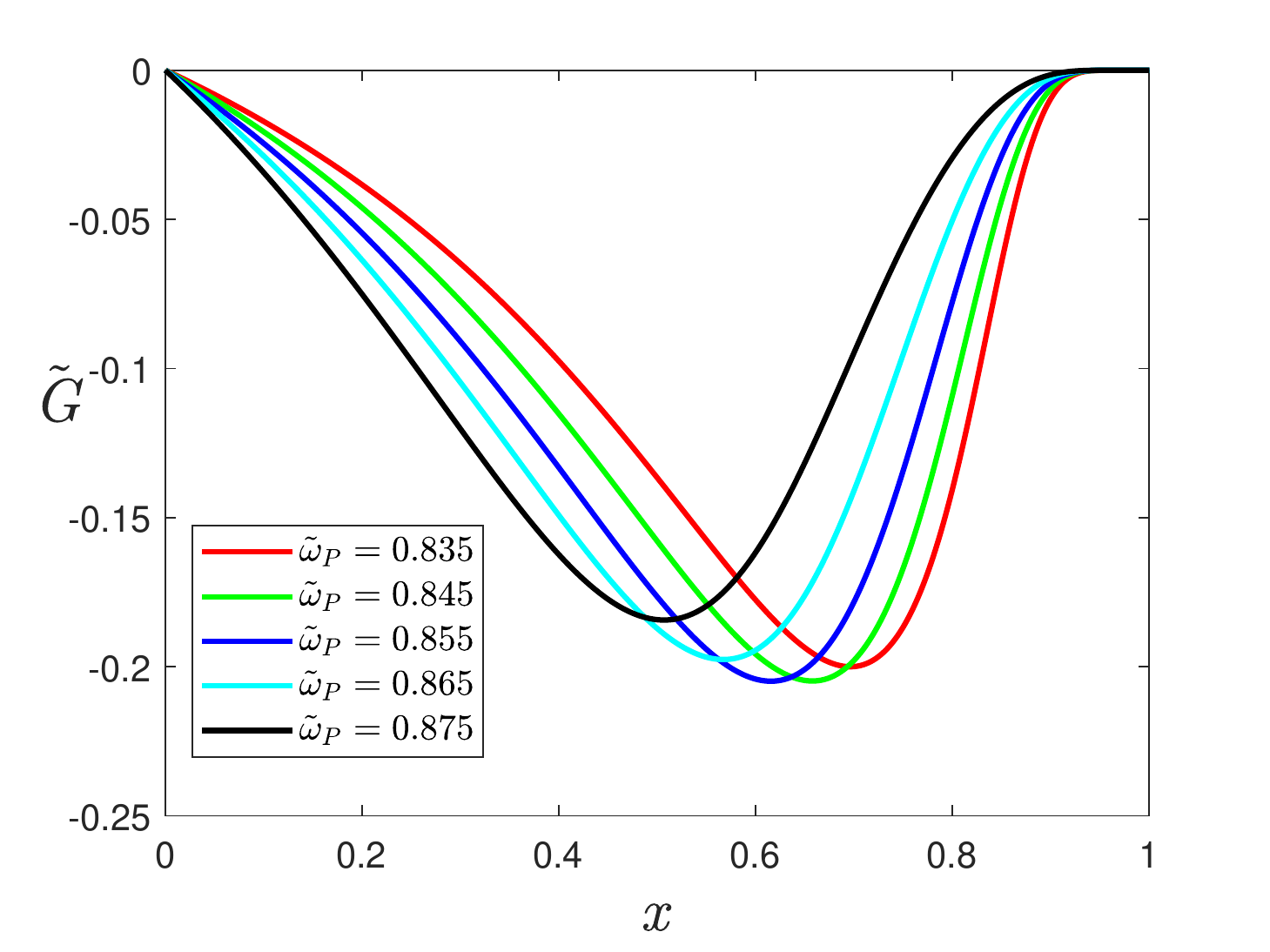}
    \includegraphics[height=.24\textheight]{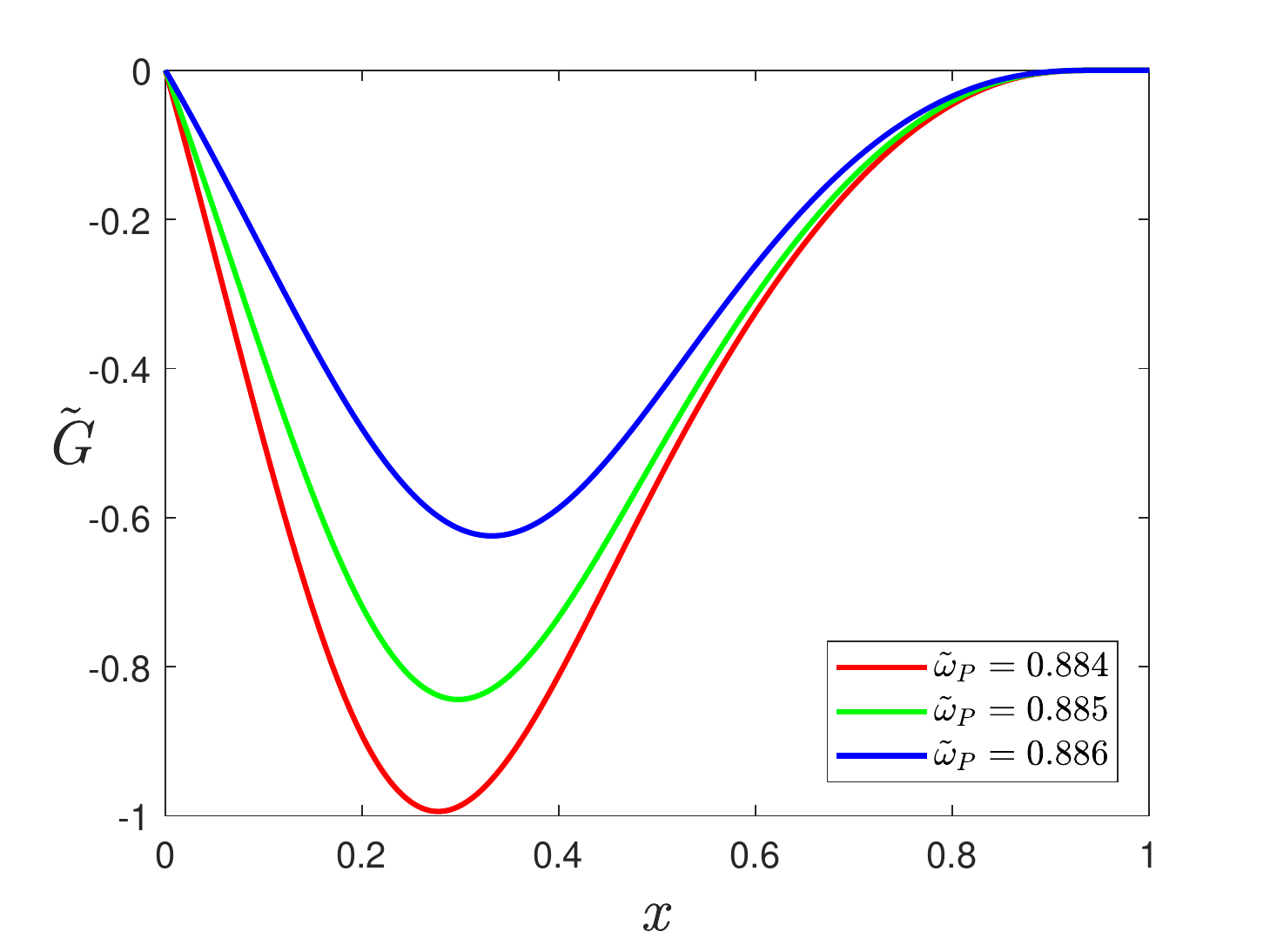}
    \includegraphics[height=.24\textheight]{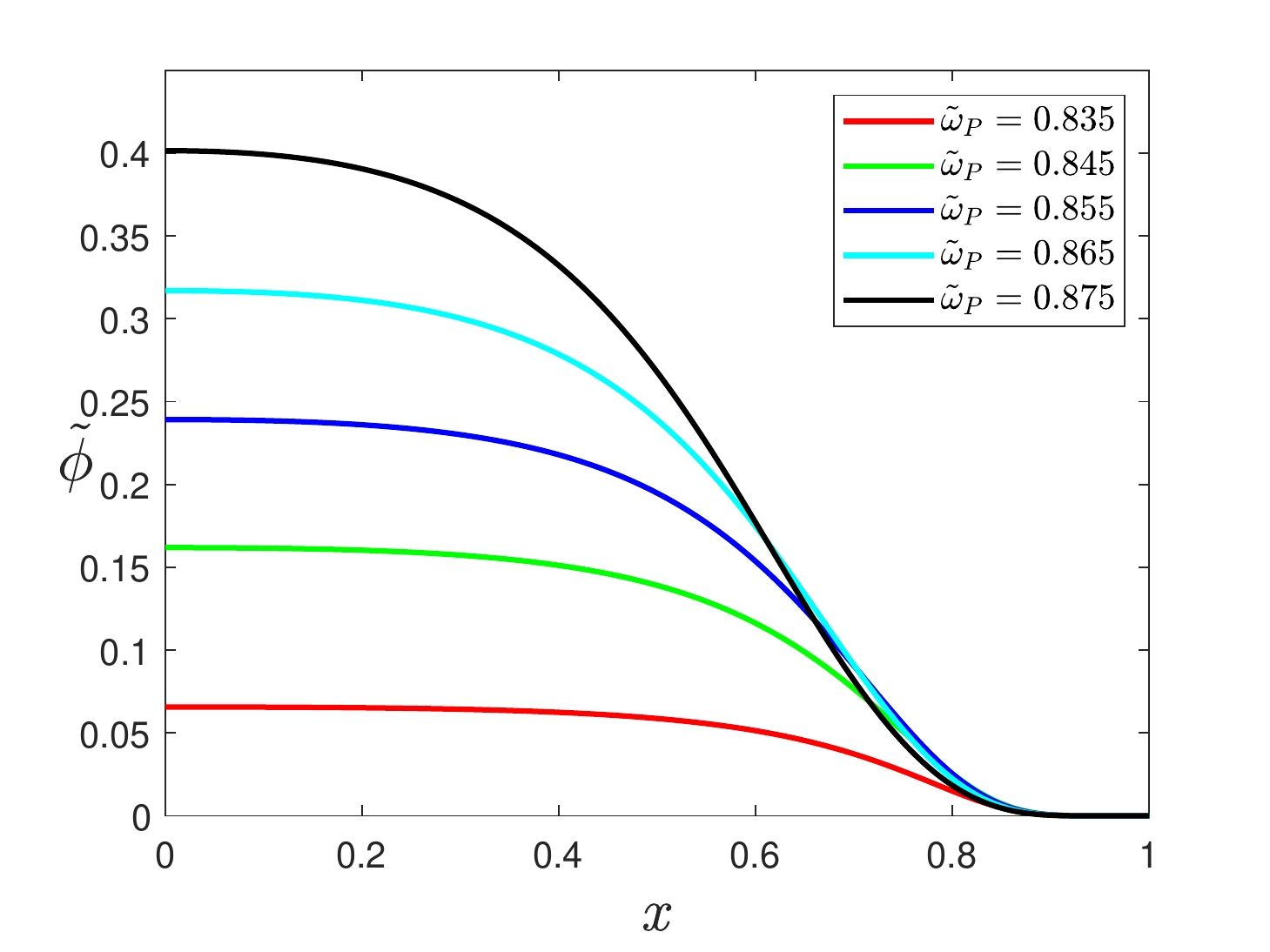}
    \includegraphics[height=.24\textheight]{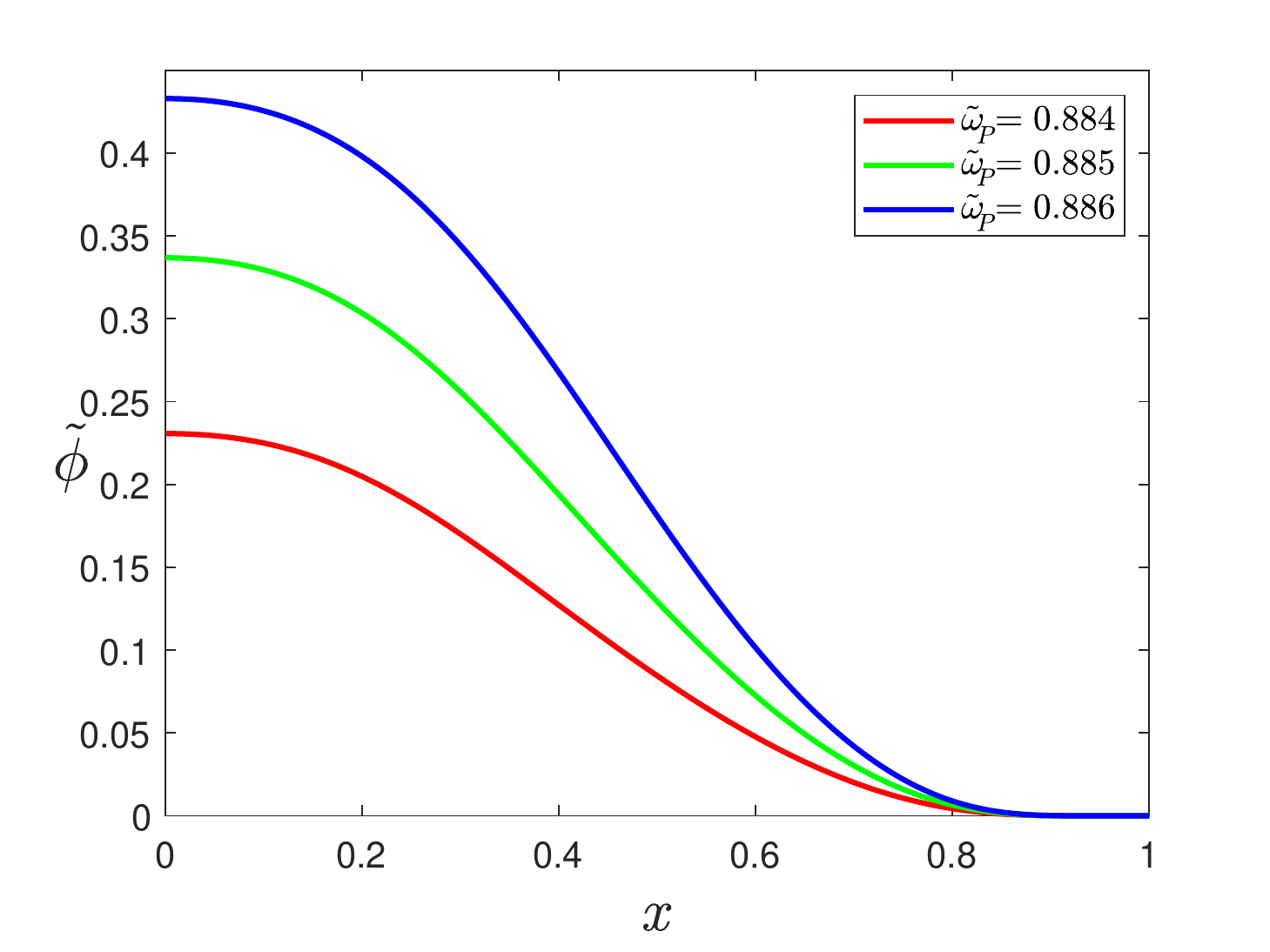}
    \end{center}
    \vspace{-2em}
    \caption{Proca field functions $\tilde{F}$(top panel) and $\tilde{G}$(middle panel)
    and scalar field function $\tilde{\phi}$(bottom panel) 
    as functions of $x$ with several values of synchronized frequency $\tilde{\omega}$, 
    where the field functions on the first and 
    second are located in the first row and second row. 
    All solutions have $\tilde{\omega}_S= 0. 76$ and $\tilde{\mu}_S=\tilde{\mu}_P=1$. }
    \label{field-multi-nonsynchronized}
\end{figure}

Fig.~\ref{field-multi-nonsynchronized} shows the relationship between the field function $\tilde{F}$, $\tilde{G}$, $\tilde{\phi}$ and the nonsynchronized frequency $\tilde{\omega}_P$ of multi-branch solutions. The graphs in the first column and the second column represent the first branch and the second two-branch respectively. For the first branch, $\left\lvert \tilde{F} \right\rvert _{max}$ and $\left\lvert \tilde{G} \right\rvert _{max}$ first increase then decrease with the frequency of nonsynchronized $\tilde{\omega}_P$, $\left\lvert \tilde{\phi} \right\rvert _{max}$ increases with the increase of nonsynchronized frequency $\tilde{\omega}_P$. For the second branch, $\left\lvert \tilde{F} \right\rvert _{max}$ and $\left\lvert\tilde{G} \right\rvert _{max}$ decrease with the increase of nonsynchronized frequency $\tilde{\omega}$, $\left\lvert \tilde{\phi} \right\rvert _{max}$ still increases as the nonsynchronized frequency $\tilde{\omega}_P$ increases. 

\begin{figure}[!htbp]
    \begin{center}
    \hbox to\linewidth{\hss
    \resizebox{9cm}{6.5cm}{\includegraphics{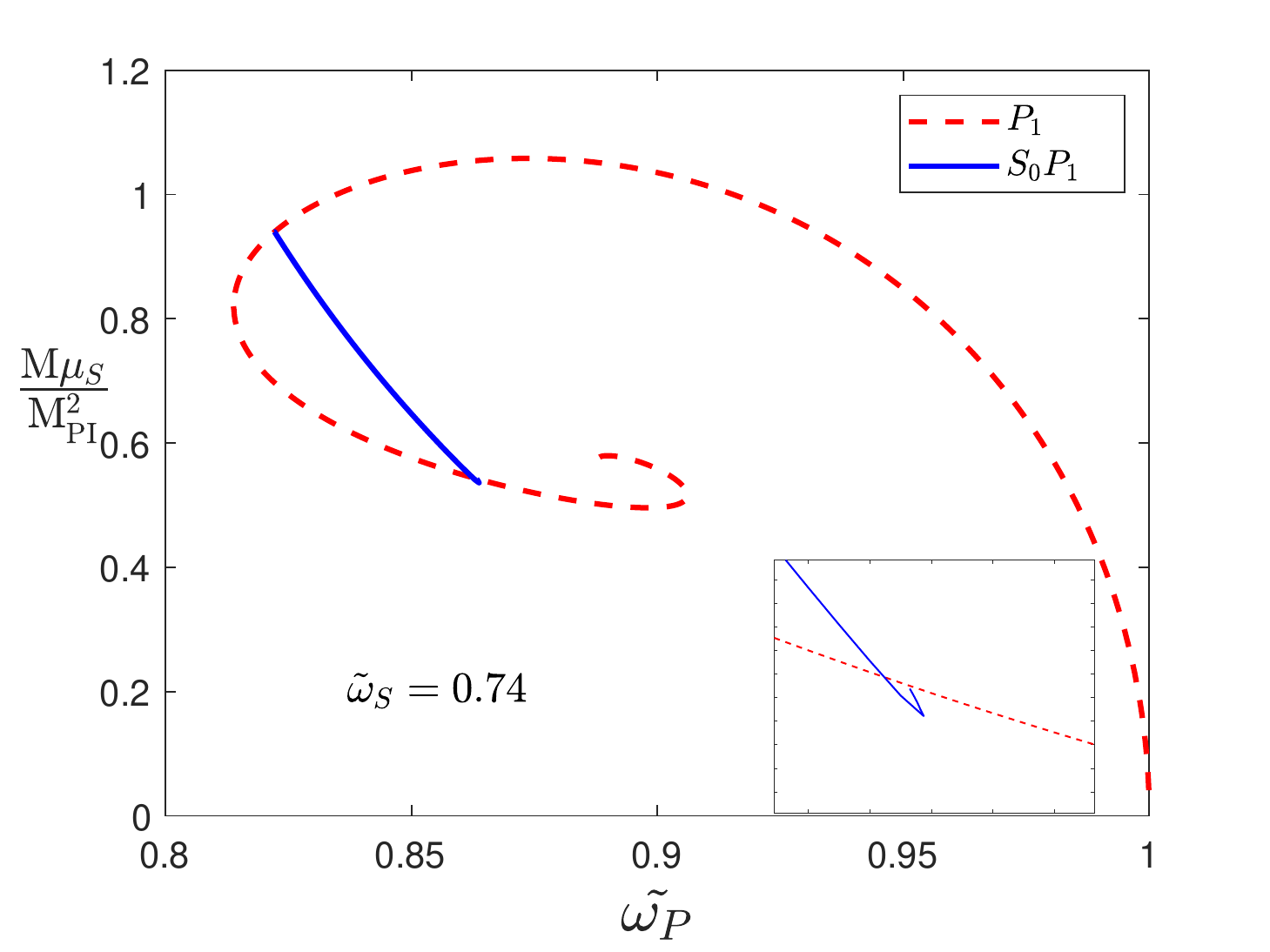}}
    \resizebox{9cm}{6.5cm}{\includegraphics{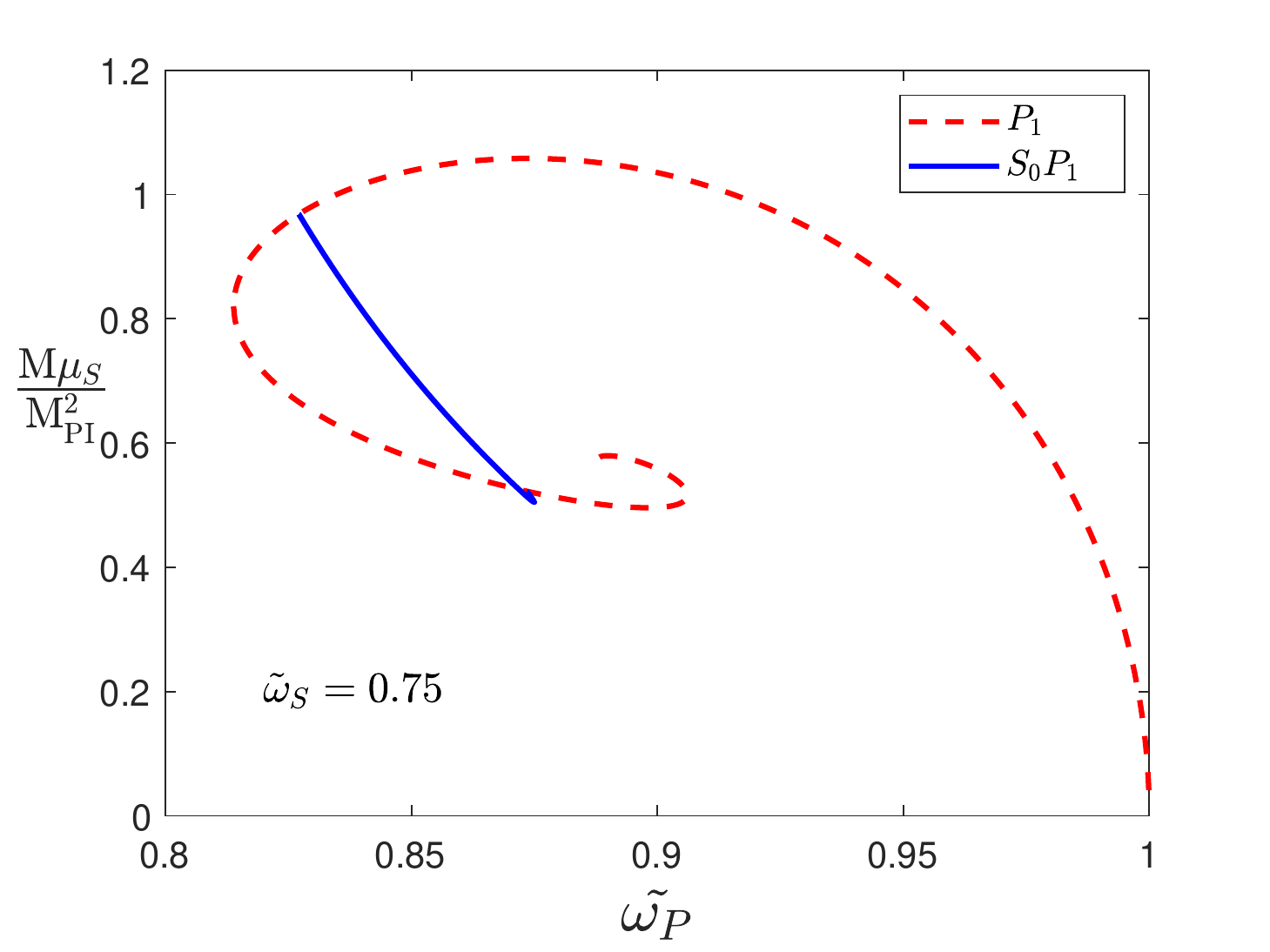}}
    \hss}
    \hbox to\linewidth{\hss
    \resizebox{9cm}{6.5cm}{\includegraphics{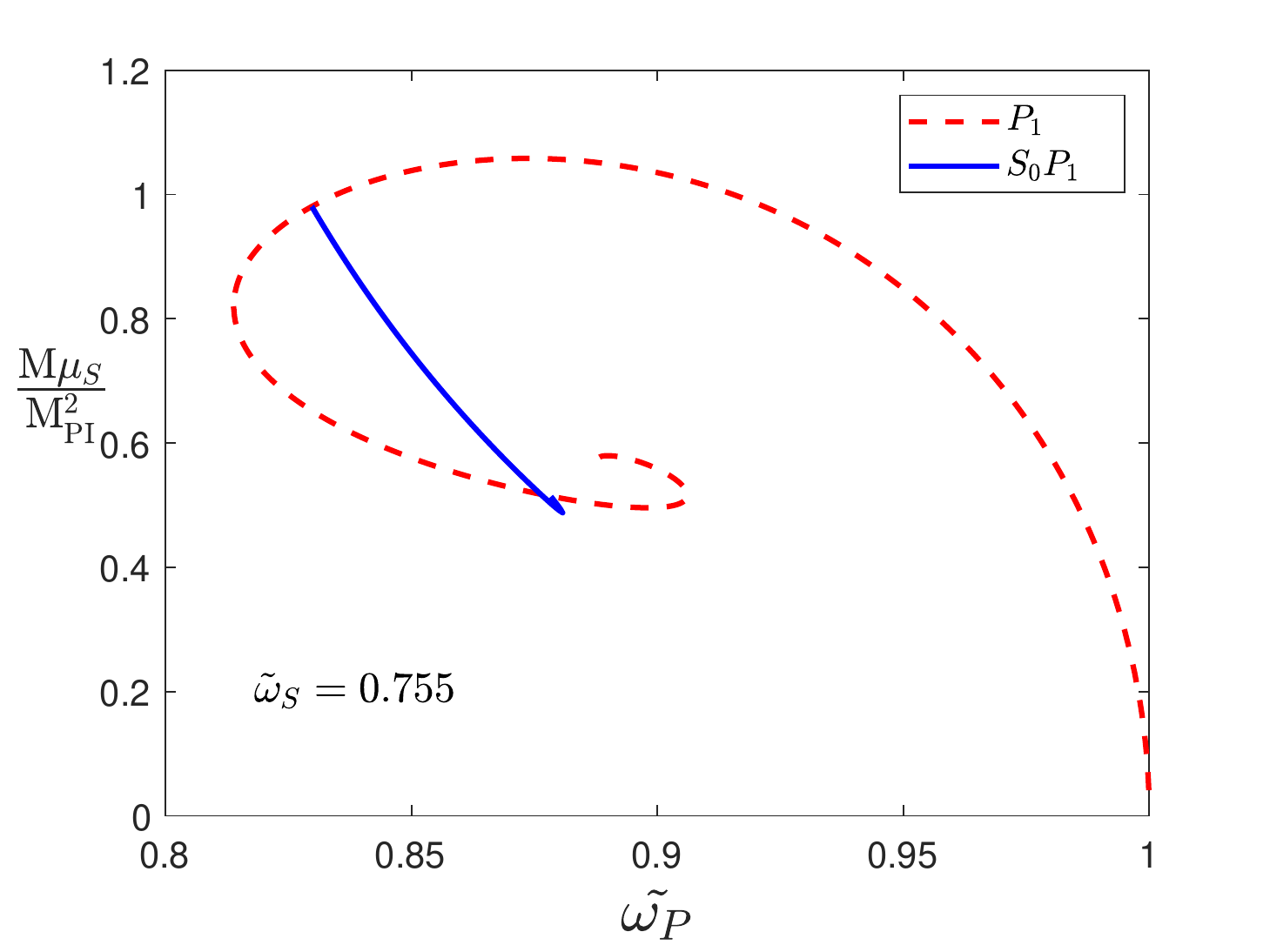}}
    \resizebox{9cm}{6.5cm}{\includegraphics{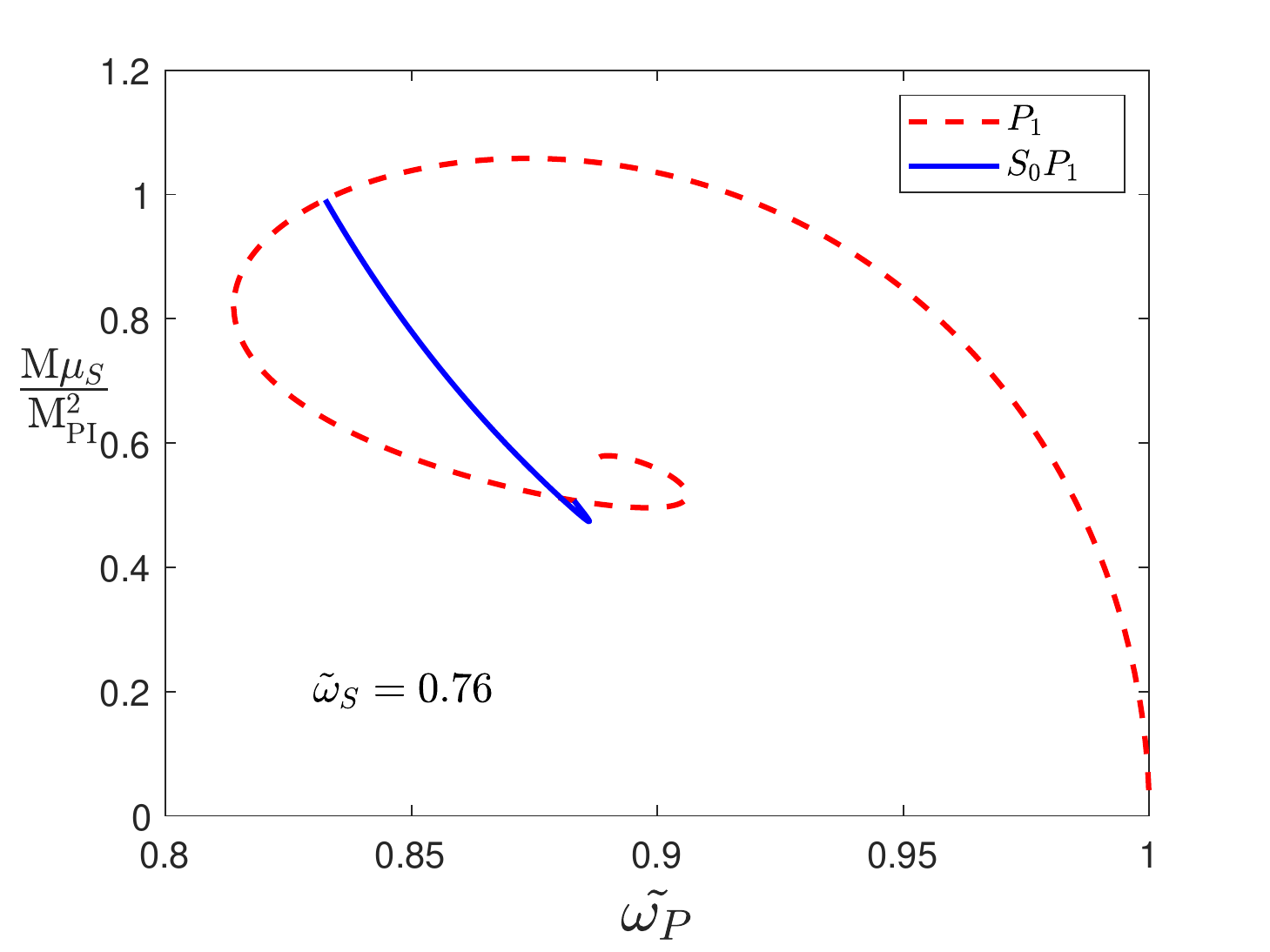}}
    \hss}
    \end{center}
    \vspace{-2em}
    \caption{The ADM mass $M$ of the Proca-boson stars as a function of the nonsynchronized frequency 
    $\tilde{\omega}_P$ for $\tilde{\omega}_S=0. 74, 0. 75, 0. 755, 0. 76$. 
    The red dashed line represents the $P_1$ state solutions, 
    and the blue line denote the coexisting state $S_0P_1$. 
    All solutions have $\tilde{\mu}_S=\tilde{\mu}_P=1$. }
    \label{ADM-multi-nonsynchronized}
\end{figure}

In Fig.~\ref{ADM-multi-nonsynchronized}, we show the relationship between the ADM mass $M$ of the mixed state solution and the nonsynchronized frequency $\tilde{\omega}_P$ when taking different values of $\tilde{\omega}_S$. It can be seen that this kind of solution is similar to the two-branch solution in the case of the synchronized frequency case. The first branch also extends from the single field curve of Proca field and returns to form the second branch at the inflection point. The end point of the second branch intersects with the single field curve of Proca field again at another place. At this time, the scalar field vanishes and the mixed star becomes a Proca star. The difference is that the two-branch solution in the synchronized frequency case is quite smooth around the inflection point, while the inflection point in the mixed state curve in the nonsynchronized frequency case is sharp. This also reflects the main difference between the synchronized frequency case and the nonsynchronized frequency case. 

In Table \ref{table5}, we show the existence domain of nonsynchronized frequency $\tilde{\omega}_P$ with different values of scalar field frequency $\tilde{\omega}_S$ and the values of $M_{max}$ and $M_{min}$. With the decrease of $\tilde{\omega}_S$, the existence domain of $B_1$ and $B_2$ gradually becomes narrower. When $\tilde{\omega}_S=0.74$, the second branch almost disappears, $M_{max}$ decreases with the decrease of $\tilde{\omega}_S$, and $M_{min}$ vice versa.

\subsubsection{One-Branch-B}
\begin{figure}[!htbp]
    \begin{center}
    \includegraphics[height=.24\textheight]{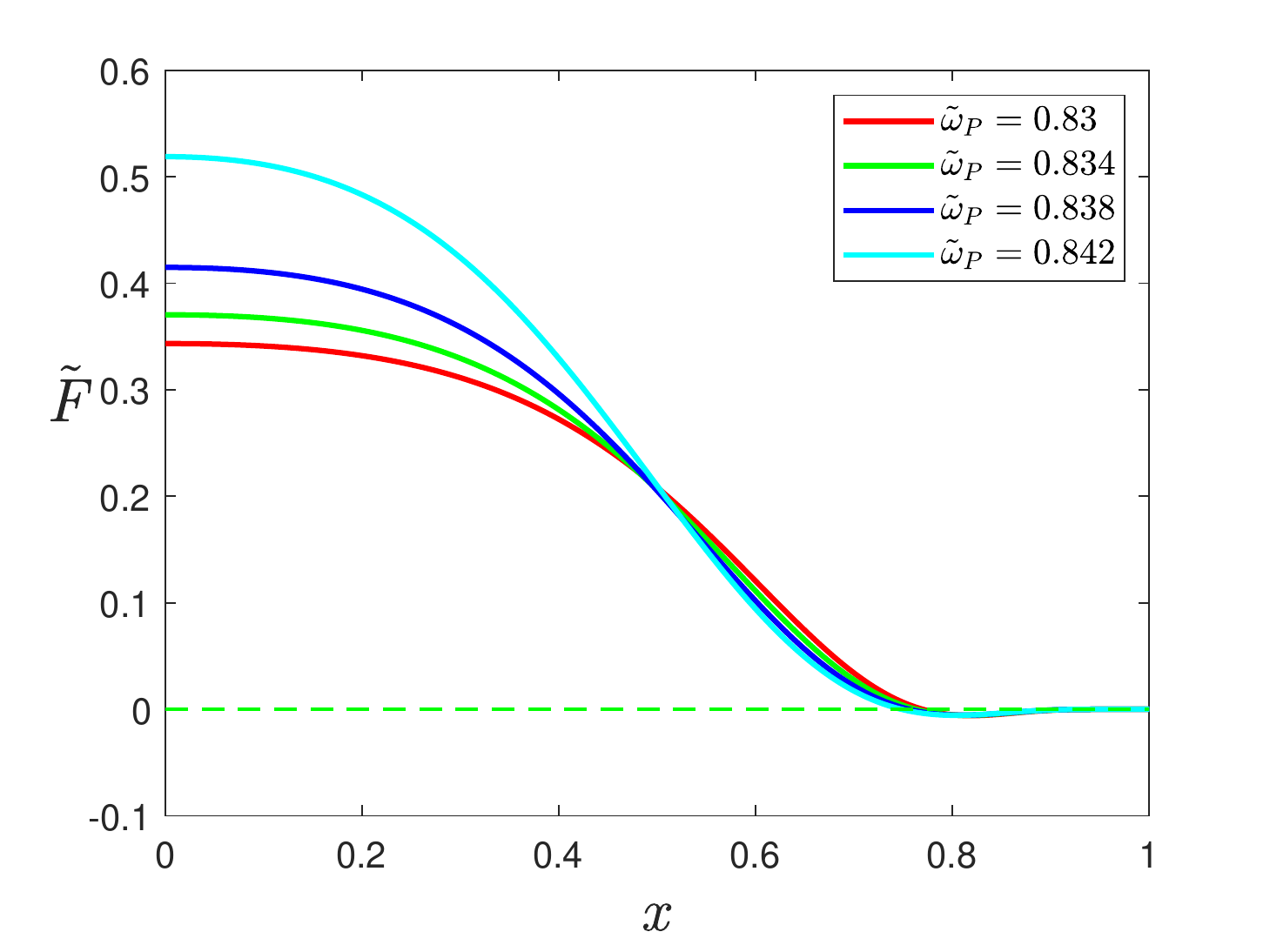}
    \includegraphics[height=.24\textheight]{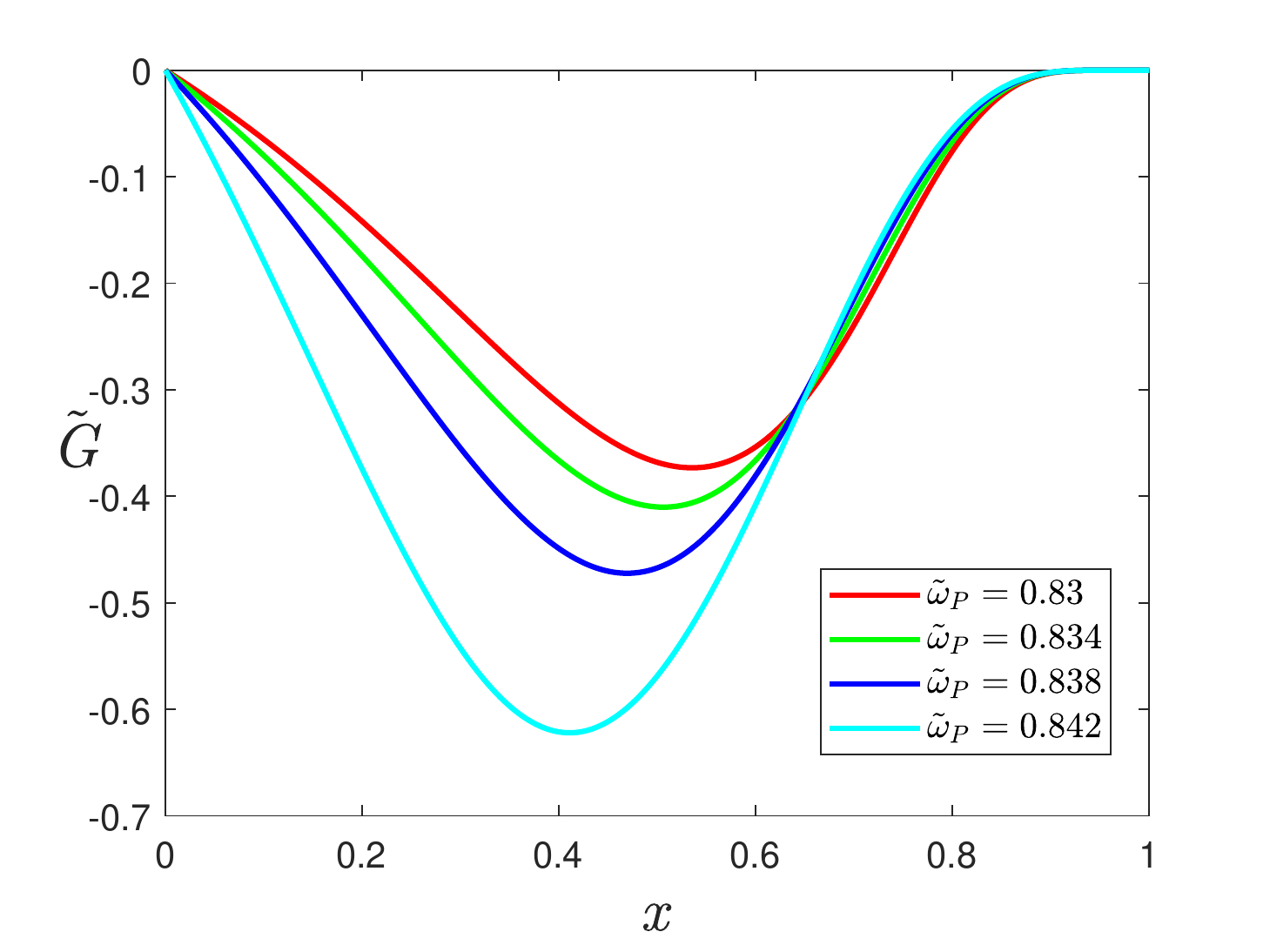}
    \includegraphics[height=.24\textheight]{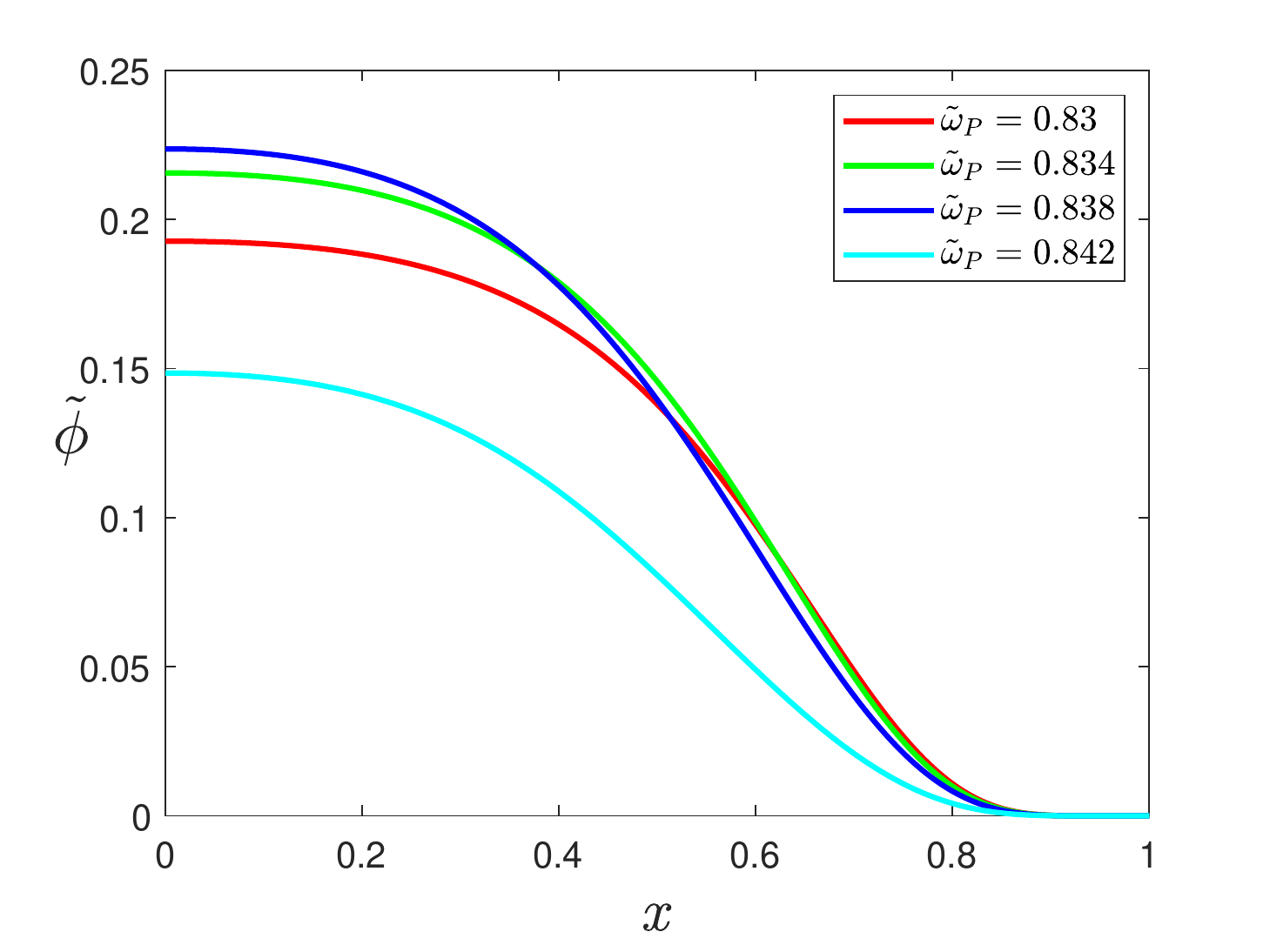}
    \end{center}
    \vspace{-2em}
    \caption{Proca field function $\tilde{F}$(left panel) and $\tilde{G}$(middle panel) 
    and scalar field functions 
    $\tilde{\phi}$(right panel) as functions of $x$ with 
    $\tilde{\omega}_P = 0. 83, 0. 834, 0. 838, 0.842$. All solutions have $\tilde{\omega}_S= 0. 722$ 
    and $\tilde{\mu}_S = \tilde{\mu}_P = 1$. }
    \label{field-one2-nonsynchronized}
\end{figure}

For one-branch-B class solutions, the image of the field functions $\tilde{F}$, $\tilde{G}$, $\tilde{\phi}$ is shown in Fig.~\ref{field-one2-nonsynchronized}. For scalar field function, $\left\lvert \tilde{\phi} \right\rvert _{max}$ increases first and then decreases with the increase of nonsynchronized frequency $\tilde{\omega}_P$. For the Proca field functions, $\left\lvert \tilde{F} \right\rvert _{max}$ and $\left\lvert \tilde{G} \right\rvert _{max}$ with nonsynchronized frequency $\tilde{\omega}_P $ increases. 

\begin{figure}[!htbp]
    \begin{center}
    \hbox to\linewidth{\hss
    \resizebox{9cm}{6.5cm}{\includegraphics{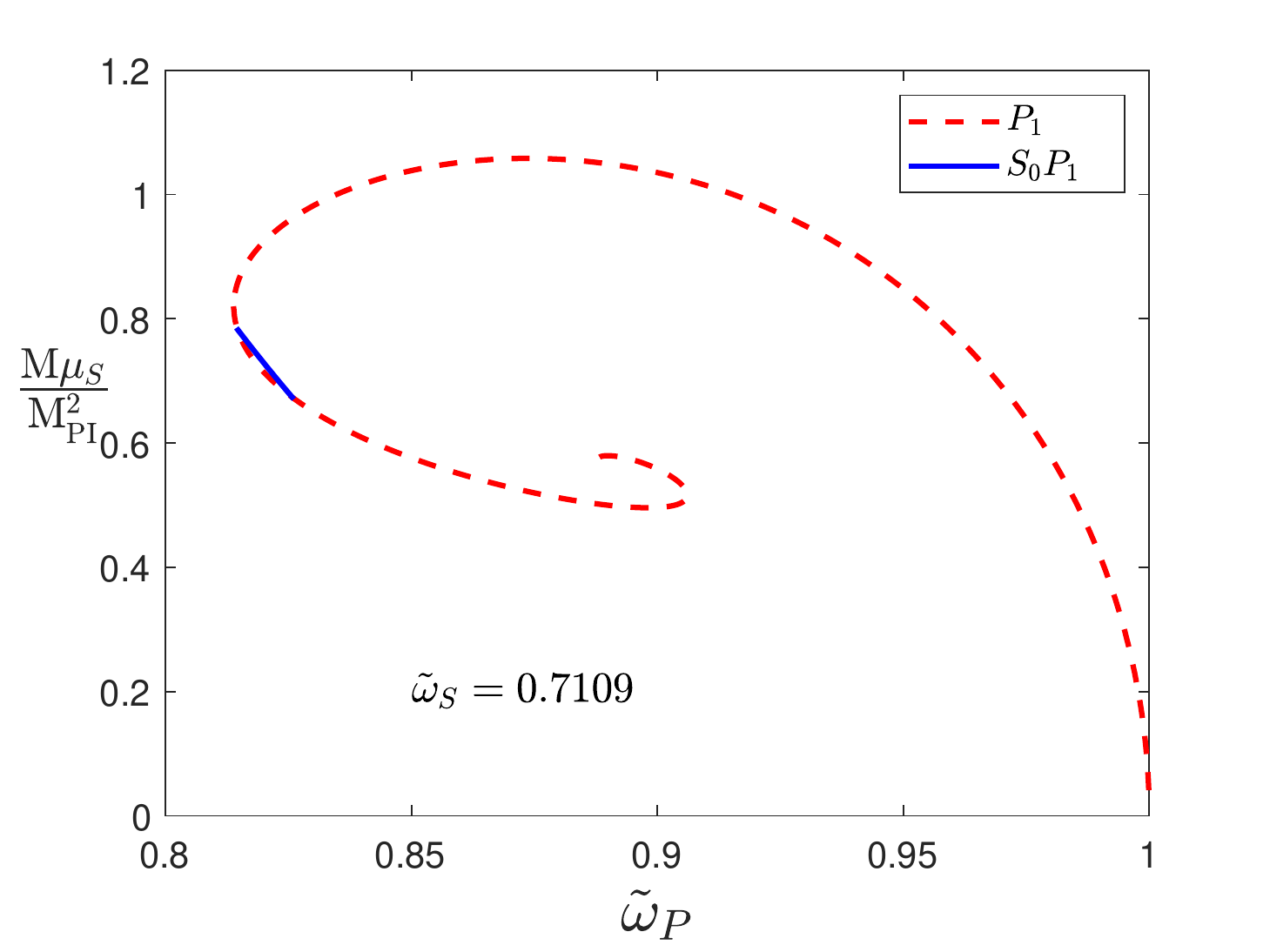}}
    \resizebox{9cm}{6.5cm}{\includegraphics{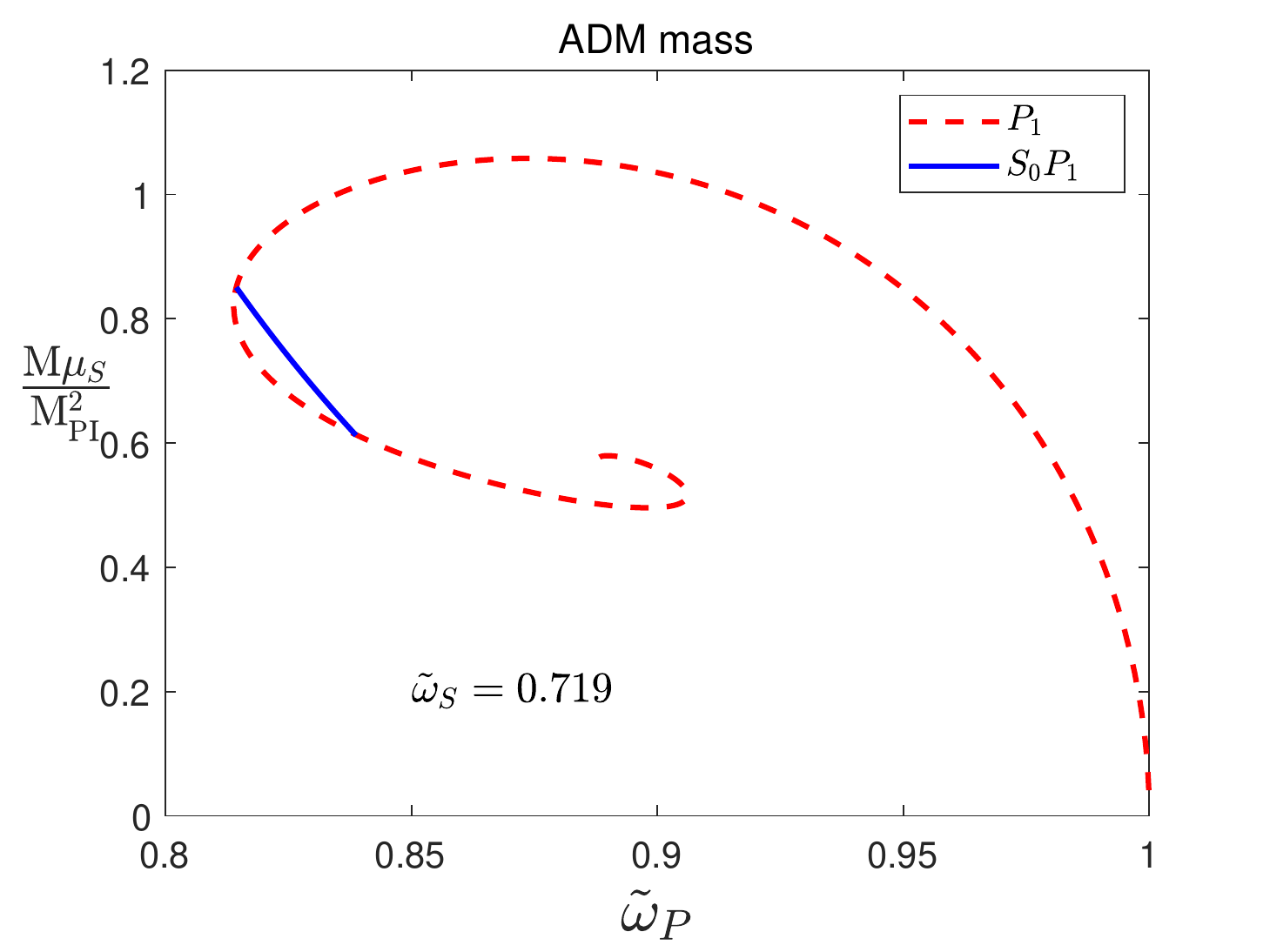}}
    \hss}
    \hbox to\linewidth{\hss
    \resizebox{9cm}{6.5cm}{\includegraphics{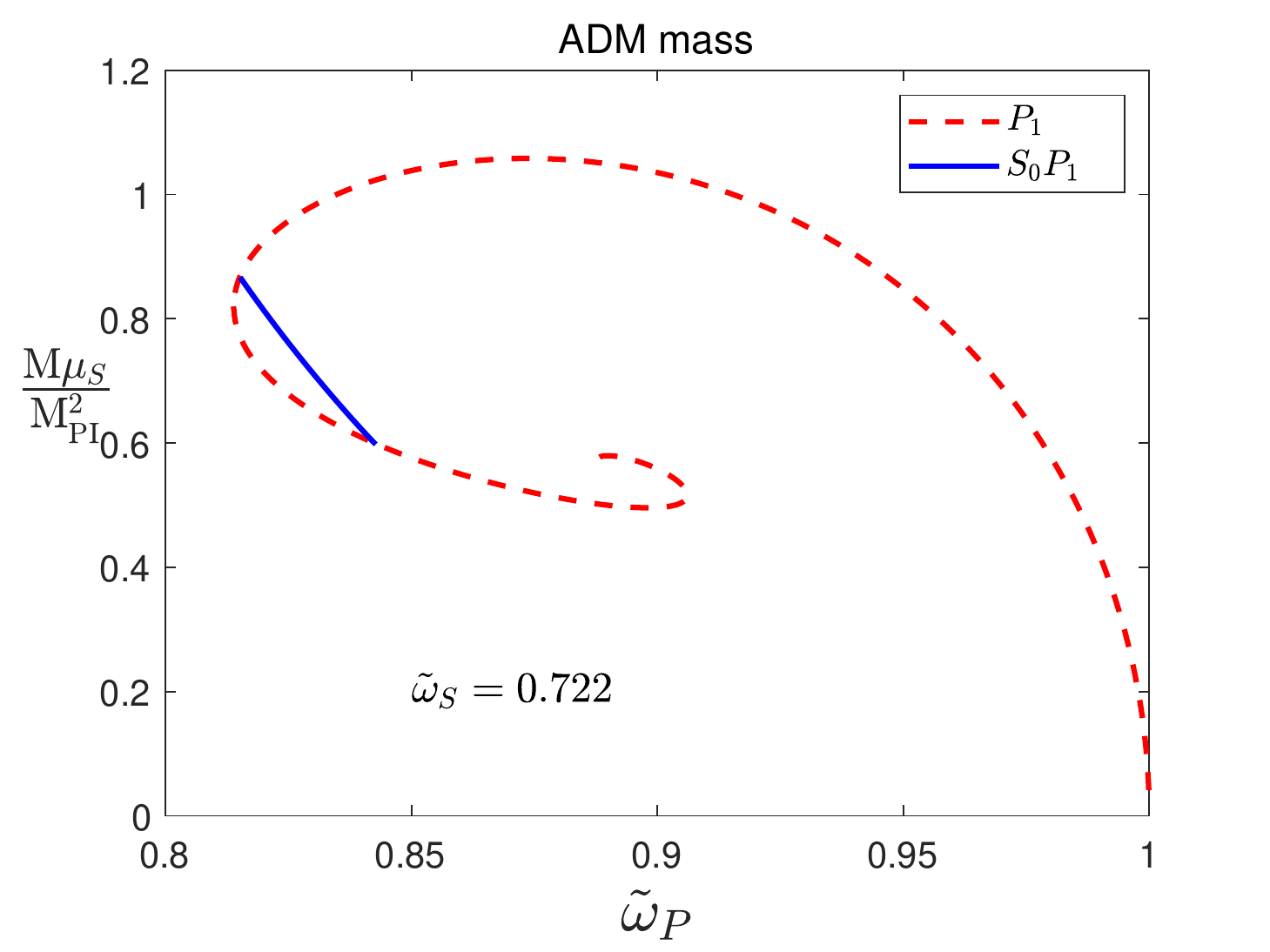}}
    \resizebox{9cm}{6.5cm}{\includegraphics{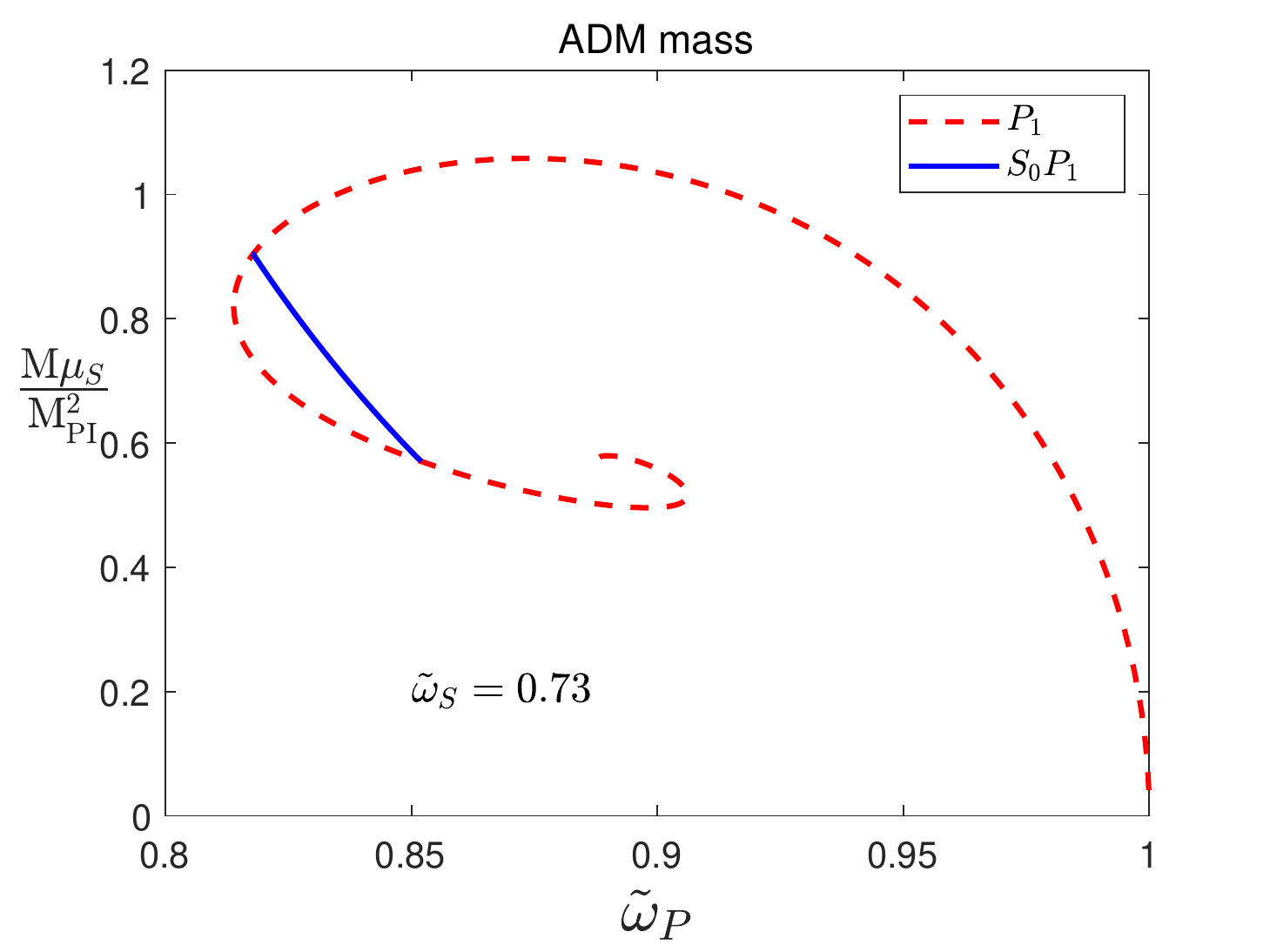}}
    \hss}    
    \end{center}
    \vspace{-2em}
    \caption{The ADM mass $M$ of the Proca-boson stars as a function of the nonsynchronized frequency 
    $\tilde{\omega_P}$ for $\tilde{\omega}_S=0. 7109, 0. 719, 0. 722, 0. 73$. 
    The red dashed line represents the $P_1$ state solutions, 
    and the blue line denote the coexisting state $S_0P_1$. All solutions 
    have $\tilde{\mu}_S=\tilde{\mu}_P=1$. }
    \label{ADM-one2-nonsynchronized}
\end{figure}

Fig.~\ref{ADM-one2-nonsynchronized} shows the relationship between the ADM mass $M$ of the mixed star and the nonsynchronized frequency $\tilde{\omega}_P$ when the scalar field frequency $\tilde{\omega}_S$ takes different values. According to Fig.~\ref{ADM-multi-nonsynchronized}, when $\tilde{\omega}_S=0. 74$, it is still two-branch solution. If $\tilde{\omega}_S$ continues to decrease, the second branch of the two-branch solution will disappear, and the mixed state will become another one-branch solution. At this time, the one-branch solution is different from the one-branch-A solution family, and the two ends of the one-branch-A solution family intersect the Proca field curve and the scalar field curve respectively. That is, the Proca field function must not disappear during the increase of $\tilde{\omega}_P$. For one-branch-B solutions, both ends of the mixed state curve are on the Proca field curve, that is, when $\tilde{\omega}_P$ reaches its minimum or maximum, the mixed star is a Proca star. As $\tilde{\omega}_P$ increases, $\left\lvert\tilde{\phi} \right\rvert _{max}$ first increases and then decreases, disappearing when $\tilde{\omega}_S$ reaches its maximum value, and the mixed star becomes Proca star again. 

In Table \ref{table6}, we show the existence domain of nonsynchronized frequency $\tilde{\omega}_P$ with different values of scalar field frequency $\tilde{\omega}_S$ and the values of $M_{max}$ and $M_{min}$. With the decrease of $\tilde{\omega}_S$, the existence domain of $\tilde{\omega}_P$ becomes narrower and $M_{max}$ gradually decreases, while $M_{min}$ is vice versa.

\subsection{Binding Energy}
\begin{figure}[!htbp]
    \begin{center}
    \includegraphics[height=.26\textheight]{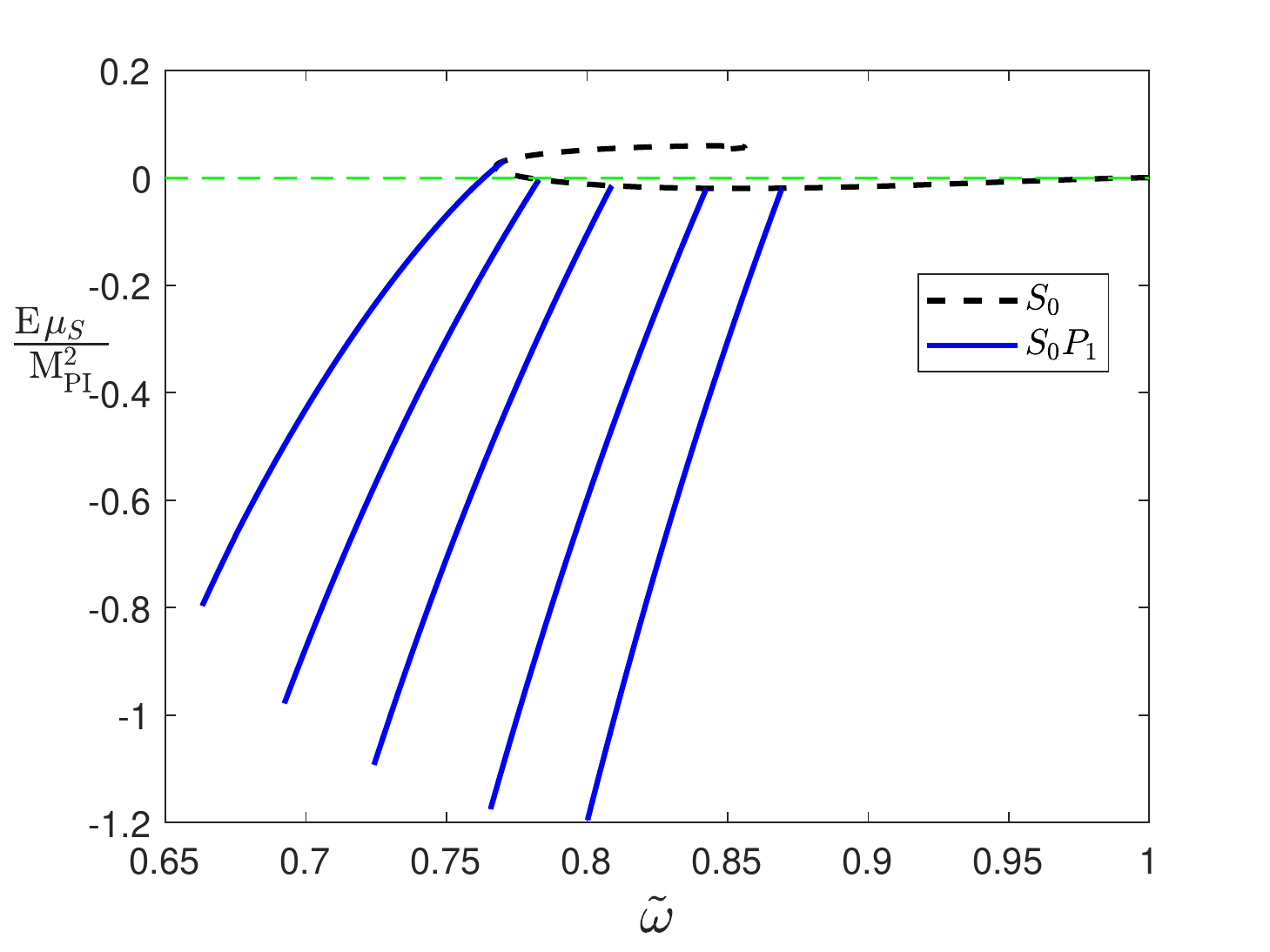}
    \includegraphics[height=.26\textheight]{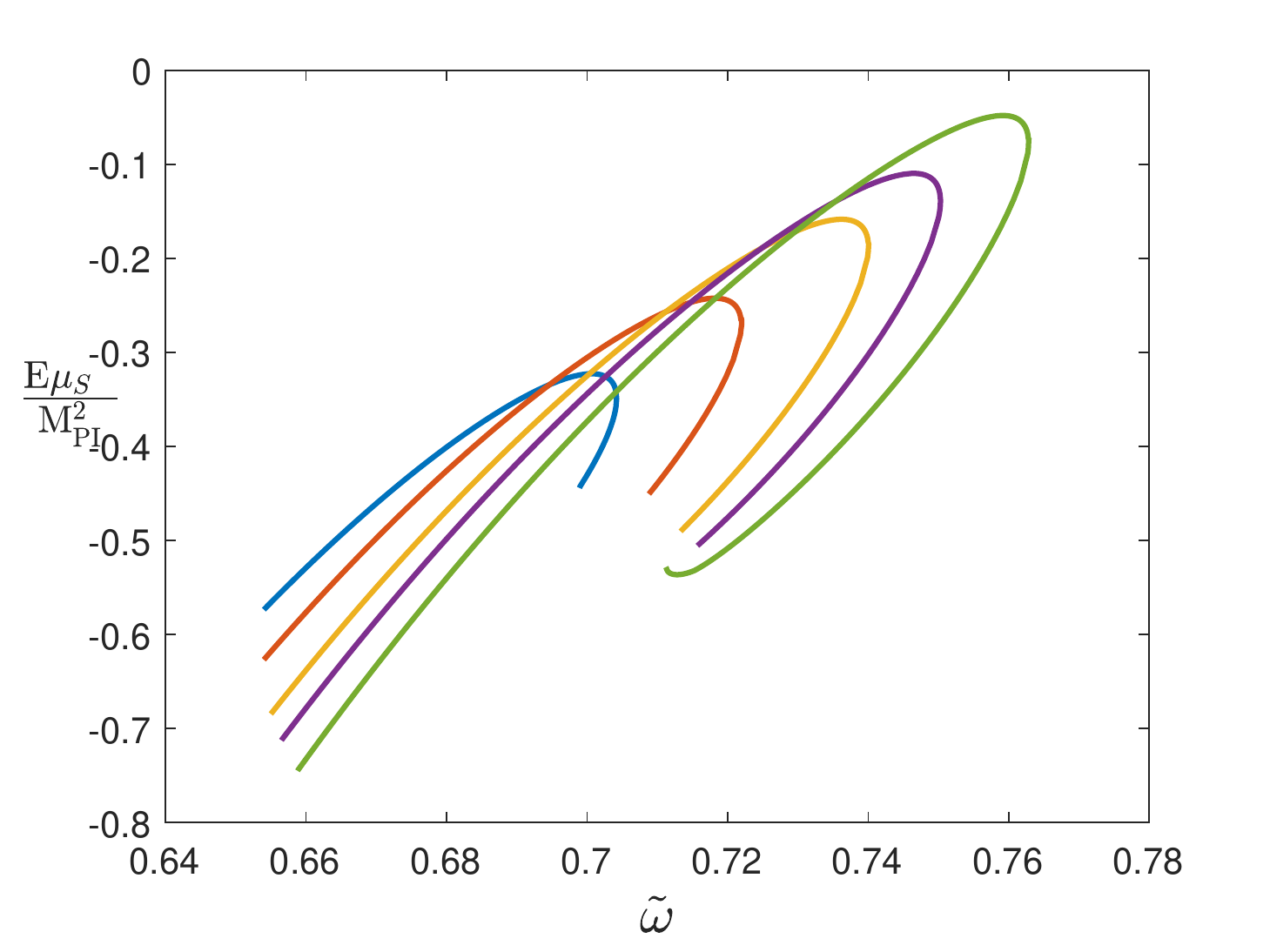}
    \includegraphics[height=.26\textheight]{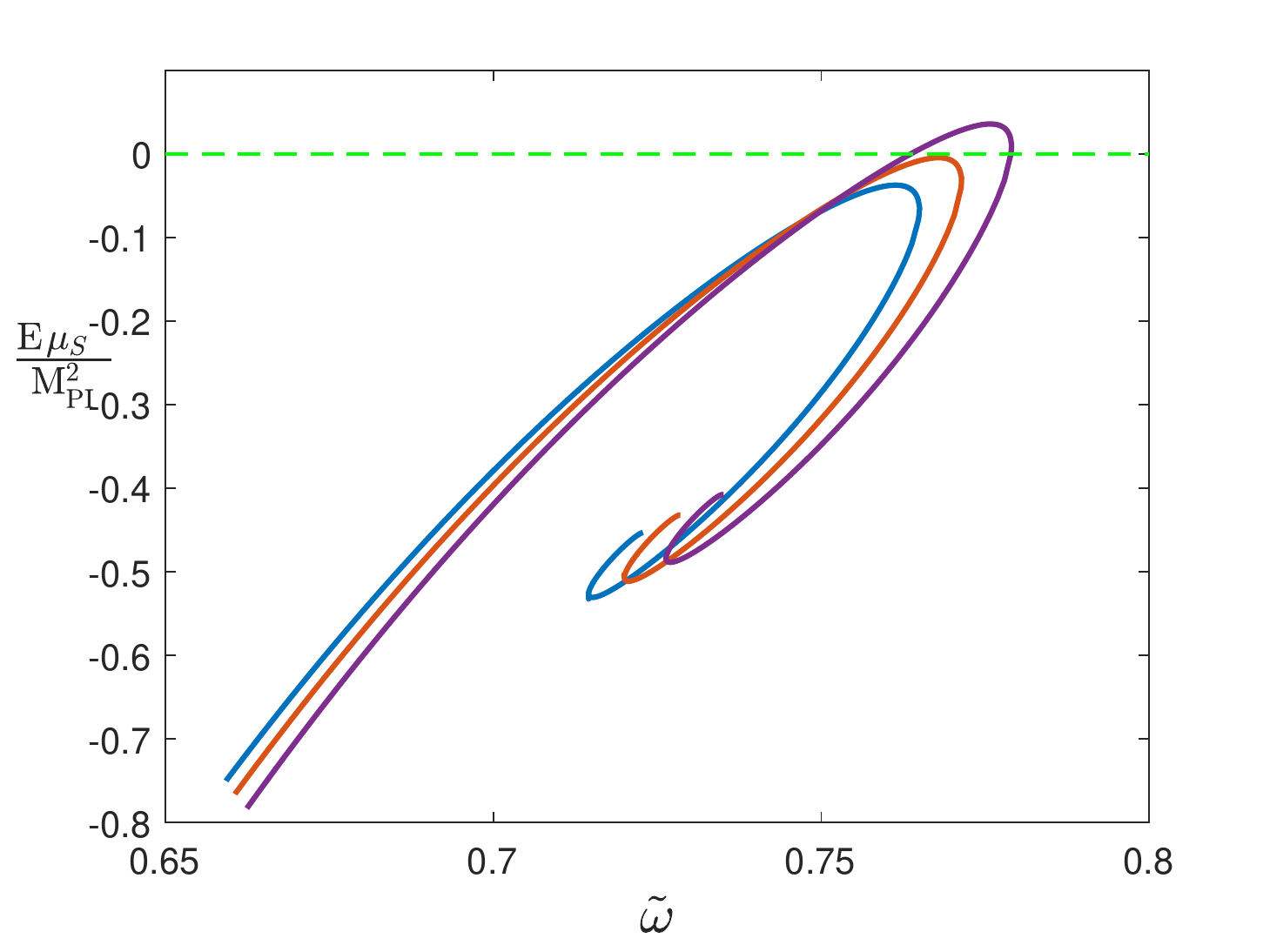}
    \end{center}
    \vspace{-2em}
    \caption{Left: The binding energy $E$ of the one-branch solution family as a 
    function of the synchronized frequency $\tilde{\omega}$ for several values of the Proca field mass 
    $\tilde{\mu}_P$. Right: Same as left panel for the multi-branch solution family. 
    The black line represents the $S_0$ state solutions. All solutions have $\tilde{\mu}_S = 1$. }
    \label{E-synchronized}
\end{figure}

At the end of this section, we discuss the binding energy $E = M-\mu_S Q_S-\mu_P Q_P$ for the six solution families obtained above. The binding energy $E$ of the coexisting state versus the synchronized frequency $\tilde{\omega}$ for several values of the mass $\tilde{\mu}_P$ are presented in Fig.~\ref{E-synchronized}. For the synchronized frequency solution, the binding energy of the one-branch solution increases with the increase of the synchronized frequency, and the mixed state solutions are stable$(E<0)$ at $0. 85\le \tilde{\mu}_P<1$. When $0. 808 < \tilde {\mu} _P < 0. 85 $, a small region of $\tilde {\omega}$ appears unstable solution ($E > 0 $); Two-branch solutions are all stable ($E < 0 $).For multi-branch solutions, when $0. 801 < \tilde{\mu} _P \le 0. 805 $, multi-state solutions are stable ($E < 0 $), when $0. 805 < \tilde{\mu} _P \le 0. 808 $, a small region of $\tilde{\omega}$ appears unstable solutions ($E > 0 $). 

\begin{figure}[!htbp]
    \begin{center}
    \includegraphics[height=.26\textheight]{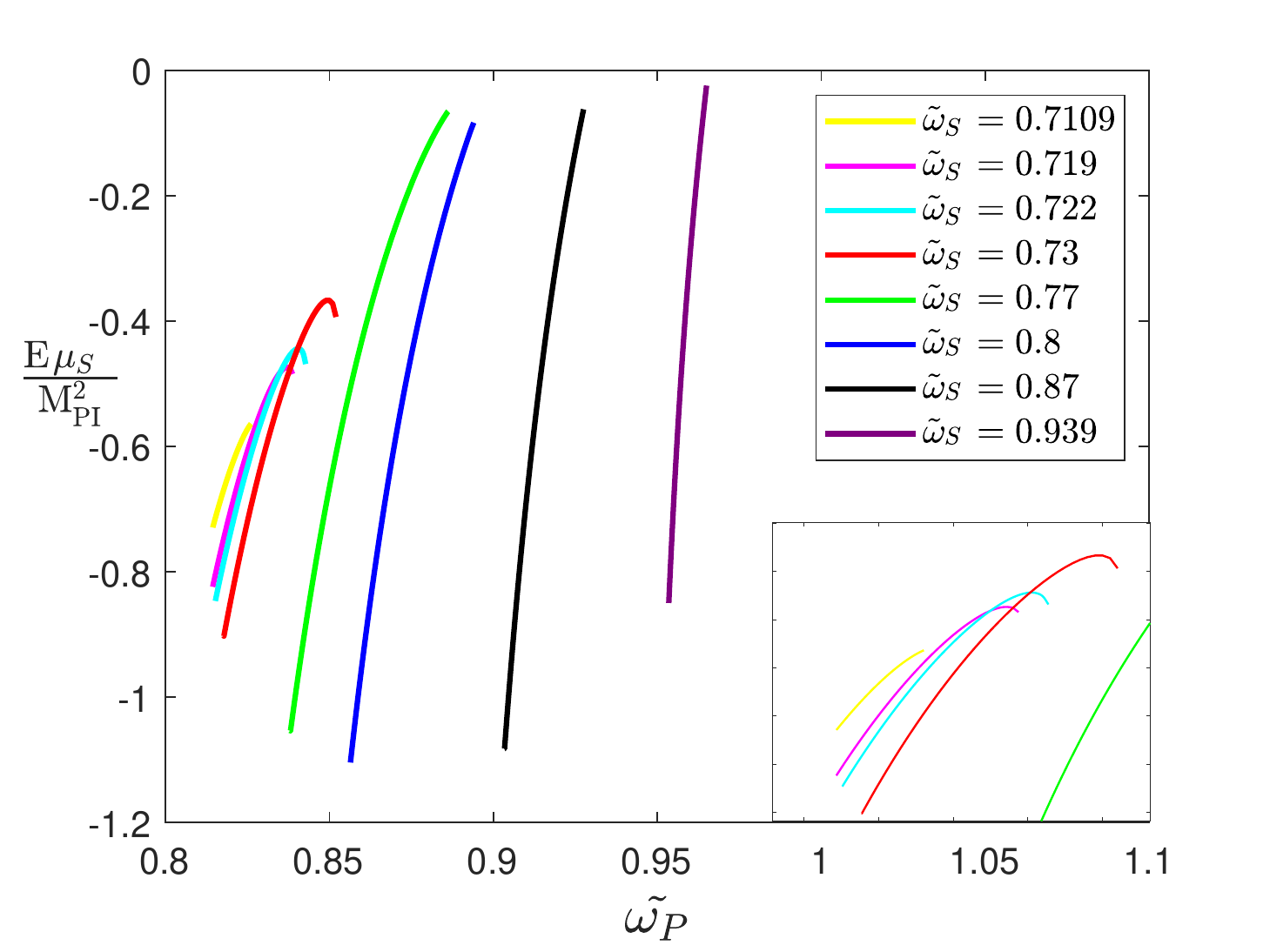}
    \includegraphics[height=.26\textheight]{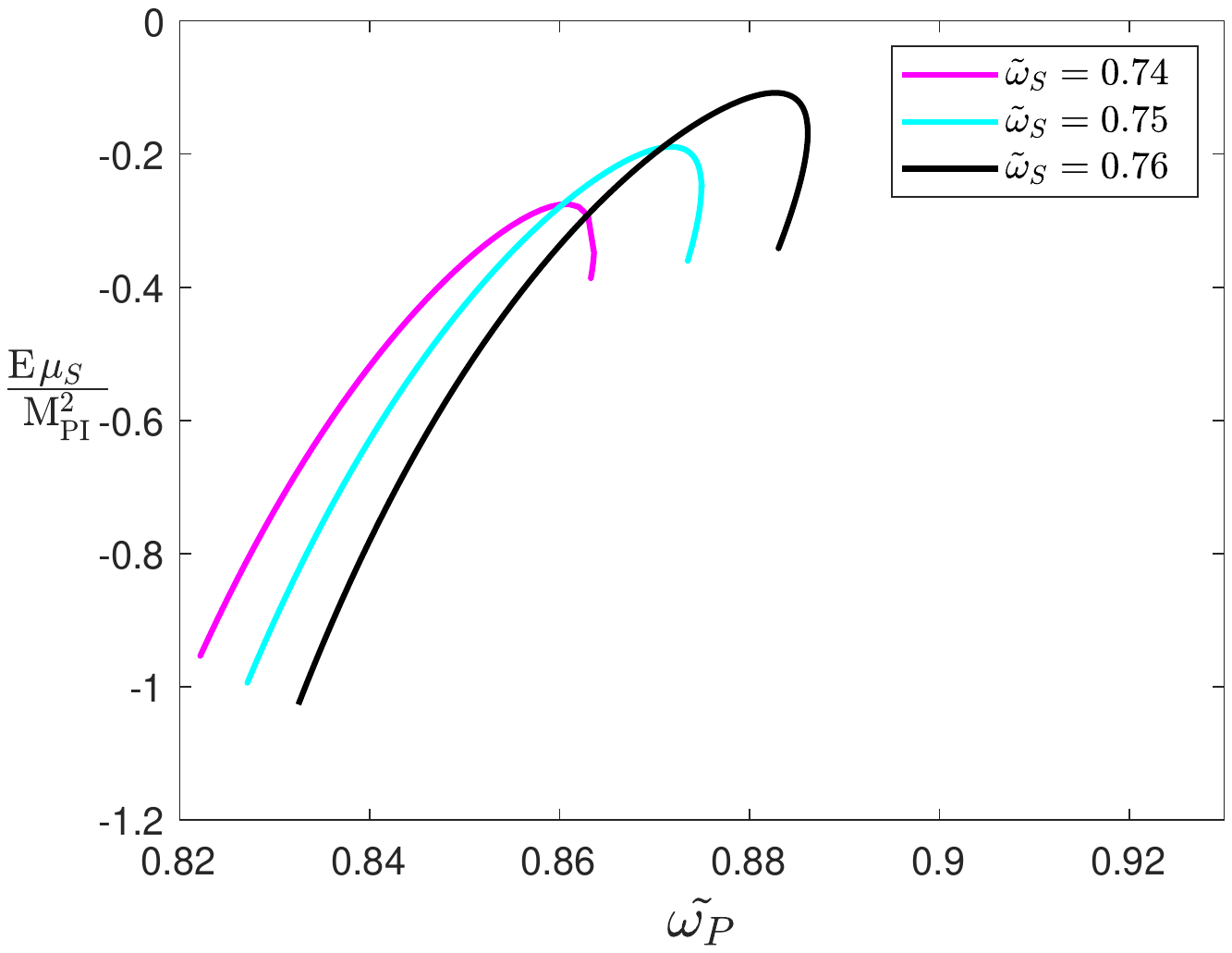}
    \end{center}
    \vspace{-2em}
    \caption{Left: The binding energy $E$ of the one-branch solution family as a 
    function of the nonsynchronized frequency $\tilde{\omega_P}$ for several values of the scalar field's 
    frequency $\tilde{\omega}_S$. Right: Same as left panel for the multi-branch solution family. 
    All solutions have $\tilde{\mu}_S = \tilde{\mu}_P = 1$. }
    \label{E-nonsynchronized}
\end{figure}

The binding energy $E$ of the coexisting state versus the nonsynchronized frequency $\tilde{\omega}_P$ for several values of the scalar field frequency $\tilde{\omega}_S$ are presented in Fig.~\ref{E-nonsynchronized}. For the \textit{one-branch-A} solution family, the binding energy $E$ increases with the increase of nonsynchronized frequency $\tilde{\omega}_P$. For the \textit{multi-branch} solution family and \textit{one-branch-B} solution family, the binding energy $E$ increases with the nonsynchronized frequency $\tilde{\omega}_P$, showing a trend of first increasing and then decreasing. But all three types of solution families are stable ($E<0$). 

\section{Conclusions}\label{Sec5}
In this paper, we present a spherically symmetric multi-field Proca-boson star model that comprises a scalar field in the ground state and a Proca field in the first excited state. We also discuss the properties of different types of mixed state solutions.

For synchronized frequency solutions, we classify the mixed-state solutions into three categories based on the number of branches: one-branch solutions, two-branch solutions, and multi-branch solutions. The field function of the first branch for these three types of solutions changes similarly. As the synchronized frequency $\tilde{\omega}$ increases, both $\left\lvert \tilde{F} \right\rvert _{max}$ and $\left\lvert \tilde{G} \right\rvert _{max}$ decrease while $\left\lvert \tilde{\phi} \right\rvert _{max}$ increases.

In the case of one-branch solutions, the mass $M$ of the mixed state decreases as the synchronized frequency $\tilde{\omega}$ increases, while the binding energy $E$ shows an opposite trend. Moreover, when we decrease the mass of the Proca field $\tilde{\mu}_P$, the existence domain of synchronized frequency solutions gradually expands.

 For the two-branch solution, the field function on the second branch changes with the synchronized frequency $\tilde{\omega}$ as well as the first branch. The image of $M$ of the mixed state with respect to the synchronized frequency $\tilde{\omega}$ is a smooth curve. Starting from a point on the Proca single-field helix corresponding to $\tilde{\mu}_P$, the frequency reaches the maximum value as the inflection point, and then decreases, the second branch appears, and finally returns to the single-field helix of Proca field when the frequency decreases to the minimum value. The $M$ of the first branch increases with the synchronized frequency $\tilde{\omega}$, first decreases and then increases slightly. The second branch is monotonically decreasing (for the critical state of $\tilde{\mu}_P=0.801$, $M$ increases slightly first and then monotonically decreasing). For multi-branch solutions, the image of $M$ with respect to the synchronized frequency $\tilde{\omega}$ becomes a spiral shape. For the first and second branches, $\left\lvert \tilde{F} \right\rvert _{max}$ and $\left\lvert \tilde{G} \right\rvert _{max}$ decrease and $\left\lvert \tilde{\phi} \right\rvert _{max}$ increases with the increase of the synchronized frequency $\tilde{\omega}$; For the third branch, the trend of $\left\lvert \tilde{F} \right\rvert _{max}$ and $\left\lvert \tilde{G} \right\rvert _{max}$ is opposite to that of the first and second branches, while $\left\lvert \tilde{\phi} \right\rvert _{max}$ is the same as that of the first and second branches. For each branch, as Proca field mass $\tilde{\mu}_P$ decreases, the synchronized frequency. The existence domain of $\tilde{\omega}$ increases. In addition, with the decrease of $\tilde{\mu}_P$, $M_{min}$ increases, and $M_{max}$ decreases first and then increases. For the multi-branch family, when $\tilde{\mu}_P\le0. 805$, the solutions are all stable $(E<0)$. when $\tilde{\mu}_P>0. 805$, some of the solutions become unstable $(E>0)$. 

For the case of nonsynchronized frequency, the mixed-state solution family is still divided into three categories, but different from the case of the synchronized frequency, they are one-branch-A solutions, multi-branch solutions and one-branch-B solutions respectively. One-branch-A is similar to the one-branch solution of the synchronized frequency. With the increase of the non-synchronized frequency $\tilde{\omega}_P$, the variation of the existence domain of $\left\lvert \tilde{F} \right\rvert _{max}$, $\left\lvert \tilde{G} \right\rvert _{max}$, $\left\lvert \tilde{\phi} \right\rvert _{max}$, mass $M$ and non-synchronized frequency $\tilde{\omega}_P$ is similar to that of the one-branch solution family of synchronized-frequency case, but the minimum value of $M$ of the non-synchronized frequency depends on $\tilde{\omega}_S$. It is equal to $M$ of the scalar field single field of $\tilde{\mu}_S=1$ when $\tilde{\omega}_S$ takes the same value, the solution is stable no matter what value $\tilde{\omega}_S$ takes $(E<0)$. 
For multi-branch solutions, $\left\lvert \tilde{F} \right\rvert _{max}$ and $\left\lvert \tilde{G} \right\rvert _{max}$ of the first branch first increase and then decrease with the increase of the non-synchronized frequency $\tilde{\omega}_P$, $\left\lvert \tilde{\phi} \right\rvert _{max}$ is monotonically increasing. The $\left\lvert \tilde{F} \right\rvert _{max}$ and $\left\lvert \tilde{G} \right\rvert _{max}$ of the second branch decrease monotonically, and $\left\lvert \tilde{\phi} \right\rvert _{max}$ is the same as that of the first branch, still increasing monotonically. For each branch, the existence domain of the non-synchronized frequency $\tilde{\omega}_P$ decreases as the frequency $\tilde{\omega}_S$ of the scalar field decreases.
 The relation between the mass $M$ and the non-synchronized frequency $\tilde{\omega}_P$ is similar to the two-branch solution in the synchronized-frequency case, but behaves differently at the inflection point. It is not smooth at the inflection point, but appears sharp bending. Finally, the non-synchronized frequency of the second branch is the lowest, and the mixed state curve falls on the Proca field. With the decrease of $\tilde{\omega}_S$ frequency, both $M_{min}$ and $M_{max}$ decrease. 
 The solutions for all branches are stable $(E<0)$. For one-branch-B, with the increase of non-synchronized frequency $\tilde{\omega}_P$, $\left\lvert \tilde{F} \right\rvert _{max}$ and $\left\lvert \tilde{G} \right\rvert _{max}$ increase,  $\left\lvert \tilde{\phi} \right\rvert _{max}$ first increase then decrease, and the mass $M$ of mixed states decreases. When the nonsynchronized frequency reaches the maximum, the curve falls on the Proca field. The existence domain of mixed state non-synchronized frequency $\tilde{\omega}_P$ decreases with the increase of scalar field frequency $\tilde{\omega}_S$. In addition, the solutions of one-branch-B are all stable $(E<0)$. 

The Proca-boson star solutions presented in this paper exhibit several new solution families that differ significantly from those found in previous studies~\cite{liang_dirac-boson_2022,li2020rotating}. While most of the new solutions discovered in~\cite{liang_dirac-boson_2022} are unstable, the majority of the newly found mixed-state solutions composed of a first excited state Proca field and a ground state scalar field are stable. In future research, we plan to investigate mixed stars composed of both fields in the first excited state. Additionally, inspired by~\cite{alcubierre_ell-boson_2018}, we may explore the superposition of multiple matter fields as another possible direction for future study.
\section*{Acknowledgements}
This work is supported by National Key Research and Development Program of China (Grant No. 2020YFC2201503) and the National Natural Science Foundation of China (Grant No.  ~12275110 and 12247101). Parts of computations were performed on the shared memory system at institute of computational physics and complex systems in Lanzhou university. 

\section*{Appendix}\label{appendix1}
\begin{table}[!htbp]
    \centering
  \setlength{\tabcolsep}{5mm}{
      \begin{tabular}{|c|c|c|c|c|c|}
      \hline
       $\tilde{\mu}_P$  & $\tilde{\omega}$  & $M_{max}$  & $M_{min}$  & $E_{max}$  & $\tilde{\omega}_0$\\
      \hline
      $0.81$  & $0.6631\sim0.7704$ & $0.894$ & $0.448$ & $0.03$ & $0.762$ \\
      \hline
      $0.85$  & $0.6923\sim0.7829$ & $1.006$ & $0.565$ & $-0.003$ & \diagbox[width=6em,height=1.8em]{}{}  \\
      \hline
      $0.862$  & $0.705\sim0.7923$ & $1.03$ & $0.584$ & $-0.009$ & \diagbox[width=6em,height=1.8em]{}{}  \\
      \hline
      $0.88$  & $0.7242\sim0.8088$ & $1.074$ & $0.609$ & $-0.015$ & \diagbox[width=6em,height=1.8em]{}{}  \\
      \hline
      $0.91$  & $0.7657\sim0.8425$ & $1.122$ & $0.632$ & $-0.02$ & \diagbox[width=6em,height=1.8em]{}{}  \\
      \hline
      $0.93$  & $0.8002\sim0.8694$ & $1.132$ & $0.63$ & $-0.02$ & \diagbox[width=6em,height=1.8em]{}{}  \\
      \hline
      $0.98$  & $0.923\sim0.955$ & $0.906$ & $0.468$ & $-0.003$ & \diagbox[width=6em,height=1.8em]{}{} \\
      \hline
      $0.999$  & $0.9955\sim0.9976$ & $0.249$ & $0.121$ & $-0.001$ & \diagbox[width=6em,height=1.8em]{}{}  \\
      \hline
      \end{tabular}}
  \caption{The existence domain of the synchronized frequency $\tilde{\omega}$, maximum and minimum ADM masses ($M_{max}$ and $M_{min}$) and maximum and minimum binding energy ($E_{max}$ and $E_{min}$) of the mixed state, for several values of the Proca field mass $\tilde{\mu}_P$. $\tilde{\omega}_0$ representsis the value of the synchronized frequency when $E = 0$ in Fig.~\ref{ADM-one-synchronized}. All solutions have $\tilde{\mu}_S=1$.}
  \label{table1}
\end{table}

\begin{table}[!htbp]
    \centering
  \setlength{\tabcolsep}{5mm}{
      \begin{tabular}{|c|c|c|c|c|c|c|}
      \hline
       $\tilde{\mu}_P$  & $B_1$    & $B_2$    & $B_3$     & $E$   & $\tilde{\omega}_0$\\
      \hline
      $0.802$  & $0.659\sim0.765$ & $0.715\sim0.765$ & $0.715\sim0.723$ &  $-0.75\sim-0.037$ & \diagbox[width=6em,height=1.8em]{}{}\\
      \hline
      $0.805$  & $0.661\sim0.771$ & $0.72\sim0.771$ & $0.72\sim0.729$ & $-0.766\sim-0.005$ & \diagbox[width=6em,height=1.8em]{}{} \\
      \hline
      $0.806$  & $0.661\sim0.774$ & $0.722\sim0.774$ & $0.722\sim0.730$ &  $-0.773\sim0.007$ & $0.766\sim0.773$\\
      \hline
      $0.808$  & $0.662\sim0.778$ & $0.725\sim0.778$ & $0.725\sim0.734$ &  $-0.783\sim0.03$ & $0.764\sim0.778$\\
      \hline
      \end{tabular}}
    \caption{The existence domain of the synchronized frequency $\tilde{\omega}$, ADM mass ($M$) and binding energy ($E$) of the mixed state, for several values of the Proca field mass $\tilde{\mu}_P$. $B_1$, $B_2$ and $B_3$ represent the first, second and third branches of in Fig.~\ref{ADM-multi-synchronized}, respectively. $\tilde{\omega}_0$ representsis the value of the synchronized frequency when $E = 0$ in Fig.~\ref{ADM-multi-synchronized}. All solutions have $\tilde{\mu}_S=1$.}
    \label{table2}
\end{table}

\begin{table}[!htbp]
    \centering
  \setlength{\tabcolsep}{5mm}{
      \begin{tabular}{|c|c|c|c|c|}
      \hline
         $\tilde{\mu}_P$   & $B_1$    & $B_2$    & $M_{max}$  & $M_{min}$ \\
      \hline
      $0.772$  & $0.654\sim0.7042$ & $0.6989\sim0.7042$ & $0.757$ & $0.612$ \\
      \hline
      $0.781$  & $0.654\sim0.722$ & $0.7088\sim0.722$ & $0.793$ & $0.577$ \\
      \hline
      $0.79$  & $0.655\sim0.74$ & $0.7133\sim0.74$ & $0.828$ & $0.536$ \\
      \hline
      $0.795$  & $0.6565\sim0.7503$ & $0.7157\sim0.7503$ & $0.846$ & $0.51$ \\
      \hline
      $0.801$  & $0.6588\sim0.7829$ & $0.7112\sim0.7829$ & $0.866$ & $0.474$ \\
      \hline
      \end{tabular}}
    \caption{The existence domain of the synchronized frequency $\tilde{\omega}$, maximum and minimum ADM masses ($M_{max}$ and $M_{min}$) and binding energy ($E$) of the mixed state, for several values of the Proca field mass $\tilde{\mu}_P$. $B_1$ and $B_2$ represent the first and second of in Fig.~\ref{ADM-two-synchronized}, respectively. $\tilde{\omega}_0$ representsis the value of the synchronized frequency when $E = 0$ in Fig.~\ref{ADM-two-synchronized}. All solutions have $\tilde{\mu}_S=1$.}
    \label{table3}
\end{table}

\begin{table}[!htbp]
    \centering
  \setlength{\tabcolsep}{5mm}{
      \begin{tabular}{|c|c|c|c|}
      \hline
         $\tilde{\omega}_S$   & $\tilde{\omega}_P$    & $M_{max}$  & $M_{min}$ \\
      \hline
      $0.77$  & $0.838\sim0.8862$ & $1.01$ & $0.516$ \\
      \hline
      $0.8$  & $0.844\sim0.8871$ & $1.026$ & $0.557$ \\
      \hline
      $0.87$  & $0.9033\sim0.9275$ & $1.029$ & $0.629$ \\
      \hline
      $0.939$  & $0.9535\sim0.965$ & $0.826$ & $0.526$ \\
      \hline
      \end{tabular}}
      \caption{The existence domain of the nonsynchronized frequency $\tilde{\omega}_P$, maximum and minimum ADM masses ($M_{max}$ and $M_{min}$) of the mixed state, for several values of the scalar field frequencies $\tilde{\omega}_S$ in Fig.~\ref{ADM-one1-nonsynchronized}. All solutions have $\tilde{\mu}_S=\tilde{\mu}_P=1$.}
      \label{table4}
\end{table}

\begin{table}[!htbp]
    \centering
  \setlength{\tabcolsep}{5mm}{
      \begin{tabular}{|c|c|c|c|c|}
      \hline
         $\tilde{\omega}_S$   & $B_1$    & $B_2$    & $M_{max}$  & $M_{min}$ \\
      \hline
      $0.74$  & $0.8222\sim0.8636$ & $0.8633\sim0.8636$ & $0.934$ & $0.538$ \\
      \hline
      $0.75$  & $0.8271\sim0.875$ & $0.8735\sim0.875$ & $0.968$ & $0.505$ \\
      \hline
      $0.755$  & $0.8298\sim0.8805$ & $0.8784\sim0.8805$ & $0.981$ & $0.49$ \\
      \hline
      $0.76$  & $0.8325\sim0.8861$ & $0.883\sim0.8861$ & $0.992$ & $0.474$ \\
      \hline
      \end{tabular}}
    \caption{The existence domain of the nonsynchronized frequency $\tilde{\omega}_P$, maximum and minimum ADM masses ($M_{max}$ and $M_{min}$) of the mixed state, for several values of the scalar field frequencies $\tilde{\omega}_S$. $B_1$ and $B_2$ represent the first and second of in Fig.~\ref{ADM-multi-nonsynchronized}, respectively. All solutions have $\tilde{\mu}_S=\tilde{\mu}_P=1$.}
  \label{table5}
\end{table}
\begin{table}[!htbp]
    \centering
  \setlength{\tabcolsep}{5mm}{
      \begin{tabular}{|c|c|c|c|}
      \hline
         $\tilde{\omega}_S$   & $\tilde{\omega}_P$    & $M_{max}$  & $M_{min}$ \\
      \hline
      $0.7109$  & $0.8144\sim0.8261$ & $0.785$ & $0.671$ \\
      \hline
      $0.719$  & $0.8144\sim0.839$ & $0.85$ & $0.612$ \\
      \hline
      $0.722$  & $0.8152\sim0.8427$ & $0.867$ & $0.598$ \\
      \hline
      $0.73$  & $0.8178\sim0.852$ & $0.904$ & $0.57$ \\
      \hline
      \end{tabular}}
      \caption{The existence domain of the nonsynchronized frequency $\tilde{\omega}_P$, maximum and minimum ADM masses ($M_{max}$ and $M_{min}$) of the mixed state, for several values of the scalar field frequencies $\tilde{\omega}_S$ in Fig.~\ref{ADM-one2-nonsynchronized}. All solutions have $\tilde{\mu}_S=\tilde{\mu}_P=1$.}
  \label{table6}
\end{table}

\bibliography{out.bib}

\begin{thebibliography}{10}

\bibitem{raveri2017partially}
{\sc Raveri, M., Hu, W., Hoffman, T., and Wang, L.-T.}
\newblock Partially acoustic dark matter cosmology and cosmological
  constraints.
\newblock {\em Physical Review D 96}, 10 (2017), 103501.

\bibitem{alcubierre_ell-boson_2018}
{\sc Alcubierre, M., Barranco, J., Bernal, A., Degollado, J.~C., Diez-Tejedor,
  A., Megevand, M., Nunez, D., and Sarbach, O.}
\newblock \${\textbackslash}ell\$-{Boson} stars.
\newblock {\em Classical and Quantum Gravity 35}, 19 (Oct. 2018), 19LT01.
\newblock arXiv:1805.11488 [astro-ph, physics:gr-qc, physics:hep-th].

\bibitem{alcubierre_dynamical_2019}
{\sc Alcubierre, M., Barranco, J., Bernal, A., Degollado, J.~C., Diez-Tejedor,
  A., Megevand, M., Núñez, D., and Sarbach, O.}
\newblock Dynamical evolutions of \${\textbackslash}ell\$-boson stars in
  spherical symmetry.
\newblock {\em Classical and Quantum Gravity 36}, 21 (Nov. 2019), 215013.
\newblock arXiv:1906.08959 [gr-qc].

\bibitem{annulli_response_2020}
{\sc Annulli, L., Cardoso, V., and Vicente, R.}
\newblock The response of ultralight dark matter to supermassive black holes
  and binaries.
\newblock {\em Physical Review D 102}, 6 (Sept. 2020), 063022.
\newblock arXiv:2009.00012 [astro-ph, physics:gr-qc, physics:hep-ph,
  physics:hep-th].

\bibitem{bernal_multi-state_2010}
{\sc Bernal, A., Barranco, J., Alic, D., and Palenzuela, C.}
\newblock Multi-state {Boson} {Stars}.
\newblock {\em Physical Review D 81}, 4 (Feb. 2010), 044031.
\newblock arXiv:0908.2435 [astro-ph, physics:gr-qc].

\bibitem{bezares_gravitational_2018}
{\sc Bezares, M., and Palenzuela, C.}
\newblock Gravitational {Waves} from {Dark} {Boson} {Star} binary mergers.
\newblock {\em Classical and Quantum Gravity 35}, 23 (Dec. 2018), 234002.
\newblock arXiv:1808.10732 [gr-qc].

\bibitem{brito_Proca_2016}
{\sc Brito, R., Cardoso, V., Herdeiro, C. A.~R., and Radu, E.}
\newblock Proca {Stars}: gravitating {Bose}-{Einstein} condensates of massive
  spin 1 particles.
\newblock {\em Physics Letters B 752\/} (Jan. 2016), 291--295.
\newblock arXiv:1508.05395 [astro-ph, physics:gr-qc, physics:hep-th,
  physics:nlin].

\bibitem{bustillo_searching_2022}
{\sc Bustillo, J.~C., Sanchis-Gual, N., Leong, S. H.~W., Chandra, K.,
  Torres-Forne, A., Font, J.~A., Herdeiro, C., Radu, E., Wong, I. C.~F., and
  Li, T. G.~F.}
\newblock Searching for vector boson-star mergers within {LIGO}-{Virgo}
  intermediate-mass black-hole merger candidates, June 2022.
\newblock arXiv:2206.02551 [gr-qc].

\bibitem{carr_primordial_2020}
{\sc Carr, B., and Kühnel, F.}
\newblock Primordial {Black} {Holes} as {Dark} {Matter}: {Recent}
  {Developments}.
\newblock {\em Annual Review of Nuclear and Particle Science 70}, 1 (2020),
  355--394.
\newblock \_eprint: https://doi.org/10.1146/annurev-nucl-050520-125911.

\bibitem{chen_new_2021}
{\sc Chen, J., Du, X., Lentz, E.~W., Marsh, D.~J., and Niemeyer, J.~C.}
\newblock New insights into the formation and growth of boson stars in dark
  matter halos.
\newblock {\em Physical Review D 104}, 8 (Oct. 2021), 083022.
\newblock Publisher: American Physical Society.

\bibitem{croon_boson_2019}
{\sc Croon, D., Fan, J., and Sun, C.}
\newblock Boson {Star} from {Repulsive} {Light} {Scalars} and {Gravitational}
  {Waves}.
\newblock {\em Journal of Cosmology and Astroparticle Physics 2019}, 04 (Apr.
  2019), 008--008.
\newblock arXiv:1810.01420 [astro-ph, physics:gr-qc, physics:hep-ph].

\bibitem{cunha_chaotic_2016}
{\sc Cunha, P. V.~P., Grover, J., Herdeiro, C., Radu, E., Runarsson, H., and
  Wittig, A.}
\newblock Chaotic lensing around boson stars and {Kerr} black holes with scalar
  hair.
\newblock {\em Physical Review D 94}, 10 (Nov. 2016), 104023.
\newblock arXiv:1609.01340 [astro-ph, physics:gr-qc, physics:hep-th].

\bibitem{delgado_rotating_2020}
{\sc Delgado, J. F.~M., Herdeiro, C. A.~R., and Radu, E.}
\newblock Rotating {Axion} {Boson} {Stars}.
\newblock {\em Journal of Cosmology and Astroparticle Physics 2020}, 06 (June
  2020), 037--037.
\newblock arXiv:2005.05982 [gr-qc, physics:hep-th].

\bibitem{delgado_kerr_2021}
{\sc Delgado, J. F.~M., Herdeiro, C. A.~R., and Radu, E.}
\newblock Kerr black holes with synchronised axionic hair.
\newblock {\em Physical Review D 103}, 10 (May 2021), 104029.
\newblock arXiv:2012.03952 [gr-qc, physics:hep-th].

\bibitem{dietrich_full_2019}
{\sc Dietrich, T., Ossokine, S., and Clough, K.}
\newblock Full {3D} {Numerical} {Relativity} {Simulations} of {Neutron} {Star}
  -- {Boson} {Star} {Collisions} with {BAM}.
\newblock {\em Classical and Quantum Gravity 36}, 2 (Jan. 2019), 025002.
\newblock arXiv:1807.06959 [astro-ph, physics:gr-qc, physics:hep-ph].

\bibitem{eby_boson_2016}
{\sc Eby, J., Kouvaris, C., Nielsen, N.~G., and Wijewardhana, L. C.~R.}
\newblock Boson stars from self-interacting dark matter.
\newblock {\em Journal of High Energy Physics 2016}, 2 (Feb. 2016), 28.

\bibitem{finster_particle-like_1999}
{\sc Finster, F., Smoller, J., and Yau, S.-T.}
\newblock Particle-{Like} {Solutions} of the {Einstein}-{Dirac} {Equations}.
\newblock {\em Physical Review D 59}, 10 (Apr. 1999), 104020.
\newblock arXiv:gr-qc/9801079.

\bibitem{garcia_charged_2016}
{\sc García, F., and Landea, I.~S.}
\newblock Charged {Proca} {Stars}.
\newblock {\em Physical Review D 94}, 10 (Nov. 2016), 104006.
\newblock arXiv:1608.00011 [astro-ph, physics:gr-qc, physics:hep-th].

\bibitem{grould_comparing_2017}
{\sc Grould, M., Meliani, Z., Vincent, F.~H., Grandclément, P., and
  Gourgoulhon, E.}
\newblock Comparing timelike geodesics around a {Kerr} black hole and a boson
  star.
\newblock {\em Classical and Quantum Gravity 34}, 21 (Nov. 2017), 215007.
\newblock arXiv:1709.05938 [astro-ph, physics:gr-qc].

\bibitem{guerra_axion_2019}
{\sc Guerra, D., Macedo, C. F.~B., and Pani, P.}
\newblock Axion boson stars.
\newblock {\em Journal of Cosmology and Astroparticle Physics 2019}, 09 (Sept.
  2019), 061--061.
\newblock arXiv:1909.05515 [astro-ph, physics:gr-qc, physics:hep-ph].

\bibitem{harrison_numerical_2002}
{\sc Harrison, R., Moroz, I., and Tod, K.~P.}
\newblock A numerical study of the {Schrodinger}-{Newton} equation 2: the
  time-dependent problem, Aug. 2002.
\newblock arXiv:math-ph/0208046.

\bibitem{hindmarsh_dark_2005}
{\sc Hindmarsh, M., and Philipsen, O.}
\newblock Dark matter of weakly interacting massive particles and the {QCD}
  equation of state.
\newblock {\em Physical Review D 71}, 8 (Apr. 2005), 087302.
\newblock Publisher: American Physical Society.

\bibitem{jetzer_stability_1989}
{\sc Jetzer, P.}
\newblock Stability of charged boson stars.
\newblock {\em Physics Letters B 231}, 4 (Nov. 1989), 433--438.

\bibitem{jetzer_charged_1993}
{\sc Jetzer, P., Liljenberg, P., and Skagerstam, B.-S.}
\newblock Charged {Boson} {Stars} and {Vacuum} {Instabilities}.
\newblock {\em Astroparticle Physics 1}, 4 (Dec. 1993), 429--448.
\newblock arXiv:astro-ph/9305014.

\bibitem{jetzer_charged_1989}
{\sc Jetzer, P., and Van Der~Bij, J.~J.}
\newblock Charged boson stars.
\newblock {\em Physics Letters B 227}, 3 (Aug. 1989), 341--346.

\bibitem{kaup_klein-gordon_1968}
{\sc Kaup, D.~J.}
\newblock Klein-{Gordon} {Geon}.
\newblock {\em Physical Review 172}, 5 (Aug. 1968), 1331--1342.

\bibitem{kleihaus_rotating_2005}
{\sc Kleihaus, B., Kunz, J., and List, M.}
\newblock Rotating {Boson} {Stars} and {Q}-{Balls}.
\newblock {\em Physical Review D 72}, 6 (Sept. 2005), 064002.
\newblock arXiv:gr-qc/0505143.

\bibitem{kling_profiles_2018}
{\sc Kling, F., and Rajaraman, A.}
\newblock Profiles of boson stars with self-interactions.
\newblock {\em Physical Review D 97}, 6 (Mar. 2018), 063012.

\bibitem{kumar_boson_2014}
{\sc Kumar, S., Kulshreshtha, U., and Kulshreshtha, D.~S.}
\newblock Boson {Stars} in a {Theory} of {Complex} {Scalar} {Fields} coupled to
  \${U}(1)\$ {Gauge} {Field} and {Gravity}.
\newblock {\em Classical and Quantum Gravity 31}, 16 (Aug. 2014), 167001.
\newblock arXiv:1605.07210 [gr-qc, physics:hep-th].

\bibitem{li2020rotating}
{\sc Li, H.-B., Sun, S., Hu, T.-T., Song, Y., and Wang, Y.-Q.}
\newblock Rotating multistate boson stars.
\newblock {\em Physical Review D 101}, 4 (2020), 044017.

\bibitem{li_self-interacting_2021}
{\sc Li, H.-B., Zeng, Y.-B., Song, Y., and Wang, Y.-Q.}
\newblock Self-interacting multistate boson stars.
\newblock {\em Journal of High Energy Physics 2021}, 4 (Apr. 2021), 42.
\newblock arXiv:2006.11281 [gr-qc].

\bibitem{liang_dirac-boson_2022}
{\sc Liang, C., Ren, J.-R., Sun, S.-X., and Wang, Y.-Q.}
\newblock Dirac-boson stars, July 2022.
\newblock arXiv:2207.11147 [astro-ph, physics:gr-qc].

\bibitem{mayer_formation_2004}
{\sc Mayer, L., and Wadsley, J.}
\newblock The formation and evolution of bars in low surface brightness
  galaxies with cold dark matter haloes.
\newblock {\em Monthly Notices of the Royal Astronomical Society 347}, 1 (Jan.
  2004), 277--294.



\bibitem{ruffini_systems_1969}
{\sc Ruffini, R., and Bonazzola, S.}
\newblock Systems of {Self}-{Gravitating} {Particles} in {General} {Relativity}
  and the {Concept} of an {Equation} of {State}.
\newblock {\em Physical Review 187}, 5 (Nov. 1969), 1767--1783.

\bibitem{ryan_spinning_1997}
{\sc Ryan, F.~D.}
\newblock Spinning boson stars with large self-interaction.
\newblock {\em Physical Review D 55}, 10 (May 1997), 6081--6091.

\bibitem{sanchis-gual_self-interactions_2022}
{\sc Sanchis-Gual, N., Herdeiro, C., and Radu, E.}
\newblock Self-interactions can stabilize excited boson stars.
\newblock {\em Classical and Quantum Gravity 39}, 6 (Mar. 2022), 064001.
\newblock arXiv:2110.03000 [gr-qc].

\bibitem{schunck_boson_2000}
{\sc Schunck, F.~E., and Torres, D.~F.}
\newblock Boson stars with generic self-interactions.
\newblock {\em International Journal of Modern Physics D 09}, 05 (Oct. 2000),
  601--618.
\newblock arXiv:gr-qc/9911038.

\bibitem{sharma_boson_2008}
{\sc Sharma, R., Karmakar, S., and Mukherjee, S.}
\newblock Boson star and dark matter, Dec. 2008.
\newblock arXiv:0812.3470 [gr-qc].

\bibitem{siemonsen_stability_2021}
{\sc Siemonsen, N., and East, W.~E.}
\newblock Stability of rotating scalar boson stars with nonlinear interactions.
\newblock {\em Physical Review D 103}, 4 (Feb. 2021), 044022.
\newblock arXiv:2011.08247 [astro-ph, physics:gr-qc, physics:hep-th].

\bibitem{silveira_boson_1995}
{\sc Silveira, V., and de~Sousa, C. M.~G.}
\newblock Boson {Star} {Rotation}: {A} {Newtonian} {Approximation}.
\newblock {\em Physical Review D 52}, 10 (Nov. 1995), 5724--5728.
\newblock arXiv:astro-ph/9508034.

\bibitem{swaters_high-resolution_2000}
{\sc Swaters, R.~A., Madore, B.~F., and Trewhella, M.}
\newblock High-{Resolution} {Rotation} {Curves} of {Low} {Surface} {Brightness}
  {Galaxies}.
\newblock {\em The Astrophysical Journal 531}, 2 (Feb. 2000), L107.

\bibitem{torres_supermassive_2000}
{\sc Torres, D.~F., Capozziello, S., and Lambiase, G.}
\newblock Supermassive boson star at the galactic center?
\newblock {\em Physical Review D 62}, 10 (Oct. 2000), 104012.
\newblock Publisher: American Physical Society.

\bibitem{vincent_imaging_2016}
{\sc Vincent, F.~H., Meliani, Z., Grandclément, P., Gourgoulhon, E., and
  Straub, O.}
\newblock Imaging a boson star at the {Galactic} center.
\newblock {\em Classical and Quantum Gravity 33}, 10 (Apr. 2016), 105015.
\newblock Publisher: IOP Publishing.

\bibitem{yoshida_rotating_1997}
{\sc Yoshida, S., and Eriguchi, Y.}
\newblock Rotating boson stars in general relativity.
\newblock {\em Physical Review D 56}, 2 (July 1997), 762--771.
\newblock Publisher: American Physical Society.

\bibitem{zeng_rotating_2021}
{\sc Zeng, Y.-B., Li, H.-B., Sun, S.-X., Cui, S.-Y., and Wang, Y.-Q.}
\newblock Rotating hybrid axion-miniboson stars, Mar. 2021.
\newblock arXiv:2103.10717 [gr-qc, physics:hep-th].

\end{thebibliography}
\bibliographystyle{acm}

\end{document}